\documentclass{article}
\usepackage{bookstyle,bm,cite}
\usepackage{graphicx}
\usepackage{graphics}
\usepackage{cite}
\usepackage[english]{babel}
\usepackage{amsfonts}
\usepackage{amsmath}
\usepackage{bm}
\usepackage{feynmf}
\newcommand{\mrm}[1]{\mathrm{#1}}
\newcommand{\lb}{\left(}
\newcommand{\rb}{\right)}
\newcommand{\be}{\begin{equation}}
\newcommand{\ee}{\end{equation}}
\newcommand{\ba}{\begin{eqnarray}}
\newcommand{\ea}{\end{eqnarray}}
\newcommand{\ie}{\textsl{i.e., }}

\newcommand{\eg}{\textsl{e.g., }}
\let\bauthor\relax
\let\fnm\relax\let\snm\relax
\let\bseries\ignorespaces
\let\btitle\relax
\let\bvolumeno\textbf
\def\bdate#1{\unskip\ (#1)}
\def\bfirstpage#1{\unskip\ #1}

\begin{document}
\author{Thibault Damour${}^1$  and Marc Lilley${}^2$}
\address{${}^1$ Institut des Hautes Etudes Scientifiques, 35 route de Chartres, F-91440 Bures-sur-Yvette, France\\
${}^2$ Institut d'Astrophysique de Paris,
98, bis Blvd. Arago, F-75014 Paris, France}
\photo{}
\title{String theory, gravity and experiment}
\runauthor{T. Damour and M. Lilley}
\title{String theory, gravity and experiment}
\frontmatter
\maketitle
\mainmatter
\maketitle

\begin{abstract}
The aim of these lectures is to give an introduction to several topics which lie at the intersection of 
string theory, gravity theory and gravity phenomenology.  One successively reviews: (i) the ``membrane'' 
approach to the dissipative dynamics of classical black holes, (ii) the current experimental tests of gravity, 
and their theoretical interpretation, (iii) some aspects of the string-inspired phenomenology of the gravitational 
sector, and (iv) some possibilities for observing string-related signals in cosmology (including a  discussion of 
gravitational wave signals from cosmic superstrings).
\end{abstract}

\section{Introduction}

The common theme of these lectures is {\it gravity}, and their aim is to discuss a few cases where string theory 
might have an interesting interplay  either with gravity theory, or with gravity phenomenology. 
We shall discuss the following topics:

\begin{itemize}
\item {\it Classical black holes as dissipative branes.}
 The idea here is to review the ``classic'' work on black holes of the seventies  which led to the picture of black 
holes as being analog to dissipative branes endowed with finite electrical resistivity, and finite surface viscosity. 
In particular, we shall review the derivation of the (classical) surface viscosity of black holes, which has recently 
acquired a new (quantum) interest in view of AdS/CFT duality.

\item {\it Hawking radiation from black holes.}
To complete our classical account of irreversible properties of black holes, we shall also give a direct derivation 
of the phenomenon of Hawking radiation, because of its crucial importance in fixing the coefficient between the area 
of the horizon and black hole ``entropy''.

\item  {\it Experimental tests of gravity.}
Before discussing possible phenomenological consequences of string theory in the gravitational sector, we find useful 
to summarize the present status of experimental tests of gravity, as well as the theoretical frameworks used to 
interpret them. In particular, we emphasize that binary pulsar experiments have already given us accurate tests 
of some aspects of strong-field (and radiative) relativistic gravity.

\item {\it String-inspired phenomenology of the gravitational sector.}
In this section we shall discuss (without any attempt at completeness) some of the ideas that have been suggested 
about observable signals possibly connected to string theory. In particular, we shall discuss 
the {\it cosmological attractor mechanism} which leads to a rather rich gravitational phenomenology that will be 
probed soon by various gravitational experiments.

\item {\it String-related signals in cosmology.}
After discussing a few alternatives to slow-roll inflation (and the possible relaxation of the Lyth bound when using 
non-linear kinetic terms for the inflaton), one discusses in some detail {\it cosmic superstrings}. We explain, 
in particular, how one computes the {\it gravitational wave burst signal} emitted by the cusps that periodically 
form during the dynamical evolution of generic string loops.
\end{itemize}
\vspace{0.25cm}
A final warning: by lack of time (and energy), no attempt has been made to give exhaustive and fair references to 
original and/or relevant work. The given references are indicative, and should be viewed as entry points into the 
relevant literature. With the modern, web-based, easy access to the scientific literature it is hoped that the 
reader will have no difficulty in using the few given references as starting points for an instructive navigation 
on the vast sea of the physics literature.
\section{Classical black holes as dissipative branes}

Early work on  (Schwarzschild, Reissner-Nordstr\"om, or Kerr-Newman) black holes (BHs) in the 1950's and 1960's  
treated them as {\it passive objects}, \ie as given geometrical backgrounds (and potential wells).  This viewpoint 
changed in the early 1970's when the study of the {\it dynamics} of BHs was initiated by Penrose \cite{P1969}, 
Christodoulou and Ruffini \cite{C1970,CR1971}, Hawking \cite{H1971}, and Bardeen, Carter and Hawking \cite{BCH1973}.  
In the works \cite{P1969,C1970,CR1971,H1971,BCH1973}, only the {\it global dynamics} of BHs was considered, \ie 
their total mass, their total angular momentum, their total irreducible mass, and the variation of these quantities. 
This viewpoint further evolved in the works of Hartle and Hawking \cite{HH1972}, Hanni and Ruffini \cite{HR1973}, 
Damour \cite{D1978,D1979,D1982}, and Znajek \cite{Z1978}, which studied the {\it local dynamics of BH horizons}. 
In this new approach (which was later called the ``membrane paradigm'' \cite{Thorne:1986iy}) a BH horizon is  
interpreted as a brane with dissipative properties, such as, for instance, an electrical resistivity $\rho$, 
equal to $377$ Ohms \cite{D1978,Z1978} independently of the type of BH, and a surface (shear) viscosity, equal 
to $\eta=\frac{1}{16 \pi}$ \cite{D1979,D1982}.  When divided by the entropy density found by Hawking
($S/A = \frac{1}{4}$), the latter shear viscosity 
yields the ratio $\frac{1}{4 \pi}$, a result which has recently raised a renewed interest in connection with 
AdS/CFT, through the work of Kovtun, Son, and Starinets \cite{KSS2005,Son:2007vk}.

\subsection{Global properties of black holes}
\label{GlobalPropertiesBH}

\begin{figure}
\includegraphics[width=10cm]{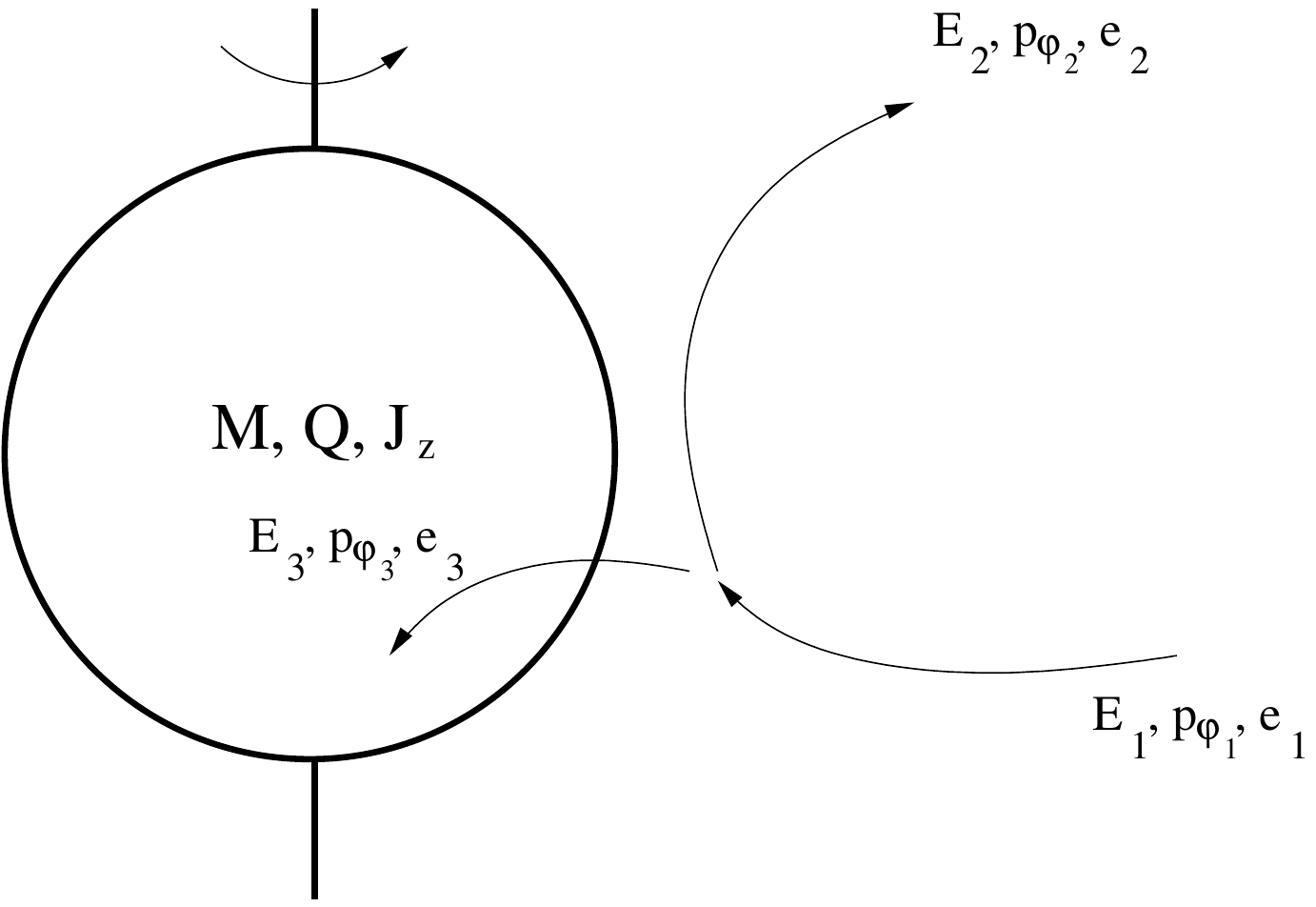}
\caption{In this figure, we schematically illustrate the ``Penrose process'', \ie the 
splitting of an ingoing particle into one that falls into the BH and 
another that exits at infinity.}
\label{BlackHoleParticles}
\end{figure}

Let us start by reviewing the study of the {\it global dynamics} of BHs.  Initially, BHs were thought of as given 
geometrical backgrounds.  In the case of a spherically symmetric object of mass $M$ without any additional attribute, 
Schwarzschild derived the first exact solution of  Einstein's equations only a few weeks after Einstein had obtained 
the final form of the field equations.  Schwarzschild's solution is as follows.  In $3+1$ dimensions, setting $G=c=1$, 
the metric for a spherically symmetric background can be written in the form
\be
\mrm{d}s^2=-A(r)\mrm{d}T^2+B(r)\mrm{d}r^2+r^2
\lb\mrm{d}\theta^2+\mathrm{sin}^2\theta \mrm{d}\varphi^2 \rb
\ee
where $T$ denotes the usual Schwarzschild-type time coordinate, and where the coefficients $A \lb r \rb$ and $B \lb r \rb$ read
\be
\begin{array}{ccc}
A(r) & = & 1-\frac{2GM}{r},\\
B(r) & = & \frac{1}{A(r)}.
\end{array}
\ee
This result was generalized in independent works by Reissner, and by Nordstr\"om (1918) for electrically charged spherically 
symmetric objects, in which case $A \lb r \rb$ and $B \lb r \rb$ are given by
\be
\begin{array}{ccc}
A(r) & = & 1-\frac{2M}{r}+\frac{Q^2}{r^2},\\
B(r) & = & \frac{1}{A(r)}.
\end{array}
\ee
We shall not review here the long historical path which led to interpreting the above solutions, as well as their later 
generalizations due to Kerr (who added to the mass $M$ the spin $J$\footnote{Note that in the case of a spinning BH, 
one often introduces the useful quantity $a=J/M$, \ie the ratio of the total angular momentum to the mass of the BH, 
which has the dimension of length.}), and  Newman {\it et al.} (mass, spin and charge), as BHs.
Up to the 1960's  BHs were viewed only as {\it passive} gravitational wells. For instance, one could think of  
adiabatically lowering a small mass $m$ at the end of a string until it disappears within the BH, thereby converting 
its mass-energy   $m c^2$ into work. More realistically, one was thinking of matter orbiting a BH and radiating away 
its potential energy (up to a maximum, given by the binding energy of the last stable circular orbit around a BH).  
This viewpoint changed in the 1970's, when BHs started being considered as {\it dynamical} objects, able to exchange 
mass, angular momentum and charge with the external world.  Whereas in the simplest case above, one uses the attractive 
potential well created by the mass $M$ without extracting energy from the BH,   Penrose \cite{P1969}, showed that energy 
could in principle be {\it extracted} from a BH itself by means of what is now called a (gedanken) ``Penrose process''
(see FIG. \ref{BlackHoleParticles}).  
Namely, if one considers a time-independent background and a BH that is more complicated than Schwarzschild's, say a Kerr 
BH,  one may  extract energy using a test particle $1$ coming in from infinity with energy $E_1$, angular momentum $p_{\varphi_1}$, 
and electric charge $e_1$.  By Noether's theorem, the time-translation, axial and $U(1)$ gauge  symmetries of the background 
guarantee the conservation of $E$, $p_\varphi$ and $e$ during the ``fall'' of the test particle.  Moreover, if, in a quantum 
process, the test particle $1$ splits, near the BH, into two particles $2$ and $3$, with $E_2$, $p_{\varphi_2}$, $e_2$, and $E_3$, 
$p_{\varphi_3}$, $e_3$ respectively,  then, under certain conditions, one finds that particle 3 can be absorbed by the BH, and 
that particle 2 {\it may come out at infinity with more energy than the incoming particle 1}. 
A detailed analysis of the efficiency of such gedanken Penrose processes by Christodoulou and Ruffini \cite{C1970,CR1971} then 
led to the understanding of the existence of a fundamental {\it irreversibility } in BH dynamics, and to the discovery of the 
BH {\it mass formula}. Let us explain these results.

\begin{figure}
\includegraphics[width=9cm]{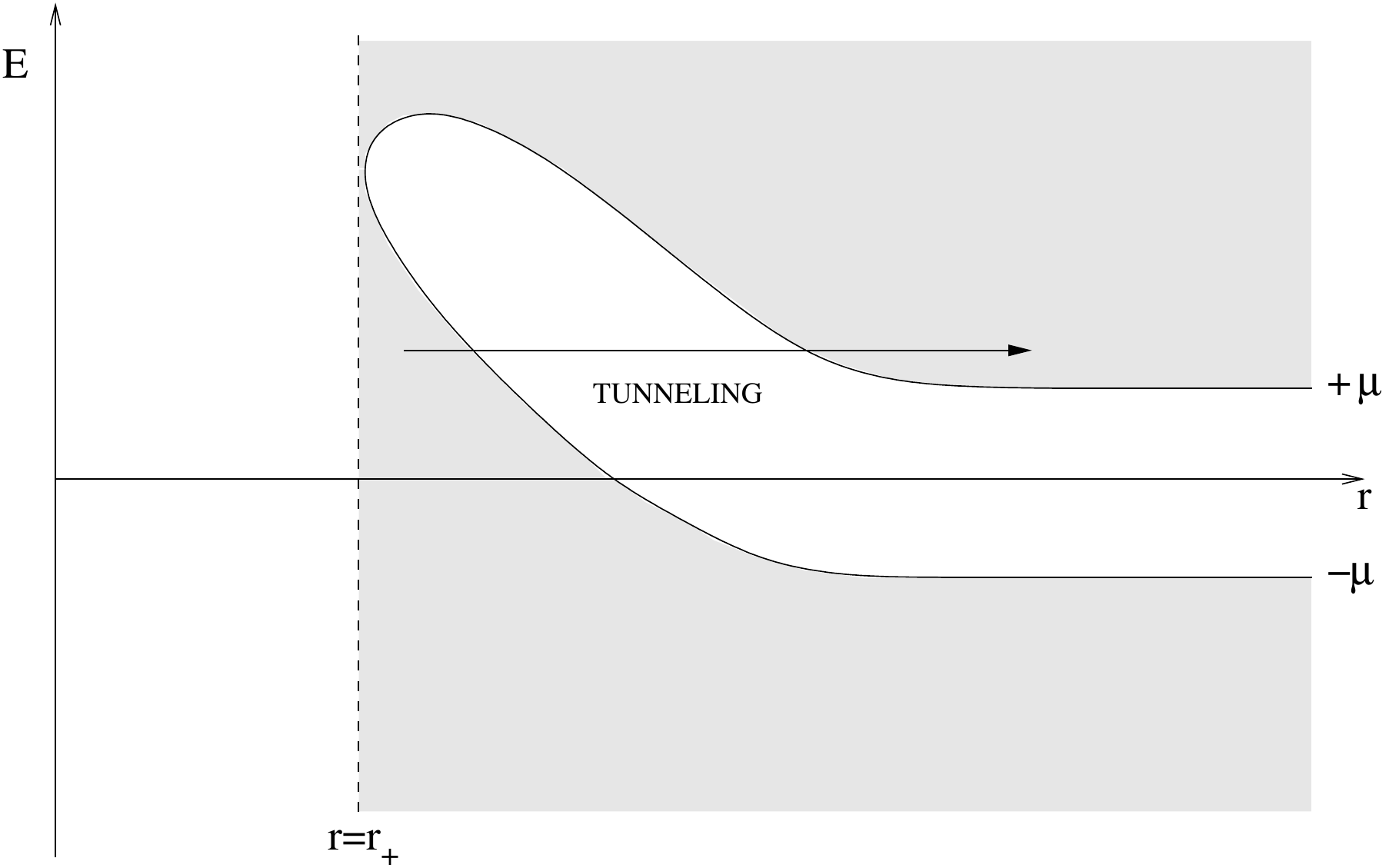}
\caption{This figure depicts the classically allowed energy levels (shaded region) as a function of radius,
for test particles in the neighborhood of a BH.  There exist positive- and negative-energy solutions, corresponding 
(after second quantization) to particles and anti-particles. Classically (as in the Penrose process) one should consider
only the ``positive-square-root'' energy levels, located in the upper shaded region. The white region is classically
forbidden. Note the possibility of tunneling (this corresponds to particle creation via the ``super-radiant'', non-thermal
mechanism briefly mentioned below).}
\label{AntiParticles}
\end{figure}

The basic idea is to explore the physics of BHs through a sequence of infinitesimal changes of their state obtained by injecting 
in them some test particles.  One starts by writing that the total mass-energy, spin and charge of the BH change, by absorption 
of particle 3, as
\be
\begin{array}{ccccc}
\delta M & = & E_3 & = & E_1-E_2,\\
\delta J & = & J_3 & = & J_1-J_2,\\
\delta Q & = & e_3 & = & e_1-e_2.
\end{array}
\ee
This preliminary result can be further exploited by making use of the Hamilton-Jacobi equation.  Considering an on-shell particle 
of mass $\mu$, and adopting the $\left(-+++ \right)$ signature, the Hamilton-Jacobi equation reads
\be g^{\mu \nu}\left(p_{\mu}-eA_{\mu}\right)
\left(p_{\nu}-eA_{\nu}\right)=- \mu^2,
\label{HJEq}
\ee
in which $p_{\mu}=\partial{S}/ \partial{x^{\mu}}$, $S$ is the action and the partial derivatives are taken w.r.t. the coordinates 
$x^{\mu}$.  The details of the splitting process will be irrelevant, as only particle 3 matters in the calculation. In an 
axially symmetric and time-independent background, $S$ can be taken as a linear function of $T$ and $\varphi$,
\be
S=-ET+p_{\varphi}\varphi+S \left(r,\theta \right).
\ee
where $E=-p_T=-p_0$ is the conserved energy, $p_{\varphi}$ is the conserved $\varphi$-component of angular momentum and the last 
term is the contribution from terms that depend on the angle $\theta$ and on the radial distance $r$.  Let us consider the case 
of a Reissner-Nordstr\"om BH, where calculations are easier: the inverse metric is easily computed and (\ref{HJEq}) can then be 
written explicitely as
\be
-\frac{1}{A(r)}\left(p_0-eA_0\right)^2+A(r)p_r^2+\frac{1}{r^2}\left(p_{\theta}^2+\frac{1}{\mrm{sin}^2
\theta} p_{\varphi}^2\right)=-\mu^2
\ee
which we re-write as
\be
\left(p_0-e A_0\right)^2=A(r)^2p_r^2+A(r)\left( \mu^2+\frac{L^2}{r^2}
\right)
\ee
The electric potential is $- A_0=+V =+Q/r$. The above expression is quadratic in $E$ (it is the generalization of the famous
flat-spacetime $E^2 = \mu^2 + {\bf p}^2$) and one finds  two possible solutions for the energy as a function of momenta 
and charge (see  FIG. \ref{AntiParticles}):
\be
E=\frac{eQ}{r}\pm\sqrt{A(r)^2p_r^2+A(r)\left(\mu^2+\frac{L^2}{r^2}\right)}.
\ee
In flat space, $A\lb r \rb=1$, so that, if we ignore charge, we recover the usual Dirac dichotomy on the choice of the $+$ or 
$-$ sign between particle and antiparticle: $E=\pm \sqrt{\mu^2 + {\bf p}^2 }$.  This shows that one 
should take the {\it plus sign} in the equation above.  We remind the reader that for a charged BH, there exists a regular 
horizon only if $Q<M$ (which can be interpreted as a BPS bound). [We have set $G=1$]. Remembering that $A(r)=1-2M/r+Q^2/r^2$, 
there exists both an outer and an inner horizon defined by $r_{\pm}=M \pm \sqrt{M^2-Q^2}$ (which are the two roots of $A(r) =0$).  
The horizon of relevance for BH physics is the outer one $r_+=M+\sqrt{M^2-Q^2}$ (it gives the usual result $2M$ when $Q=0$).  
As particle 3 is absorbed by the BH, we can compute its (conserved) energy when it crosses the horizon, \ie in the limit where 
the radial coordinate $r$ is equal to $r_+$. This simplifies the expression of $E_3$ to
\be
E_3=\frac{e_3 \, Q}{r_+}+|p^r|,
\ee
where we have introduced the contravariant component $p^r=g^{rr}p_r=A(r)p_r$, which has a finite limit on the horizon. Note the 
presence of the {\it absolute value } of $p^r$ (coming from the limit of a positive square-root).  The change in the mass of 
the BH is equal to the energy $E_3$ of the particle absorbed, \ie particle $3$.  Using $e_3 = \delta Q$, this yields
\be
\delta M=\frac{Q \delta Q}{r_+(M,Q)}+|p^r|.
\ee
{}From the positivity of $|p^r|$ we deduce that
\be
\delta M \ge \frac{Q \delta Q}{r_+(M,Q)}.
\label{RNIrreversibilityEq}
\ee
We have derived an {\it inequality} and have thereby demonstrated (by following Christodoulou and Ruffini) the irreversibility 
property of BH energetics.  There exist two types of processes, the {\it reversible} ones with an `$=$' sign in 
(\ref{RNIrreversibilityEq}), and the {\it irreversible} ones with an `$>$' sign. The former ones are  reversible because 
if a BH first absorbs a particle of charge $+e$ with vanishing $|p^r|$ (thereby changing its mass by $\delta' M = e Q/r_+(M,Q)$ 
and its charge by $\delta' Q=e$), and then a particle of charge $-e$ with vanishing $|p^r|$ (thereby changing its mass by 
$\delta'' M = -e Q/r_+(M,Q)$ and its charge by $\delta'' Q=-e$), it will be left, at the end, in the same state as the original 
one (with mass $M + \delta' M + \delta'' M = M$ and charge $Q + \delta' Q + \delta'' Q = Q$).  Evidently, such reversible 
transformations are delicate to perform, and one expects that irreversibility will occur in most BH processes. The situation 
here is clearly similar to the relation between reversible and irreversible processes in thermodynamics.

The same computation as for the Reissner-Nordstr\"om BH can be performed for the Kerr-Newman BH. One obtains in that case, 
by a slightly more complicated calculation,
\be
\delta M - \frac{a\delta J+r_+Q\delta Q}{r_+^2+a^2}=\frac{r_+^2+a^2 \mrm{cos}^2 \theta}{r_+^2+a^2}|p^r|.
\label{KNIrreversibilityEq}
\ee
in which $r_+(M,J,Q)=M+\sqrt{M^2-Q^2-a^2}$. We recall that $a=J/M$, and that one has the bound $Q^2+(J/M)^2 \le M^2$.

The idea now is to consider an infinite sequence of infinitesimal reversible changes ({\it i.e.}, $p^r \rightarrow 0$), and 
to study the BH states which are reversibly connected to some initial BH state with given mass $M$, angular momentum $J$ and 
charge $Q$. This leads to a  partial differential equation for $\delta M$,
\be
\delta M=\frac{a \delta J+r_+Q \delta Q}{r_+^2+a^2},
\ee
which is found to be integrable. Integrating it, one finds the
Christodoulou-Ruffini mass formula \cite{CR1971}
\be
M^2=\left(M_{\rm irr}+\frac{Q^2}{4 M_{\rm irr}}\right)^2+\frac{J^2}{4
M_{\rm irr}^2}.
\ee
Here the {\it irreducible mass} $M_{\rm irr}=\frac{1}{2}\sqrt{r_+^2+a^2}$ appears as an integration constant. The mass squared 
thus appears as a function of three contributions, with one term containing the square of the sum of the irreducible mass 
and of the Coulomb energy, and the other one containing the rotational energy. Inserting this expression into 
Eq.~(\ref{KNIrreversibilityEq}), one finds 
\be
\delta M_{\rm irr}\ge 0
\ee
with $\delta M_{\rm irr}=0$ under reversible transformations and $\delta M_{\rm irr} > 0$ under irreversible transformations. 
The irreducible mass $M_{\rm irr}$ can only increase or stay constant. This behaviour is certainly reminiscent of the second 
law of thermodynamics.  The free energy of a BH is therefore $M-M_{\rm irr}$, \ie this is the maximum extractable energy. In 
this view, BHs are no longer passive geometrical backgrounds but contain stored energy that can be extracted. Actually, 
the stored energy can be enormous because a BH can store up to 29 \%  of its mass as rotational energy, and up to 50 \% as Coulomb energy!

The irreducible mass is related to the area of the horizon of the BH, by $A=16 \pi M_{\rm irr}^2$ so that in a reversible process 
$\delta A = 0$, while in an irreversible one $\delta  A > 0$. Hawking showed \cite{H1971} that this irreversible evolution of 
the area of the horizon was a general consequence of Einstein's equations, when assuming the weak energy condition.  He also 
showed that in the merging of two BHs of area $A_1$ and $A_2$, the total final area satisfied $A_{\mrm{tot}}\ge A_1+A_2 $.  

 Such results evidently evoque the second law of thermodynamics.  The analog of the first law  
[$ \mrm{d}E(S,\mrm{extensive\; parameters})=\mrm{d}W+\mrm{d}Q$, where the work $\mrm{d}W$ is linked to the variation of extensive 
parameters (volume, etc.) and where $\mrm{d}Q = T \mrm{d}S$ is the heat exchange] reads, for BH processes,
\be
\mrm{d}M\left(Q,J,A\right)=V\mrm{d}Q+\Omega \mrm{d}J+\frac{g}{8
\pi}\mrm{d}A.
\label{FirstLawTD}
\ee
Comparing this result with expression (\ref{KNIrreversibilityEq}), one has
\be
\begin{array}{ccc}
\displaystyle V & \displaystyle = & \displaystyle \frac{Qr_+}{r_+^2+a^2},\\
\displaystyle \Omega & \displaystyle = & \displaystyle \frac{a}{r_+^2+a^2},
\end{array}
\ee
and
\be
g=\frac{1}{2}\frac{r_+-r_-}{r_+^2+a^2},
\ee
which, in the Kerr-Newman case, is given by
\be
g=\frac{\sqrt{M^2-a^2-Q^2}}{r_+^2+a^2}.
\ee
$V$ is interpreted as the electric potential of the BH, and $\Omega$ as its angular velocity. Expression (\ref{FirstLawTD}) 
resembles the usual form of the first law of thermodynamics in which the area term has to be interpreted as some kind of 
entropy. The parameter $g$ is called the ``surface gravity''. [In the Schwarschild case, it reduces to $M/r_+^2$ 
(in $G=1$ units), \ie the usual formula for the surface gravitational acceleration $g=G M/R^2$.].  In the Les Houches 
Summer School of 1972, a more general version of the first law was derived, that included the presence of matter around 
the BH, and energy exchange  \cite{BCH1973}. An analog of the  zeroth law  was also derived \cite{C1973}, in the sense 
that the surface gravity $g$ (which is analog to the temperature) was found to be uniform on the surface of a BH in equilibrium.  

In 1974,  Bekenstein went further in taking seriously (and no longer as a simple analogy) the thermodynamics of BHs.  
First, note that one can write the formal BH ``heat exchange'' term in various ways
\be
\mrm{d}Q=T\mrm{d}S=\frac{g\mrm{d}A}{8 \pi}=4 g M_{\mrm{ irr}}\mrm{d}M_{\mrm{ irr}}.
\ee
In light of this, is the appropriate physical analog of the entropy the irreducible mass or the area of a BH?  Is the 
analog of temperature proportional to the surface gravity $g$ or to the product $M_{\mrm{irr}}\,g$? Can one give a physical 
meaning to the temperature and entropy of a BH ?  To address such questions, Bekenstein used several different approaches.

In particular, he used Carnot-cycle-type arguments.  For instance, one may extract work from a BH by slowly lowering into it a box 
of radiation of infinitesimal size.  In fact, in this ideal case, one can theoretically convert all the energy of the box 
of radiation, $mc^2$, into work.  The efficiency of Carnot cycles is defined in terms of both a hot and a cold source as
\be
\eta=1-\frac{T_{\mrm{cold}}}{T_{\mrm{hot}}}.
\ee
{}From what we just said, it would seem that the efficiency of classical BHs as thermodynamic engines is 100\%, $\eta=1$. 
This would then correspond to a BH temperature (= the cold source) $T_{\mrm{BH}}=T_{\mrm{cold}}=0$. The point made by Bekenstein was
 that this classical result will be modified by quantum effects. Indeed, one expects (because of the uncertainty principle) that
 a box of thermal radiation at temperature $T$ (made of typical wavelengths $\lambda \sim 1/T$) cannot be made infinitesimally 
small, but will have a minimum finite size $\sim \lambda$. From this limit on the size of the box, Bekenstein then deduced an 
upper bound on the efficiency $\eta$, and therefore a lower bound on the BH temperature  $T_{\mrm{BH}} \ne 0$.

Let us indicate another reasoning (of Bekenstein) which suggests that the absorption of a single particle by a BH augments its surface by a finite 
amount proportional to $\hbar$.  As we said above the change of BH energy as it absorbs a particle is
(when $a=0$, for simplicity)
\be
E_3=\frac{eQ}{r_+}+\mrm{lim}_{r\rightarrow r_+}|p^r|
\ee
We also showed that the transformation will be reversible (\ie will {\it not} increase the surface area of the BH) only if 
$\mrm{lim}_{r\rightarrow r_+}|p^r| = 0$. However, for this to be true {\it both} the (radial) position and the (radial) 
momentum of the particle must be exactly fixed: namely, $r=r_+$ and $p^r=0$.  This would clearly be in contradiction with 
the Heisenberg uncertainty principle.  Technically, we must consider the conjugate momentum to the position $r$ which is 
the {\it covariant} component $p_r$ of the radial momentum (instead of the {\it contravariant} component $p^r$  used in the 
equation above).  The uncertainty relation therefore reads
\be
\delta r \delta p_r \geq \frac{1}{2} \hbar.
\ee
Near the horizon (i.e. when $\delta r \equiv r - r_+$ is small), the contravariant radial momentum reads (using $g^{rr} = 1/g_{rr}= A(r)$)
\be
\begin{array}{ccc}
\displaystyle p^r & \displaystyle = & \displaystyle A(r) p_r\\
\displaystyle & \displaystyle = & \displaystyle \frac{\left(r-r_+\right)\left(r-r_-\right)}{r^2}p_r\\
\displaystyle & \displaystyle \simeq & \displaystyle \delta r  \frac{\left(r_+-r_-\right)}{r_+^2}p_r\\
\displaystyle & \displaystyle \simeq & \displaystyle \left(\frac{\partial A}{\partial r}\right)_{r_+}\delta r p_r,
\end{array}
\ee
so that Heisenberg's uncertainty relation yields a {\it lower bound} for $p^r$.  We can reexpress this lower bound in terms 
of the BH surface gravity $g$ introduced above by noting that the partial derivative of $A$ w.r.t. $r$, 
$\left(\frac{\partial A}{\partial r}\right)_{r_+}$, entering the last equation, is proportional to $g$: 
\be
\left(\frac{\partial A}{\partial r}\right)_{r_+}=2g.
\ee
This then gives
\be
p^r \simeq 2 g \delta r \delta p_r \geq g \hbar
\ee
>From the relation $\delta M=\frac{Q \delta Q}{r_+\left(M,Q\right)}+|p^r|_{r_+}$, we finally obtain
\be
\delta M-\frac{Q \delta Q}{r_+}=|p^r|\geq g \hbar,
\ee
which can be rewritten as
\be
\delta A \geq 8 \pi \hbar.
\ee
In other words, quantum mechanics tells us that when one lets a particle fall into a BH, one cannot do so in a 
perfectly reversible way. The area must increase by a quantity of order $\hbar$.  If (still following Bekenstein) 
one considers that the irreversible absorption of a particle by a BH corresponds to the loss of one bit of information 
(for the outside world), we are led to the idea of attributing to a BH an entropy (in the sense of ``negentropy'') 
equal (after re-introducing the constants $c$ and $G$) to \cite{B1973}
\be
S_{\mrm{BH}}=\hat{\alpha}\frac{c^3}{\hbar G}A,
\ee
with a dimensionless numerical coefficient equal to $\hat{\alpha}= \ln 2/ 8 \pi$ according to the reasoning just made. More generally, 
Bekenstein suggested that the above formula should hold with a dimensionless coefficient $\hat{\alpha}\approx \mathcal{O}\lb 1 \rb$, 
without being able to fix in a unique, and convincing, manner the value of $\hat{\alpha}$.  This result in turn implies 
(by applying the law of thermodynamics) that one should attribute to a BH a temperature equal to
\be
T_{BH}=\frac{1}{8 \pi \hat{\alpha}}\frac{\hbar}{c}g.
\ee
This attribution of a finite temperature to a BH looked rather strange in view of the  definition of a BH has being ``black'', \ie
 as allowing no radiation to come out of it. In particular, Stephen Hawking resisted this idea, and tried to prove it wrong by 
studying quantum field theory in a BH background. However, much to his own surprise, he so discovered (in 1974) the phenomenon 
of quantum radiation from BH horizons (see below) which remarkably vindicated the physical correctness of Bekenstein's suggestion.  
Hawking's calculation also unambiguously fixed the numerical value of $\hat{\alpha}$ to be $\hat{\alpha}=\frac{1}{4}$ \cite{H1975}.  
[We shall give below a simple derivation (from Ref.~\cite{DR1976}) of Hawking's radiation.]
 
Summarizing so far: The results on BH dynamics and thermodynamics of the early 1970's modified the early view of BHs as 
passive potential wells by endowing them with {\it global} dynamical and thermodynamical quantities, such as mass, charge, 
irreducible mass, entropy, and temperature. In the following section, we shall review the further changes in viewpoint brought 
by work in the mid and late 1970's (\cite{HH1972,D1978,Z1978,D1979,D1982}) which attributed {\it local} dynamical and thermodynamical 
quantities to BHs, and led to considering BH horizons as some kind of {\it dissipative branes}.  Note  that, in the following section,
 we shall no longer consider only  Kerr-Newman BHs (\ie stationary BHs in equilibrium, which are not distorted by sources at infinity).  
We shall consider more general  non-stationary BHs distorted by outside forces.

\subsection{Black hole electrodynamics}

The description of BHs we give from here on is essentially  ``holographic'' in nature since it will consist of excising the interior 
of a BH, and replacing the description of the interior BH physics by quantities and phenomena taking place entirely on the ``surface 
of the BH'' (\ie the horizon). The surface of the BH is defined as being  a null hypersurface, \ie a surface everywhere tangent to 
the lightcone, separating the region inside the BH from the region outside. As just said, we ignore the region inside, including
 the spacetime singularity, and consider the physics in the outside region, completing it with suitable ``boundary effects''  on 
the horizon.  These boundary effects are fictitious, and do not really exist on the BH surface but play the role of representing, 
in a holographic sense, the physics that goes on inside. In the end, we shall have a horizon, a set of surface quantities on the 
horizon and a set of bulk properties outside the horizon.  
We first consider Maxwell's equations, namely $F_{\mu \nu}= \partial_{\mu} A_{ \nu}- \partial_{\nu} A_{ \mu}$, and 
\be
\begin{array}{ccc}
\displaystyle \nabla_{\nu} F^{\mu \nu} & \displaystyle = & \displaystyle 4 \pi J^{\mu},\\
\displaystyle \nabla_{\mu}J^{\mu} & \displaystyle = & \displaystyle 0.
\end{array}
\ee
A priori, the electromagnetic field $F_{\mu \nu}$ permeates the full space time, existing both inside and outside the horizon, 
and the current, \ie the source term of $F_{\mu \nu}$ that carries charge, is also distributed both outside and inside the BH.  
In order to replace the internal electrodynamics of the BH by surface effects,  we replace the real 
$F_{\mu \nu}(x)$ by $F_{\mu \nu}(x) \Theta_{\mrm{H}}$, where $\Theta_H$ is a Heaviside-like step function, equal to $1$ outside 
the BH and $0$ inside. Then we consider what equations are satisfied by this   $\Theta_H$-modified electromagnetic field. 
The corresponding modified Maxwell equations contain two types of source terms,
\be
\begin{array}{ccc}
\nabla_{\nu}\lb F^{\mu \nu} \Theta \rb & = & \lb \nabla_{\nu} F^{\mu \nu} \rb \Theta+F^{\mu \nu}\nabla_{\nu}\Theta\\
& = & 4 \pi \lb J^{\mu} \Theta+j_{\nu}^{\mu} \rb,
\end{array}
\ee
where we have introduced a {\it BH  surface current}  $j_{H}^{\mu}$ as
\be
 j_{H}^{\mu}=\frac{1}{4 \pi}F^{\mu \nu}\nabla_{\nu}\Theta.
\ee
This surface current contains a Dirac $\delta$-function which restricts it to the horizon. Indeed,
let us consider a scalar function $\varphi \lb x \rb$ such that $\varphi \lb x \rb=0$
on the horizon, with $\varphi \lb x \rb< 0$ inside
the BH, and $\varphi \lb x \rb> 0$ outside it. The BH $\Theta$-function introduced above is
simply equal to $ \Theta_H = \theta (\varphi \lb x \rb)$, where $\theta$
denotes the standard step function of one real variable. Therefore, the
 gradient of $ \Theta_{\mrm{H}}$ reads
\be
\partial_{\mu}\Theta_H = \partial_{\mu} \theta\lb \varphi \lb x \rb \rb=\delta \lb \varphi \lb x \rb \rb \partial_{\mu} \varphi,
\ee
where $\delta$ is the (one dimensional) usual Dirac delta, so that 
$\delta \lb \varphi \lb x \rb \rb$ is a delta function with support on the horizon.
Morally, the gradient $\partial_{\mu} \varphi$ yields a vector ``normal to the horizon''.  In the case of a BH (by contrast to 
the usual case of a hypersurface in Euclidean space), there exists an extra subtlety in the exact definition of the normal to 
the horizon.  The horizon is a null hypersurface which by definition is normal to a null covariant vector $\ell_{\mu}$ satisfying 
both $\ell_{\mu} \ell^{\mu}=0$ and $\ell_{\mu}\mrm{d}x^{\mu}$ for any infinitesimal displacement $\mrm{d}x^{\mu}$ within the 
hypersurface.  Since $\ell_{\mu}$ is null, it cannot be normalized in the same way as
in Euclidean space.  This leads to an ambiguity in the physical observables related to $\ell_{\mu}$.  In stationary-axisymmetric 
spacetimes, one uniquely normalizes  $\ell_{\mu}$ by demanding that
the corresponding directional gradient  $\ell^{\mu}\partial_{\mu}$ be of the form $\partial/\partial t
+ \Omega \partial/\partial \phi$ (with a coefficient one in front of the time-derivative
term). We shall assume (in the general non-stationary case) that $\ell_{\mu}$ is normalized
so that its normalization is compatible with the usual normalization when considering the
limiting case of stationary-axisymmetric spacetimes.
Anyway, given any normalization, there exists a scalar $\omega$ such that
\be
\ell_{\mu}=\omega \partial_{\mu} \varphi,
\ee
and we can then define an ``horizon $\delta$-function''
\be
\delta_{H}=\frac{1}{\omega}\delta \lb \varphi \rb,
\ee
such that
\be
\partial_{\mu} \Theta_H=\ell_{\mu}\delta_H.
\ee
This leads to defining a ``BH surface current density''
\be
\label{Kmu}
K^{\mu}=\frac{1}{4 \pi}F^{\mu \nu}\ell_{\nu}.
\ee
With this definition, the BH current $j_{H}^{\mu}$  reads
\be
j_{H}^{\mu}=K^{\mu}\delta_{\mrm{H}},
\ee
and satisfies 
\be
\nabla_{\mu}\lb \Theta_H J^{\mu}+K^{\mu}\delta_H \rb = 0,
\ee
which is a conservation law for the sum of the outside bulk current $\Theta_H J^{\mu}$ and of the boundary current $K^{\mu}\delta_H $. In 
picturesque terms, the surface current $K^{\mu}\delta_H $ effectively ``closes'' the external current lines penetrating the 
BH (analogously to the case of external currents being injected in a perfect conductor and leading to currents flowing on 
its surface).  In addition, Eq. (\ref{Kmu}) shows that this surface current is 
linked to the electromagnetic fields which are on the horizon.  We have thus 
endowed the horizon with surface quantities, defined uniquely and locally on the horizon.

Before we proceed, we introduce a convenient coordinate system to describe the physics on the horizon of a general BH.  We 
assume some regular ``slicing'' of the horizon and its neighbourhood by some (advanced) Eddington-Finkelstein-like time 
coordinate  $t=x^0$. Then we assume that the first coordinate  $x^1$ is such that it is equal to zero on the horizon (like $r - r_+$ 
in the Kerr-Newman case). Finally $x^{A}$ for $A=2,3$ denote some angular-like coordinates on the two-dimensional spatial slice 
$S_t$ ($x^0 = t$) of the horizon.  In this coordinate system, we normalize $\ell^{\mu}$ such that
\be
\label{velocity}
\ell^{\mu}\partial_{\mu}=\frac{\partial}{\partial t}+v^A\frac{\partial}{\partial x^A}.
\ee
Here, we have used the fact that the ``normal'' vector $\ell^{\mu}$, being null, is also  {\it tangent} to the horizon, so that 
$\ell^{\mu}\partial_{\mu}$  is a general combination of $\partial / \partial t$ and $\partial / \partial x^A$ but has no component 
along the ``radial'' (or ``transverse'') coordinate $x^1$.  Because $\ell^{\mu}$ is a vector tangent to the hypersurface, we can
 consider its integral lines $\ell^{\mu} = d x^{\mu}/dt$, which lie within the horizon. These integral curves are called the
 {\it generators} of the horizon. They are null geodesics curves, lying entirely within the horizon.

Expression (\ref{velocity}) for the directional gradient along $\ell^{\mu}$ suggests that $v^A$  be interpreted as the 
velocity of  some ``fluid particles'' on the horizon, which are the ``constituents'' of a null  membrane.  Similarly to the 
usual  description of the motion of a fluid, one has to keep track of the changes in the distance between two fluid particles 
as the fluid  expands and shears. For a usual fluid, one considers the gradient of the velocity field, splitting it into its 
symmetric and anti-symmetric parts. The antisymmetric part is simply a local rotation which has no incidence on the physics 
and can be ignored. The symmetric part is further split into its trace and tracefree parts, namely
\be
\frac{1}{2}\lb \partial_i v_j+\partial_j v_i \rb =\sigma_{ij}+ \frac{1}{d}\partial \cdot v \delta_{ij}
\ee
where $d$ is the spatial dimension of the considered fluid (which will be $d=2$ in our case). Here the first term describes the 
shear, and the second describes the rate of expansion. We will see later how the BH analogs of these quantities are defined.  
For the moment let us consider the distances on the horizon. They are measured by considering the restriction to the horizon of 
the spacetime metric (which is assumed to satisfy Einstein's equations). As we are considering a null hypersurface, we have
\be
\mrm{d}s^2|_{x^1=0}=\gamma_{AB}\lb t,x^C \rb \lb \mrm{d}x^A-v^A\mrm{d}t \rb \lb \mrm{d}x^B-v^B\mrm{d}t \rb
\ee
where $v^A=\frac{\mrm{d}x^A}{\mrm{d}t}$.  Note that $\mrm{d}s^2$ is a degenerate metric: indeed, on a (three-dimensional) null 
hypersurface, there is no real time direction ($\mrm{d}s^2$ vanishes along the generators). One has  only two positive-definite space 
dimensions along, \eg the spatial slices $S_t$.  This metric describes the geometry on the horizon from which one can compute 
the area element of the spatial sections $S_t$
\be
\mrm{d}A=\sqrt{\mrm{det}\gamma_{AB}}\mrm{d}x^2\wedge \mrm{d}x^3.
\ee
One can decompose the current density $K^{\mu}$ into a time component $\sigma_H=K^0$, and two spatial components $K^A$ tangent to
 the spatial slices  $S_t$ ($t =$ const.) of the horizon,
\be
K^{\mu}\partial_{\mu}=\sigma_H\partial_t+K^A\partial_A
\ee
in which $\partial_t=\ell^{\mu}\partial_{\mu}-v^A\partial_A$ so that
\be
K^{\mu}\partial_{\mu}=\sigma_H \ell^{\mu}+(K^A-\sigma_H v^A)\partial_A
\ee
The total electric charge of the spacetime is defined by a surface integral at $\infty$, say
\be
Q_{\rm tot}=\frac{1}{4 \pi} {\oint}_{S_{\infty}} \frac{1}{2}F^{\mu \nu}\mrm{d}S_{\mu\nu}.
\ee
This result can be re-written as the sum of a surface integral on the horizon and a volume integral in between the horizon and 
$\infty$.  The volume integral is simply the usual charge contained in space, so that we can 
define the BH charge $Q_H $ as
\be
Q_H=\frac{1}{4 \pi} {\oint}_H \frac{1}{2}F^{\mu \nu}\mrm{d}S_{\mu\nu},
\ee
where the tensorial horizon surface element reads $\mrm{d}S_{\mu \nu}=\frac{1}{2}\varepsilon_{\mu \nu \rho \sigma} \mrm{d}x^{\rho}\wedge 
\mrm{d}x^{\sigma}=\lb n_{\mu} \ell_{\nu}- n_{\nu} \ell_{\mu} \rb \mrm{d}A$. Here, $n^{\mu}$ is a second null vector, which is transverse 
to the horizon, and which is orthogonal to the spatial sections $S_t$. It is normalized such that $n^{\mu} \ell_{\mu}= + 1$.  Using 
the definitions above of the BH surface current, one easily finds that the total BH charge can be rewritten as
\be
Q_H= {\oint}_H \sigma_H  dA,
\ee
where $\sigma_H $ is the time component of the BH surface current introduced above. Though it is a priori only the integrated BH 
charge which has a clear physical meaning, it is natural to consider the density $\sigma_H $ appearing in the above surface integral 
as defining a charge distribution on the horizon. Then the link
\be
\sigma_H = K^{\mu} n_{\mu} = \frac{1}{4 \pi} F^{\mu \nu} n_{\mu} l_{\nu}
\ee
can be thought of as being analog to the result $\sigma = \frac{1}{4 \pi} E^i n_i$ giving the electric charge distribution on a 
metallic object. This can again be viewed as part of a holographic approach in which the interior of the BH is replaced by boundary 
effects.  This analogy extends to the (spatial) currents flowing along the surface of the BH. Indeed, using the conservation law 
$\nabla_{\mu}\lb \Theta_H J^{\mu}+K^{\mu}\delta_H \rb = 0$, which is just a Bianchi identity, one has
\be
\frac{1}{\sqrt{\gamma}}\frac{\partial}{\partial t}\lb \sqrt{\gamma} \sigma_H \rb+\frac{1}{\sqrt{\gamma}}\frac{\partial}{\partial x^A}\lb 
\sqrt{\gamma} K^A \rb=-J^{\mu}\ell_{\mu}.
\ee
This shows, in a mathematically precise way, how an external current injected ``normally'' to the horizon ``closes'' onto a combination 
of currents flowing along the horizon, and/or of an increase in the local horizon charge density.  One can also introduce the electromagnetic 
2-form and restrict it on the horizon. It then defines the electric and magnetic fields on the horizon according to
\be
\frac{1}{2}F_{\mu \nu}\mrm{d}x^{\mu} \wedge \mrm{d}x^{\nu}|_{\mrm{H}} = E_A \mrm{d}x^A \wedge \mrm{d}t+B_{\perp}\mrm{d}A.
\ee
Taking the exterior derivative of the left-hand-side then gives
\be
\nabla \times \vec{E} = -\frac{1}{\sqrt{\gamma}}\partial_t \lb \sqrt{\gamma} B_{\perp} \rb.
\ee
which relates the electric and magnetic fields on the horizon.

>From the various formal definitions above, one also gets the following relation
\be
E_A+\epsilon_{AB}B_{\perp}v^B=4 \pi \gamma_{AB}\lb K^B-\sigma_H v^B \rb,
\ee
or
\be
\vec{E}+\vec{v}\times \vec{B_{\perp}}=4 \pi \lb \vec{K}-\sigma_H \vec{v} \rb.
\ee
We recognize here a BH analog of the usual Ohm's law relating the electric field to the current (especially in the case where 
$v \rightarrow 0$, \ie in the absence of the various ``convection effects'' linked to the horizon ``velocity'' $\vec{v}$).  From this form of 
Ohm's law, we can read off that BHs have a {\it surface electric resistivity} equal to $\rho=4 \pi=377\,$ Ohm \cite{D1978,Z1978}.

Let us give an example in which this BH Ohm's law can be ``applied'' to a specific system.  We consider for simplicity the case of a 
Schwarzschild BH and set up an electric circuit ``on the surface of the BH'' by injecting on the North pole (through an electrode 
penetrating the horizon under a polar angle $\theta_1$, with, say, $\theta_1 \ll 1$) an electric current $I$, and letting it 
escape\footnote{Actually, as (classical) charges cannot escape from a BH, we need to inject in the South electrode a flow of negative 
charges (while injecting a flow of positive charges down the North pole).} from the South pole (via an electrode penetrating the 
horizon under a polar angle $\theta_2$, with, say, $\pi - \theta_2 \ll 1$).  When viewing the BH as a membrane with surface resistivity 
$\rho$, this set up will give rise to a fictitious current flow on the horizon, closing the circuit between the North and the South poles. 
Associated to the current flow on the horizon, there will be a potential drop $V$ between the poles. This 
potential drop is simply given by the usual Ohm's law, $V = R I$, \ie the product of the current $I$ by a ``resistance'' $R$:
\be
\displaystyle V=-A_0\lb \theta_1 \rb+A_0 \lb \theta_2 \rb=RI. 
\ee
The BH resistance $R$ can be computed in two different ways, either by solving Maxwell's equations in a Schwarzschild background, 
or by computing, in usual Euclidean space, the total resistance of a spherical metallic shell with a uniform surface resistivity 
$\rho = 4 \pi$ (by decomposing the problem in many elementary resistances, some being in parallel, and others in series). Both methods 
give the same answer, namely
\be 
\displaystyle R=\displaystyle 2 \ln \frac{\mrm{tan}\frac{\theta_2}{2}}{\mrm{tan}\frac{\theta_1}{2}},
\ee
expressed in units of $30\, \Omega$. \footnote{Indeed, in CGS-Gaussian units (as used, say, in the treatise of Landau and Lifshitz) 
$30$ ohms is equal to the velocity of light (or its inverse, depending on whether one uses esu or emu).  Then, when using (as we 
do here) units where $c=1$, $30 \Omega = 1$.}  This result is saying that the typical total resistivity of a BH is of the order 
of $30\, \Omega$.  In addition, if one considers a rotating BH placed in a magnetic field out of alignment with its axis of rotation  
(a field uniform at $\infty$, but distorted on the horizon), one expects to find eddy currents on the horizon, currents which dissipate 
the energy. These currents exist, can be computed and do indeed brake the rotation of the BH. In such a situation, one also finds a 
torque which acts to restore the alignment of the BH with the field \cite{D1978}.

\subsection{Black hole viscosity}

In the previous section, we introduced the electromagnetic dissipative properties of a BH, using a holographic approach which kept 
the physics outside up to infinity, and replaced the physics inside the BH by defining suitable quantities on the horizon, and then 
showed that they satisfied equations similar to well-known ones (such as Ohm's law).  We now turn to the viscous properties of BHs 
and show how suitably defined ``surface hydrodynamical'' quantities satisfy a sort of Navier-Stokes equation.  Technically, we would 
like to do, for the gravitational surface properties, something similar to what we did for the electrodynamic properties. Namely, we 
would like to replace the spacetime connection, say $\omega$, by some sort of ``screened connection'' $\Theta_H \omega$, and see what 
kind of quantities and physics will be so induced on the surface of the BH.   However, Einstein's equations being nonlinear, one 
cannot simply use a BH step function $\Theta_H$ as was done for BH electrodynamics.  We shall therefore motivate the definition of suitable 
``surface quantities'' related to  $\omega$ in a slightly different way and then study the evolution of these surface 
quantities and their connection to the physics outside the horizon, up to $\infty$. Our presentation will be sketchy; for technical details,
see~\cite{D1979,D1982,Gourgoulhon:2005ch,Gourgoulhon:2005ng}.

Let us start by considering an axisymmetric spacetime.  Then there exists a Killing vector $\vec{m}=m^{\mu} \partial/ \partial x^{\mu}
=\partial / \partial \varphi$, to which, by Noether's theorem, one can associate a conserved total angular momentum, which can be 
written as a surface integral at $\infty$. The total angular momentum $J_z$ w.r.t. $\varphi$ reads
\be
J_{\infty}=-\frac{1}{8 \pi}\int_{S_{\infty}}
\frac{1}{2}\nabla^{\nu}m^{\mu}\mrm{d}S_{\mu \nu},
\ee
where $\mrm{d}S_{\mu \nu}=\frac{1}{2}\varepsilon_{\mu \nu \rho \sigma} \mrm{d}x^{\rho} \wedge \mrm{d}x^{\sigma}$, $\nabla^{\nu}$ denotes 
a covariant derivative, and the surface integral is performed over the 2-sphere, $S_{\infty}$.  This starting point is the analog 
of the surface-integral expression for the total electric charge used above to motivate the definition of a BH surface charge distribution.

In a way similar to what was done in the electromagnetic case, we can use Gauss'theorem to rewrite this integral as the sum of two 
contributions: (i) a volume integral (over the 3-volume contained between the horizon and infinity) measuring the angular momentum 
of the matter present outside the horizon, and (ii) a surface integral over a (topological) 2-sphere $S_{H}$ at the horizon, 
representing what we can call the BH angular momentum  $J_H$, \ie
\be
 J=J_{\mrm{matter}}+J_H,
\ee
where $J_H$ is given by the same surface-integral formula as  $J_{\infty}$, except for the replacement of $S_{\infty}$ by $S_H$ as 
integration domain.

The horizon being tangent to the lightcone, one defines on the horizon, as above, a null vector $\ell_{\mu}$ both normal and tangent 
to it. $\ell_{\mu}$ can in turn be complemented by another null vector $n_{\mu}$ such that $\ell^{\mu}n_{\mu}=1$ and such that the surface 
element $dS_{\mu \nu}$ can then be re-expressed as $\lb n_{\mu} \ell_{\nu}-n_{\nu}\ell_{\mu} \rb \mrm{d}A$. Remembering that the Killing 
symmetry preserves the generators of the horizon \ie the commutator $\small[ \vec{\ell}, \vec{m} \small]=0$, one has $\ell^{\nu} 
\nabla_{\nu}m^{\mu}=m^{\nu}\nabla_{\nu}\ell^{\mu}$, so that we can re-express the BH angular momentum $J_H$ as the following surface 
integral
\be
J_H=-\frac{1}{8 \pi} \int_{S_H} n_{\mu}m^{\nu}\nabla_{\nu}\ell^{\mu}\mrm{d}A.
\ee
This result involves the directional (covariant) derivative of the horizon {\it normal vector} $\vec{\ell}$ along a vector $\vec{m}$ 
which is {\it tangent} to the horizon. The crucial point now is to realize that, very generally, given any hypersurface, the parallel 
transport along some  {\it tangent} direction, say $\vec{t}$, of the  (normalized) vector $\vec{\ell}$ normal to the hypersurface yields 
{\it another tangent vector}. The technical proof of this fact consists of starting from the fact that $\vec{\ell} \cdot \vec{\ell} = 
\epsilon$, where $\epsilon$ is a {\it constant} which is equal to $\pm 1$ in the case of a time-like or spacelike hypersurface, and to 
$0$ in the case (of interest here) of a null hypersurface.  Then, taking the directional gradient of this starting equality along an 
arbitrary tangent vector $\vec{t}$ yields $\displaystyle \small( \nabla_{\vec{t}} \vec{\ell} \small) . \vec{\ell}=0$. From this result, one deduces
 that the vector $\small( \nabla_{\vec{t}} \vec{\ell} \small)$ must be {\it tangent} to the hypersurface. Therefore, there exists a 
certain linear map $K$, acting in the tangent plane to the hypersurface, such that $ \nabla_{\vec{t}} \vec{\ell} = K(\vec{t})$. For a
 usual (time-like or space-like) hypersurface, the linear map $K$ is called the ``Weingarten map'' and is simply the mixed-component
 $K_j^i$ version of the extrinsic curvature of the hypersurface (usually thought of a being a symmetric covariant tensor $K_{i j}$).
 On the other hand, in the case of a null hypersurface, there is no unique way to define the analog of the covariant tensor $K_{i j}$
 (where the indices $i,j$ are ``tangent'' to the hypersurface), but it is natural, and useful, to consider the mixed-component tensor$K_j^i$,
 intrinsically defined as the Weingarten map $K$ in $ \nabla_{\vec{t}} \vec{\ell} = K(\vec{t})$.

To explicitly write out the various components of the linear map $K$ (acting on the hypersurface tangent plane), we need to  define a basis
 of vectors tangent to the horizon.  This basis contains the null vector $\vec{\ell}$ (which is both normal and tangent to the horizon),
 and two spacelike vectors. Using a coordinate system $x^0, x^1, x^A$ ($A =2,3$) of the type already introduced (with the horizon being located
 at $x^1 = 0$), we can choose, as two spacelike horizon tangent vectors, the vectors $\vec{e}_A = \partial_A$.  Then one finds that the Weingarten map $K$
 is fully described by the set of equations
\be
\begin{array}{ccc}
\displaystyle \nabla_{\vec{\ell}}\, \vec{\ell} & \displaystyle = & \displaystyle g \, \vec{\ell},\\
\displaystyle \nabla_A\vec{\ell} & \displaystyle = & \displaystyle \Omega_A \vec{\ell}+D_A^B \vec{e}_B.
\end{array}
\ee
The first equation follows from the fact that $\vec{\ell}$ is tangent to a null geodesic lying within the null hypersurface. [In turn,
 this follows from the fact that $\vec{\ell}$ is proportional to the gradient of some scalar, say $\varphi$ (satisfying the eikonal equation
 $(\nabla \varphi)^2 =0$).] The coefficient $g$ entering the first equation defines (in the most general manner) the {\it surface gravity}
 of the BH. We see that it represents one component of the Weingarten map $K$. The other components are the two-vector $\Omega_A$, and the
 mixed two-tensor $ D_A^B$.  One can show that the  components $ D_A^B$ are the mixed components of a symmetric two-tensor $ D_{A B}$, which
 measures the ``deformation'', in time, of the geometry of the horizon.  We remind the reader of the expression of the horizon metric,
 introduced above,  $\mrm{d}s^2|_H=\gamma_{AB}\lb t,\vec{x} \rb \lb\mrm{d}x^A-v^A \mrm{d}t\rb \lb \mrm{d}x^B-v^B\mrm{d}t \rb$.  Here,
 $\gamma_{AB}\lb t,\vec{x} \rb$  is a symmetric rank 2 tensor \ie a time-dependent 2-metric such that the horizon may by viewed as a
 2-dimensional brane. In addition, we have the generators, which are the vectors tangent to $\vec{\ell}$. When decomposing $\vec{\ell}=
 \partial_t + v^A \partial_A$ w.r.t. our coordinate system, they appear to have a ``velocity'' $v^A$ which can also be viewed as the
 velocity of a fluid particle on the horizon.  $D_{AB}$ is then defined as the deformation tensor of the horizon geometry, namely
 $D_{AB}=\gamma_{B C} D_A^C=\frac{1}{2} \frac{D\gamma_{AB}}{\mrm{d}t}$, where $D/dt$ denotes the {\it Lie derivative} along $\vec{\ell}=
 \partial_t + v^A \partial_A$. It is explicitly given by
\be
\begin{array}{ccc}
\displaystyle D_{AB} & \displaystyle = & \displaystyle \frac{1}{2}\lb \partial_t \gamma_{AB}+
v^C \partial_C \gamma_{AB}+ \partial_A v^C\gamma_{CB}+\partial_B v^C \gamma_{AC}\rb\\
\displaystyle & \displaystyle = & \displaystyle \frac{1}{2} \lb \partial_t \gamma_{AB}+v_{A|B}+v_{B|A}\rb
\end{array}
\ee
where `$_|$' denotes a covariant derivative w.r.t. the Christoffel symbols of the 2-geometry $\gamma_{AB}$.  Note the contribution from
 the ordinary time derivative of $\gamma_{AB}$, and that from the variation of the generators of velocity $v^A$ along the horizon.  It 
is then convenient to split the deformation tensor  $D_{AB}$ into a tracefree part and a trace, \ie $D_{AB}=\sigma_{AB}+\frac{1}{2}\theta 
\gamma_{AB}$, where the tracefree part $\sigma_{AB}$ is the ``shear tensor'' and the trace, $\theta=D_A^A=\frac{1}{2}\gamma^{AB}\partial_t
 \gamma_{AB}+v^A_{|A}$, the ``expansion''. The remaining component of the Weingarten map, namely the 2-vector $\Omega_A$, is defined
 as $\Omega_A=\vec{n}.\nabla_A \vec{\ell}$ with $\vec{\ell}.\vec{n}=1$.  Its physical meaning can be seen from looking at the BH angular
 momentum $J_H$.

 Indeed, from the definition above of $J_H$, one finds that the total BH angular momentum is the projection of $\Omega_A$ on the direction
 of the rotational Killing vector $\vec{m} = \partial_{\varphi}$ introduced at the beginning of this section, so that we have
\be
J_H=-\frac{1}{8 \pi}\oint_S m^A\Omega_A \mrm{d}A,
\ee
where $m^A\Omega_A$ is the $\varphi$-component of $\Omega_A$.  It is therefore natural to define, for a BH,  a ``surface density of
 linear momentum'' as $\pi_A=-\frac{1}{8 \pi}\Omega_A=-\frac{1}{8 \pi}\vec{n} \cdot \nabla_A\vec{\ell}$. With this definition, one has
\be
 J_H=\int_S \pi_{\varphi}\mrm{d}A,
\ee
which is similar to the result above giving the BH electric charge as the surface integral of the ``charge surface density'' $\sigma_H$.

Having so defined some (fictitious) ``hydrodynamical'' quantities on the surface of a BH (fluid velocity, linear momentum density,
 shear tensor, expansion rate, etc.), let us now see what evolution equations they satisfy as a consequence of Einstein's equations. 
 By contracting Einstein's equations with the normal to the horizon, we can relate the quantities just defined to the flux of the 
energy-momentum tensor $T_{\mu \nu}$ into the horizon.  For instance, by projecting Einstein's equations
along $\ell^{\mu}e_A^{\nu}$, one finds
\be
\frac{D\pi_A}{\mrm{d}t}=-\frac{\partial}{\partial x^A} \lb \frac{g}{8 \pi} \rb+\frac{1}{8 \pi}\sigma_{A\,|B}^B -\frac{1}{16 \pi}\partial_A
 \theta-\ell^{\mu}T_{\mu A}
\ee
where
\be
\begin{array}{ccc}
\displaystyle \frac{D\pi_A}{\mrm{d}t} & \displaystyle = & \displaystyle \lb \partial_t+\theta \rb \pi_A+v^B \pi_{A|B}+v_{\,|A}^B\pi_B, \\
\displaystyle \sigma_{AB} & \displaystyle = & \displaystyle \frac{1}{2} \lb \partial_t \gamma_{AB}+v_{A|B}+v_{B|A} \rb 
-\frac{1}{2} \theta \gamma_{AB}, \\
\displaystyle \theta & \displaystyle = & \displaystyle \frac{\partial_t \sqrt{\gamma}}{\sqrt{\gamma}}+v_{\, |A}^A
\end{array}
\ee
correspond to a convective derivative, a shear and an expansion rate respectively.  Let us recall that the Navier-Stokes equation for a
 viscous fluid reads
\be
\lb \partial_t+\theta \rb \pi_i+v^k\pi_{i,k}=-\frac{\partial}{\partial x^i}p+2\eta \sigma_{i\, ,k}^k+\zeta \theta_{,i}+f_i,
\ee
where $\pi_i$ is the momentum density, $p$ the pressure, $\eta$ the shear viscosity, 
$\sigma_{ij}=\frac{1}{2}\lb v_{i,j}+v_{j,i}\rb-\mrm{Trace}$,
 the shear tensor, $\zeta$ the bulk viscosity, $\theta=v_{\, ,i}^i$ the expansion rate, and $f_i$ the external force density. The two
 equations are remarkably similar.  This suggests that a BH can be viewed as a (non-relativistic\footnote{The non-relativistic
 character of the BH hydrodynamical-like equations may seem surprising in view of the ``ultra-relativistic'' nature of a BH.
 This non-relativistic-looking character is due to our use of an adapted ``light-cone frame'' $(\ell, n, e_A)$. It is well-known
 that light-cone-gauge results have a distinct ``non-relativistic'' flavour.}) 
 brane with (positive\footnote{This is
 consistent with the idea that the BH surface pressure must counteract the self-gravity.}) surface pressure $p= + \frac{g}{8 \pi}$, 
external force-density $f_A=-\ell^{\mu}T_{\mu A}$ which corresponds to the flow of external linear momentum, surface shear viscosity 
$\eta= +\frac{1}{16 \pi}$, and surface bulk viscosity $\zeta=-\frac{1}{16 \pi}$ (in units where $G=1$). Note, finally, that both the 
surface shear viscosity and the surface bulk viscosity apply to any type of deformed non-stationary BH.

\subsection{Irreversible thermodynamics of black holes}

In previous sections, we have introduced some electrodynamic and fluid dynamical quantities associated to a kind of dissipative dynamics
 of BH horizons. In addition, following Bekenstein, we would like to endow a BH with a surface density of entropy equal to a dimensionless 
constant $\hat{\alpha}$ (in units where $\hbar=G=c=1$).  Any dissipative system verifying Ohm's law and the Navier-Stokes equation is also 
expected to satisfy corresponding thermodynamic dissipative equations, namely  Joule's law and the usual expression of the viscous heat
 rate proportional to the sum of the squares of the shear tensor and of the expansion rate.  More precisely, we would expect to have a 
``heat production rate'' in each surface element $\mrm{d}A$ of the form
\be
\dot{q}=\mrm{d}A\left[ 2 \eta \sigma_{AB}\sigma^{AB}+\zeta \theta^2+\rho \lb \vec{\mathcal{K}}-\sigma_H \vec{v} \rb ^2 \right],
\ee
where $\rho$ is the surface resistivity, and $\eta$ and $\zeta$ the shear and bulk viscosities. In addition, one expects that this  heat 
production rate should be associated with a corresponding local increase of the entropy $ s = \hat{\alpha} \mrm{d}A $ contained in any local 
surface element of the form
\be
\frac{\mrm{d}s}{\mrm{d}t} = \frac{\dot{q}}{T}.
\ee
with a local temperature $T$ expected to be equal to $T=\frac{g}{8 \pi \hat{\alpha}}$.

Remarkably, the ``scalar'' ($\ell^{\mu} \ell^{\nu}$) projection of Einstein's equations, (\ie, the Raychauduri equation) yields an 
evolution law for the entropy $s=\hat{\alpha}\mrm{d}A$ of a local surface element which is very analogous to what one would expect. Indeed,
 it yields
\be
\frac{\mrm{d}s}{\mrm{d}t}-\tau
\frac{\mrm{d^2}s}{\mrm{d}t^2}=\frac{\mrm{d}A}{T}\left[2 \eta
\sigma_{AB}\sigma^{AB}+\zeta \theta^2+\rho \lb
\vec{\mathcal{K}}-\sigma_H \vec{v}\rb ^2\right],
\ee
where  $T=\frac{g}{8 \pi \hat{\alpha}}$, where $2 \eta \sigma_{AB}\sigma^{AB}+\zeta \theta^2$ are exactly the expected  viscous contributions,
 and where $\rho \lb \vec{\mathcal{K}}-\sigma_H \vec{v}\rb ^2$ is Joule's law.\\
The only unexpected term in this result is the second term on the l.h.s.; this term goes beyond usual near-equilibrium thermodynamics
 (which involves only the first order time derivative of the entropy), and is proportional to the second time derivative of the entropy
 density and to a time scale $\tau=\frac{1}{g}$. It is interesting to note that,  for the value $\hat{\alpha}=1/4$, corresponding to the
 Bekenstein-Hawking entropy density, $\tau$ is equal to $\frac{1}{2 \pi T}$, \ie the inverse of the lowest ``Matsubara frequency''
 associated to the temperature $T$ (one also notes that $\tau =D = 2 \cal{D}$, where $D$, $\cal{D}$ are
  the diffusion constants of \cite{Son:2007vk}).  The minus
 sign in front of this new term is also a particularity of BH physics.  In the approximation of a constant  $\tau$, and in solving for
 the non-equilibrium second law of thermodynamics, one finds that the rate of increase of entropy is given by
\be
\frac{\mrm{d}s}{\mrm{d}t}=\int_t^{\infty}\frac{\mrm{d}t'}{\tau}e^{-\frac{\lb t'-t\rb}{\tau}}\lb \frac{\dot{q}}{T}\rb \lb t'\rb,
\ee
\ie it is defined not as the value of the heat dissipated instantaneously, nor as an integral over the past heat dissipation, but as an
 integral over the {\it future}.  This highlights the acausal nature of BHs, \ie a BH is defined as a  null hypersurface which {\it will
 become stationary in the far future}. As such,  it has to anticipate any external perturbation.  Failing this, the null hypersurface 
would generically tend either to collapse, or blow up toward $\infty$.\\

We also note that the ratio of the shear viscosity $\eta= 1/(16 \pi)$ to the entropy density $ \hat{s} = s/dA = \hat{\alpha}$ is given by
\be
\frac{\eta}{\hat{s}}= \frac{1}{\hat{\alpha} 16 \pi}= \frac{1}{4 \pi}
\ee
where, in the second equality, we have used the Bekenstein-Hawking value for $ \hat{s} = s/dA = \hat{\alpha}= \frac{1}{4}$. It is 
interesting to note that the result $\frac{\eta}{\hat{s}}= \frac{1}{4 \pi}$ is indeed the ratio found by Kovtun, Son and Starinets in
 the gravity duals of strongly coupled gauge theories \cite{KSS2005,Son:2007vk}.\\

Finally, let us note another remarkable agreement between BH dissipative dynamics and a rather general property of ordinary 
(near-equilibrium) irreversible thermodynamics. This agreement concerns what Prigogine has called the ``minimum entropy production 
principle''.
Let us consider the total ``dissipation function''
\be
D=\oint_S\dot{q}
\ee
as a functional of the velocity field $v^{\mrm{A}}\lb x^2,x^3\rb$ and of the electric potential $\phi \lb x^2,x^3\rb$ in the presence of 
given external influences such as an external magnetic field or tidal forces acting on the rotating BH. Then,  $D[\phi]$ or $D[v^{\mrm{A}}]$
 (imposing $v_{|\mrm{A}}^{\mrm{A}}= 0$ as a constraint), reach a {\it minimum} when (and only when)  the lowest order (Einstein-Maxwell)
 dynamical equations for $\phi$ or $v^{\mrm{A}}$ are satisfied.

\subsection{Hawking Radiation}

In this section we discuss the phenomenon of Hawking radiation, first obtained in Ref.~\cite{H1975} (we shall follow here the derivation
 of Ref.~\cite{DR1976}). For simplicity, we will consider a 3+1 dimensional spherically symmetric BH.   We remind the reader that the
 coefficient $A \lb r \rb$, associated to the time coordinate (here denoted as $T$), goes to zero on the horizon, so that the horizon is
 an {\it infinite redshift surface}. It is also a {\it Killing} horizon, \ie the (suitably normalized) normal vector
 $\vec{\ell} = \partial/\partial t + \Omega \partial/\partial \phi$ is a Killing vector. These points  will be crucial in the following.  Since the coefficient of
 the radial coordinate is defined by $B \lb r \rb=\frac{1}{A \lb r \rb}$ (see Section \ref{GlobalPropertiesBH}), it is singular on
 the horizon. To get a good coordinate system on the horizon, we first factorize $A \lb r \rb$,
\be
\begin{array}{ccc}
\displaystyle \mrm{d}s^2 & \displaystyle = & \displaystyle -A\lb r \rb \mrm{d}T^2+\frac{\mrm{d}r^2}{A\lb r \rb}+r^2\mrm{d}\Omega^2\\
\displaystyle & \displaystyle = & \displaystyle -A\lb r \rb \lb \mrm{d}T^2-\lb \frac{\mrm{d}r}{A\lb r \rb}\rb ^2 \rb+ r^2 \mrm{d}\Omega^2,
\end{array}
\ee
and then introduce a new radial coordinate, the so-called {\it tortoise} coordinate, defined by
\be
r_*=\int \frac{\mrm{d}r}{A\lb r \rb}.
\ee 
Note that as $r \rightarrow r_+$, $A\lb r \rb \simeq \lb \frac{\partial A}{\partial r} \rb_{r_+}\lb r-r_+\rb$, 
where $\lb \frac{\partial A}{\partial r} \rb_{r_+}$ is (as mentioned above) equal to twice the surface gravity $g$. This implies
\be
\begin{array}{ccc}
\displaystyle r_* & \displaystyle \simeq & \displaystyle \int \frac{\mrm{d}r}{\lb r-r_+ \rb \lb \frac{\partial A}{\partial r} \rb_{r_+}}\\
\displaystyle & \displaystyle \simeq & \displaystyle \frac{\ln \lb r-r_+ \rb}{2 g}
\end{array}
\ee
such that as $r\rightarrow + \infty$, $r_* \simeq r+2 M \ln r$ and as $r \rightarrow r_+$, $r_* \simeq \frac{\ln \lb r-r+ \rb}{2 g}$. 
 The line element can thus be re-written as
\be
\mrm{d}s^2=-A\lb \mrm{d}T-\mrm{d}r_* \rb \lb \mrm{d}T+\mrm{d}r_* \rb +r^2\mrm{d}\Omega^2.
\ee
We now switch to the so-called {\it Eddington-Finkenstein} coordinates, $(t,r,\theta,\varphi)$, where  the combination $t=T+r_*$ of $T$
 and $r_*$  remains regular across the (future) horizon~\footnote{Given that the horizon is an infinite redshift surface, it takes an 
infinite time $T$ to fall into it, while  $r_* \simeq \ln \lb r-r_+ \rb/2 g$ goes to $-\infty$ at the horizon.The sum of the two remains,
 however, finite.} and define $t=T+r_*$. The time translation Killing vector $\partial/\partial t$ co\"incides with the usual one
 $\partial/\partial T$.  In terms of these new coordinates the metric reads
\be
\mrm{d}s^2=-A(r) dt^2 + 2 dt dr +r^2\mrm{d}\Omega^2.
\ee
This metric is now regular (with a non-vanishing determinant) as the radial coordinate $r$ penetrates within the horizon, \ie becomes
 smaller than $r_+$.

\begin{figure}
\includegraphics[width=10cm]{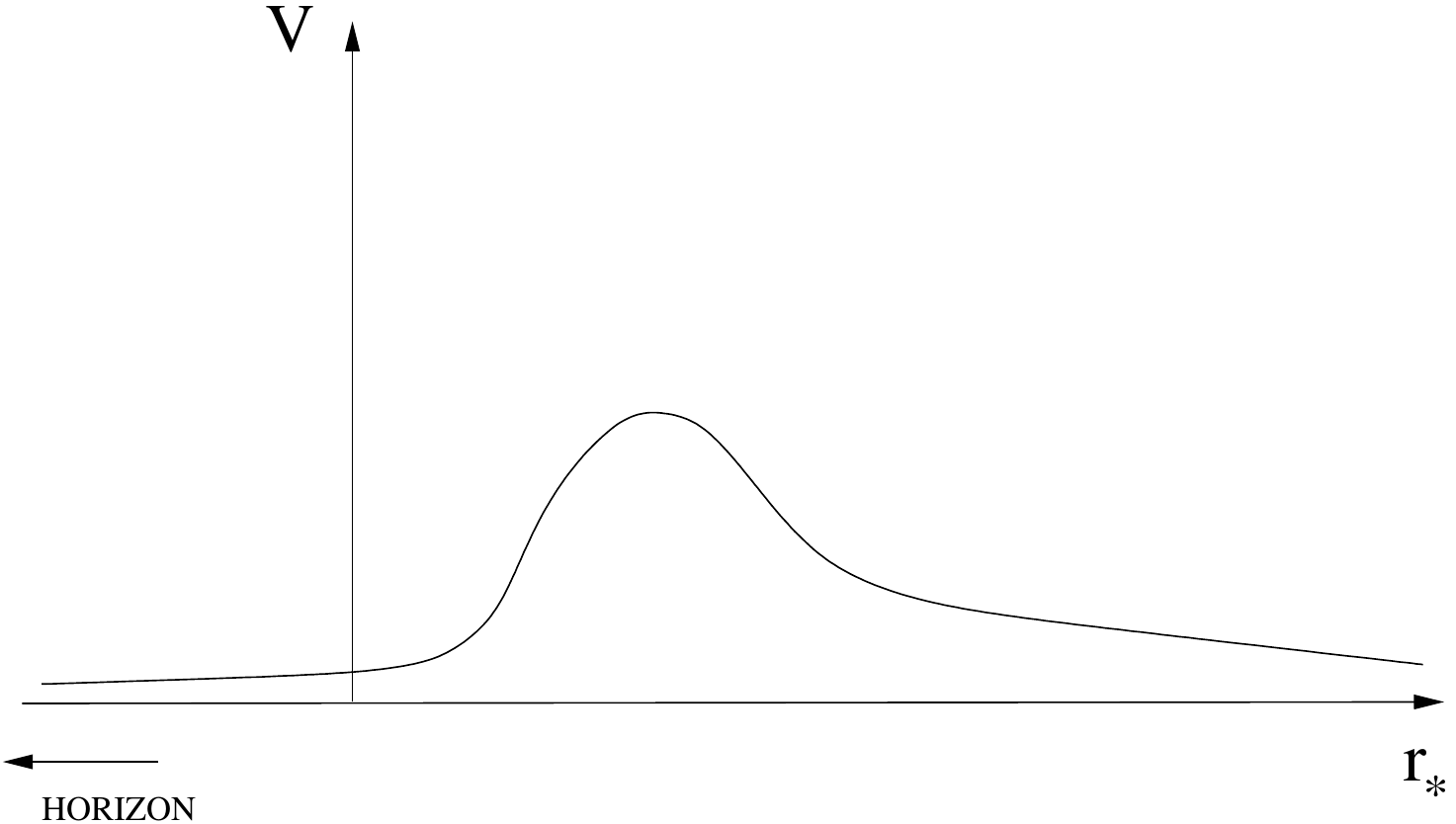}
\caption{This figure is a schematic representation of the effective gravitational potential in the neighborhood of a BH.  Note that as
 far as the particles are concerned, the spacetime is essentially flat both at infinity and near the horizon. The tidal-centrifugal 
barrier that separates the horizon
from infinity gives rise to the grey body factor.}
\label{BlackHolePotential}
\end{figure}

Having defined a regular coordinate system, we now consider a massless scalar field, the dynamics of which is given by the massless
 Klein Gordon equation
\be
0=\Box_g \varphi=\frac{1}{\sqrt{g}} \partial_{\mu} \lb \sqrt{g} g^{\mu \nu} \partial_{\nu} \varphi \rb
\ee
The solutions to this equation in a spherically symmetric and time-independent background are given by mode functions which are themselves
 given simply by products of a Fourier decomposition into frequencies, spherical harmonics and a radial dependence and thus read
\be
\varphi_{\omega \ell m}\lb T,r,\theta,\varphi \rb=\frac{e^{-i\omega T}}{\sqrt{2 \pi |\omega|}} 
\frac{u_{\omega \ell m}\lb r \rb}{r}Y_{\ell m}\lb \theta, \varphi \rb.
\ee
The problem is then reduced to solving a radial equation with the radial coordinate $r_*$,
\be
\frac{\partial ^2 u}{\partial r_*^2}+\left[ \omega^2-V_{\ell}\lb r\lb r_* \rb \rb \right] u=0.
\ee
In the case of the Schwarzschild metric (\ie when $A(r) = 1 - 2M/r$) the effective radial potential $V_{\ell}\lb r \rb$ is given by
\be
V_{\ell}\lb r \rb = \lb 1-\frac{2 M}{r} \rb \lb \frac{\ell \lb \ell+1 \rb}{r^2} \rb.
\ee
An essential point to note is that the effective potential vanishes both at $\infty$ (like a massless centrifugal potential
 $\ell(\ell+1)/r^2$), {\it and} at the horizon (where it is proportional to $A \lb r \rb$). Therefore, in these two regimes
 (which correspond to $r_* \rightarrow \pm \infty$), the solution of the wave equation behaves essentially as in flat space
 (see FIG. \ref{BlackHolePotential}).  The effect of the coupling to curvature is non-negligible only in an intermediate region,
 where the combined effect of curvature and centrifugal effects yield a {\it positive} potential barrier. In turn, this potential
 barrier yields a {\it  grey body factor} which diminishes the amplitude of the quantum modes considered below, \ie those generated
 near the horizon and which must penetrate through the potential barrier on their way towards $\infty$.  Far from the potential barrier,
 the general solution for $\varphi$ is
\be
\label{outgoingmodes}
\varphi_{\omega \ell m} \sim \frac{e^{-i \omega \lb T \pm r_* \rb}}{\sqrt{2 \pi |\omega|}}\frac{1}{r}Y_{\ell m}\lb \theta,\varphi \rb.
\ee
The quantification of the scalar field $\varphi$ is rather standard (and similar to what one does when studying  the amplification of
 quantum fluctuations during cosmological inflation).   The quantum operator for the scalar field is decomposed into mode functions,
 \ie the eigenfunctions of the Klein Gordon equation, with coefficients given by creation and annihilation operators. The subtlety lies,
 however, in the definition of  positive and negative frequencies.

Let us start by formally considering the simpler case of a quantum scalar field $\hat{\varphi}(x)$ in a background spacetime which becomes
 stationary {\it both} in the infinite past, and in the infinite future. In that case, one can define positive and negative frequencies in
 the usual way, in the two asymptotic regions $ t \to \pm \infty$.   The coefficient of the positive frequency modes, say $p(x)$, then 
defines an annihilation operator $\hat{a}$.  However, there are two sorts of  positive-frequency ($p(x)$), and negative-frequency ($n(x)$)
 modes.  The {\it in} ones $p_i^{in},n_i^{in}$ (defined in the asymptotic region $ t \to - \infty$, and then extended everywhere by solving
 the Klein-Gordon equation), and the {\it out} ones $p_i^{out},n_i^{out}$ (defined in the asymptotic region $ t \to + \infty$). The 
operator-valued coefficients of these modes define some corresponding {\it in} and {\it out} annihilation or creation operators, so
 that we can write the field operator $\hat{\varphi}(x)$ as (here $i$ is a label that runs over a basis of modes)
\be
\begin{array}{ccc}
\displaystyle \hat{\varphi}\lb x \rb & \displaystyle = & \displaystyle \sum_i \hat{a}_i^{in} p_i^{in} \lb x \rb + \lb \hat{a}_i^{in} \rb ^{\dagger} n_i^{in}\lb x \rb, \\
\displaystyle & \displaystyle = & \displaystyle \sum_i \hat{a}_i^{out} p_i^{out} \lb x \rb + \lb \hat{a}_i^{out} \rb ^{\dagger} n_i^{out}\lb x \rb.
\end{array}
\ee
Here, the modes are normalized as $\lb p_i,p_j \rb = \delta_{ij}$ and $\lb n_i,n_j \rb = -\delta_{ij}$, where `$\lb \, \, , \, \rb$' is
 the Klein-Gordon scalar product, \ie $\sim i \int d\sigma^{\mu} \lb \varphi_1^* \partial_{\mu} \varphi_2 - \partial_{\mu}\varphi_1^*
 \varphi_2 \rb$. The operators $a_i$ and $a_j^{\dagger}$ are correspondingly normalized in the usual way as $\left[ a_i,a_j^{\dagger}
 \right]=\delta_{ij}$.  One defines both an {\it in vacuum} $|in\rangle$ and an {\it out vacuum} $|out\rangle$ as the states that
 are respectively annihilated by $a_i^{in}$ or $a_i^{out}$.  Then, the phenomenon of particle creation corresponds to the fact that the 
{\it out} vacuum differs from the {\it in} one. More quantitatively, the expectation value of the number of {\it out} particles, in the mode labelled by
 $i$, which will be observed when the quantum field is in the {\it in} vacuum state is given by
\be
\begin{array}{ccc}
\label{particlecreation}
\displaystyle \langle N_i \rangle & \displaystyle = & \displaystyle \langle in | \lb a_i^{out} \rb ^{\dagger} a_i^{out} | in \rangle\\
\displaystyle & \displaystyle = & \displaystyle \sum_j |T_{ij}|^2
\end{array}
\ee
where we have introduced the transition amplitude $T_{ij}=\lb p_i^{out},n_j^{in} \rb $ from an initial negative frequency mode $n_j^{in}$
 into a final positive frequency one $p_i^{out}$. These transition amplitudes (also called Bogoliubov coefficients) enter the calculation
 because, as is easily deduced from the double expansion of the field $\hat{\varphi}\lb x \rb$ above, they give the part of $a_i^{out}$ 
which is proportional to $(a_j^{in})^{\dagger}$.
The application of the previous general formalism to the BH case is delicate since a BH background is not asymptotically stationary
 in the infinite future (because of the BH interior where the Killing vector $\partial/\partial t$ is spacelike), and is asymptotically
 stationary in the infinite past only if we do consider explicitly the collapse leading to the formation of a BH from an initially 
stationary star.  However, Hawking showed how to essentially bypass these difficulties by focussing on two types of modes:
\begin{itemize}
\item The high-frequency modes coming from the infinite past, which reach the horizon with practically no changes (because of their high-frequency
 nature) and,
\item the outgoing modes, viewed in the asymptotically flat region and in the far future.
 \end{itemize}

 Concerning the outgoing modes, they can be unambiguously decomposed in positive- and negative-frequency parts, because, as explained above,
their asymptotic behaviour is given by a sum of essentially flat-spacetime modes, (\ref{outgoingmodes}). One then defines the outgoing 
$p_i^{out}$'s as being proportional to $e^{-i \omega \lb T - r_* \rb}$ with a {\it positive} $\omega$.

Let us now focus on the definition of positive- and negative-frequency modes  near the horizon.  We recall that, as mentioned at the 
beginning of this section, there is a physically infinite redshift between the surface of the horizon and asymptotically flat space at
 infinity. If one is interested in particle creation with a finite given frequency, as observed at infinity, the corresponding wave 
packets will have very high frequency near the horizon and can therefore be approximated by very localized wave packets.   Given that
 the spacetime geometry in the vicinity of the horizon is regular, with a finite radius of curvature, it can be regarded as a piece
 of flat spacetime locally if one looks in a small enough region.  In this approximation, the calculation can be performed in a single step.

We wish to compute the average number of final outgoing particles\footnote{Note that the ``out'' label, in the general discussion of
 particle creation above, referred to ``final'' particles (as defined in the final, asymptotic, stationary spacetime background). In
 the case of a BH background, the ``final'' spacetime is made  of two separate asymptotic regions: (i) the outgoing wave region at 
spatial infinity, and (ii) the vicinity of the (spacelike?) singularity within the BH.  The definition of positive- and negative-frequency
 modes in the latter region is ill-defined. However, luckily, the calculation of the physically relevant flux of final, outgoing modes can
 be performed without worrying about the physics near the BH interior singularity. In other words, it is enough to consider as ``out''
 positive-frequency modes $p_i^{out}$ only the ones outgoing at spatial infinity (\ie on ``scri$^+$''), though they do not constitute a
 complete basis of  final modes.} seen in the {\it in} vacuum.  Then the average number of outgoing particles of type $p_i^{out}$ is 
given by  $\sum_j |T_{ij}|^2$, where $T_{ij}=\lb p_i^{out},n_j^{in} \rb $ is the transition amplitude from an initial negative frequency
 mode $n_j^{in}$ into a final {\it outgoing} positive frequency one $p_i^{out}$ (recorded at spatial infinity). To compute this transition
 amplitude, we need to describe what is an initial negative frequency mode $n_j^{in}$. As said above, Hawking suggested that only 
high-frequency initial modes are important, and that they essentially look the same (some kind of  WKB wave) in the real {\it in} 
region (in the far past, before the formation of the BH) as in the vicinity of the horizon. Our technical problem is then reduced 
to characterizing what is a negative frequency mode $n_j^{in}$ as seen in a small neighborhood of the horizon, which looks like the 
Minkowski vacuum.

To do this, it is convenient to have a technical criterion for characterizing positive and negative frequency modes in (a local) 
Minkowski spacetime. Locally, one can perform a Fourier decomposition of the wave packet and use the mathematical fact that the 
Fourier space properties are mapped onto analytic continuation properties in $x$-space. This relation can then be used to define 
positive and negative frequency modes. This is easy to see.  Consider a general {\it negative frequency}  wave packet 
in Minkowski spacetime. It has the form $\varphi_{-} \lb x \rb = \oint_{C^{-}} d^4k\tilde{\varphi} \lb k \rb e^{i k_{\mu}x^{\mu}}$ where 
$k_{\mu}$ is {\it timelike-or-null} and {\it past-directed}, \ie $k^{\mu} \in C^-$. We now perform a complex shift of the spacetime coordinate, $x_{\mu} \rightarrow 
x^{\mu} + i y^{\mu}$, where $y^{\mu}$ is timelike-or-null and  {\it future-directed } ({\it i.e.}, $y^{\mu} \in C^+$), then, the $e^{i k_{\mu}x^{\mu}}$ 
term will be suppressed by a $e^{-k_{\mu}y^{\mu}}$ term, where the scalar product $k_{\mu}y^{\mu}$ is {\it positive} because it involves two
 timelike vectors that point in opposite directions (we use the ``mostly plus'' signature). This ensures that a negative-frequency 
wavepacket can indeed be analytically continued to complex spacetime points of the form $x^{\mu} + i y^{\mu}$, with  $y^{\mu} \in C^+$.

The strategy for applying this criterion to characterizing  negative-frequency modes $n_j^{in}$ in the vicinity of the horizon is then
 the following. One starts from a wavepacket which is not purely a ``negative-frequency'' one near the horizon, but which has the 
property of evolving into an {\it outgoing} positive-frequency wave packet (so that it will have a non-zero transition amplitude to
 some $p_i^{out}$). Then, one modifies the initial wavepacket so that it becomes a purely negative-frequency mode $n_j^{in}$ near the horizon.

When looking at a  wavepacket of the form of $ \varphi_{\omega}^{out}\lb t,r \rb \propto e^{-i \omega \lb T-r_*\rb}$ (with positive $\omega$)
 just outside the horizon, one must first switch to well-defined coordinates to examine its physical content. We therefore replace the 
Schwarzschild-type time coordinate $T$ by the Eddington-Fikenstein time coordinate, $t=T+r_*$ which is regular on the horizon. After
 rearranging terms according to $T-r_*=\lb T+r_* \rb -2 r_*$ the previous (outgoing, positive-frequency) wave reads
\be
\begin{array}{ccc}
\displaystyle \left[ \varphi_{\omega}^{out}\lb t,r \rb \right]_{r_+}\propto e^{-i \omega \lb T-r_*\rb} & \displaystyle = & \displaystyle e^{-i \omega t}e^{2 i \omega r_*}\\
\displaystyle & \displaystyle = & \displaystyle e^{-i \omega t}e^{i\frac{\omega}{g}\ln \lb r-r_+ \rb}\\
\displaystyle & \displaystyle = & \displaystyle e^{-i \omega t}\lb r-r_+ \rb ^{\frac{i \omega}{g}}.
\end{array}
\ee
This describes the behaviour, just outside the horizon, of a wavepacket which will become (modulo some grey-body factor) an
 outgoing positive frequency wavepacket at $\infty$.  However, locally on the horizon, it is neither a positive nor a
 negative frequency wavepacket because, at this stage, it is defined only outside the horizon, but not inside. Let us now
 show how one must continue this wavepacket inside the horizon, so that it becomes a genuine  negative-frequency wavepacket
 ``straddling'' the horizon.  Using the criterion explained above, we can ``continue'' the wavepacket inside the horizon by
 a suitable {\it analytic continuation}. More precisely, we need an analytic continuation of the form $x^{\mu} \rightarrow x^{\mu} + i y^\mu$,
 where $y ^{\mu}$ belongs to the future lightcone to ensure that we shall then be dealing with a local negative frequency wavepacket. 
 It is easy to see, from a spacetime diagram of the  lightcone on the horizon, that the vector $\partial / \partial r$ is everywhere
 null and past-directed, such that $r \rightarrow r - \varepsilon$, where $\varepsilon >0$, is everywhere null and future  directed.
  The analytic continuation of $ \varphi_{\omega}^{out}\lb t,r \rb$ to $r \rightarrow r - i \varepsilon$ will therefore define for us
 a good local negative-frequency mode $n_j^{in} = n_{\omega \ell m}^{in}$.  One easily sees that this analytic continuation in $r$ generates
 a new component to the wavepacket which is located inside the BH (\ie for $r < r_+$).  More precisely, a one line calculation yields 
\be
\begin{array}{ccc}
\label{nin}
\displaystyle n_{\omega \ell m}^{in}\lb r,t \rb & \displaystyle = & \displaystyle N_{\omega}\varphi_{\omega}^{\mrm{out}}\lb t, r-i\varepsilon \rb \\
\displaystyle & \displaystyle = & \displaystyle N_{\omega} \Big[ \theta \lb r-r_+ \rb \varphi_{\omega}^{\mrm{out}}\lb r-r_+ \rb + \\
\displaystyle & \displaystyle  & \displaystyle e^{\frac{\pi \omega}{g}} \theta \lb r_+-r \rb \varphi_{\omega}^{\mrm{out}} \lb r_+-r \rb \Big],
\end{array}
\ee
where the second term is the wavefunction inside the horizon that has acquired an additional exponential factor due to the rotation
 $e^{-i \pi}$ in the complex plane from $r>r_+$ to $r<r_+$.  The overall factor $N_{\omega}$ is a normalization factor (needed because
 we have extended the mode inside the BH),  such that
\be
\langle n_{\omega \ell m}^{in}\lb r,t \rb n_{\omega ' \ell ' m'}^{in}\lb r,t \rb \rangle =\delta\lb \omega-\omega ' \rb \delta_{\ell
 \ell '}\delta_{m m'},
\ee
from which we obtain (when remembering that  $\varphi_{\omega}^{\mrm{out}}$ was correctly normalized)
\be
\displaystyle |N_{\omega \ell m}|^2=\displaystyle \frac{1}{e^{2 \pi \omega/g}-1}.
\ee
The physical meaning of  equation (\ref{nin}) is the description of  the splitting of the in mode $n_j^{in} = n_{\omega \ell m}^{in}$ into
 a positive-frequency wave outgoing from the horizon and a wave falling from the horizon towards the singularity. 
 One can read off from it  the needed transition amplitude  $T_{ij}=\lb p_i^{out},n_j^{in} \rb $. 
It is essentially given by the factor $N_{\omega}$, which must, however, be corrected by a grey-body factor $\sqrt{ \Gamma_{\ell}\lb 
\omega \rb}$  taking into account the attenuation of the outgoing wave $e^{-i \omega \lb T-r_*\rb}$ as it crosses the curvature + 
centrifugal potential barrier $V_{\ell}\lb r \rb$ on its way from the horizon to $\infty$.
Then (using Fermi's golden rule), one easily finds that the general result (\ref{particlecreation}) yields a rate of particle
 creation given by
\be
\frac{\mrm{d}\langle N \rangle}{\mrm{d}t}=\sum_{\ell, m} \int \frac{\mrm{d}\omega}{2 \pi}\frac{\Gamma_{\ell}\lb \omega \rb}
{e^{\frac{2 \pi \omega}{g}}-1}.
\ee
One recognizes here a thermal (Planck) spectrum (corrected by a grey-body factor).  From the Planck factor, one  reads off the Hawking
 temperature, $T=\hbar \frac{g}{2 \pi}$.  This result  fixes the dimensionless coefficient $\hat{\alpha}$ in the Bekenstein entropy to the
 famous result $\hat{\alpha} = \frac{1}{4}$, \ie
\be
S_{BH}=\frac{A}{4 G \hbar}  \; .
\ee

Let us end by two final comments.  First, the generalisation of the Hawking radiation to a more general rotating and/or charged BH is
 given essentially by replacing in the result above the frequency $\omega$ by $\omega - \omega_0$ where $\omega_0 = m \Omega + e V$
 exhibits the couplings of the created particles to the angular velocity $\Omega$ and the electric potential $V$ of the BH.  Then, in
 astrophysically realistic conditions, the ``Hawking'' part of the particle creation (\ie the thermal aspect) is too small to be relevant,
 while the combined effect of the grey body factor and of the zero-temperature limit of  $(e^{\frac{2 \pi (\omega - \omega_0)}{g}}-1)^{-1}$
 yield potentially relevant particle creation phenomena in Kerr-Newman BHs, associated to the ``superradiance'' of modes with frequencies
 $\mu < \omega < \omega_0$, where $\mu$ is the mass of the created particle (see, \eg \cite{DR1976} for more details and references).
  We conclude by noting that the situation just described is not only technically similar to the one in the inflationary scenario
 for cosmological perturbations, but also physically similar in that, in both cases, {\it transplanckian frequency modes} in the ultraviolet are
 redshifted to a finite, observable frequency.

\section{Experimental tests of gravity}

Before discussing various possibilities of string-inspired phenomenology (and of possible string-inspired deviations from Einstein's
 theory of General Relativity) we give an overview of what is known experimentally about the gravitational sector.

\subsection{Universal coupling of matter to gravity}

The standard model of gravity is Einstein's General Relativity (GR).  In GR, all fields of the standard model of particle physics (SM)
 are  universally coupled to gravity by replacing the flat spacetime metric $\eta_{\mu \nu}$ by a curved spacetime one $g_{\mu \nu}$. In ``standard GR''
one also assumes that gravity is the only long range coupling (apart from electromagnetism). We shall see below, how the presence of 
 other long range interactions (coupled to bulk matter) modify the usual ``pure GR'' phenomenology.  The action for the matter sector,
 $S_{SM}$, has the structure
\be
\begin{array}{ccc}
\displaystyle S_{SM} & \displaystyle = & \displaystyle \int \mrm{d}^4 x \Big[ -\frac{1}{4}\sum \sqrt{g}g^{\mu \alpha}g^{\nu \beta} 
F_{\mu \nu}^a F_{\alpha \beta}^a-\sum \sqrt{g}\bar{\psi}\gamma^{\mu}D_{\mu}\psi-\\
\displaystyle & \displaystyle & \displaystyle \frac{1}{2}\sqrt{g}g^{\mu \nu}\overline{D_{\mu}H}D_{\nu}H-\sqrt{g}V\lb H \rb \\
\displaystyle & \displaystyle & \displaystyle  -\sum \lambda \sqrt{g}\bar{\psi} H \psi -\sqrt{g}\rho_{\mrm{vac}}\Big],
\end{array}
\ee
where $D$ denotes a (gauge and gravity) covariant derivative, while the dynamics of $g_{\mu \nu}$ is described by the Einstein-Hilbert
 action, $S_{EH}$,
\be
S_{EH}=\int \mrm{d}^4 x\frac{c^4}{16 \pi G}\sqrt{g}g^{\mu \nu}R_{\mu \nu}\lb g \rb.
\ee
The total action is therefore given by
\be
S=S_{EH}\left[g_{\mu \nu} \right]+S_{SM}\left[\psi,A_{\mu},H;g_{\mu \nu} \right],
\ee
and its  variation w.r.t.  $g_{\mu \nu}$ yields the well-known Einstein field equations
\be
R_{\mu \nu}-\frac{1}{2}R \, g_{\mu \nu}=\frac{8 \pi G}{c^4}T_{\mu \nu}
\ee
where $T^{\mu \nu}=\frac{2}{\sqrt{g}}\frac{\delta \mathcal{L}_{SM}}{\delta g_{\mu \nu}}$.  The universal coupling of any type 
of particle to $g_{\mu \nu}$ is made manifest in $S_{SM}$ while $S_{EH}$ contains all the information on the propagation of 
gravity. For instance, expanding $S_{EH}$ in powers of $h_{\mu \nu}$
(where $g_{\mu \nu} \equiv \eta_{\mu \nu} +h_{\mu \nu} $), one obtains, at 
quadratic order in $h_{\mu \nu}$, the spin-two Pauli-Fierz Lagrangian.  Higher orders in  $h_{\mu \nu}$ contain an infinite series of 
nonlinear self-couplings of gravity: $\partial \partial hhh, \partial \partial hhhh,$ etc. As we shall see, this nonlinear structure 
has been verified experimentally to high accuracy (both in the weak-field regime, where the cubic vertex $\partial \partial hhh$ has 
been checked, and in the strong-field regime of binary pulsars, where the fully nonlinear GR dynamics has been confirmed). In the 
following, we discuss, successively, (i) the experimental tests of the coupling of matter to gravity, 
and (ii) the tests of the dynamics of the gravitational field: kinetic terms (describing the propagation of gravity), and 
cubic  and higher gravitational vertices.

The universal nature of matter's coupling to gravity, \ie the coupling of matter to a universal deformation of spacetime, has many 
experimental consequences.  These experimental consequences can be derived by using a simple theorem by Fermi and Cartan.  Given any
 pseudo-Riemannian manifold, for instance a curved spacetime endowed with a metric $g_{\mu \nu}$, and given any worldline $\mathcal{L}$
 in this spacetime (not assumed to be a geodesic),   there always exists a coordinate system such that, all along $\mathcal{L}$,
 $g_{\mu \nu}\lb x^{\lambda} \rb = \eta_{\mu \nu} +\mathcal{O}\lb \vec{x}^2\rb$, where $\vec{x}$ denotes the {\it spatial} deviation
 away from the central worldline $\mathcal{L}$. It is important to note that there is no linear term in $\vec{x}$, but only $\vec{x}^2$
 effects, \ie tidal effects. There exists a very simple and intuitive demonstration of this Fermi-Cartan theorem.   Let us view the
 curved manifold as being some ``brane'' embedded within a {$\it flat$} ambient auxiliary manifold. For instance, consider an ordinary
 2-surface $\Sigma$ within a three-dimensional flat euclidean space. Given any (smooth) curve $\mathcal{L}$ traced on $\Sigma$, we can
 take a flat sheet of paper and progressively
``apply'' (or ``fit'') this sheet on $\Sigma$ along the curve $\mathcal{L}$. The orthogonal projection
of $\Sigma$ onto this applied flat sheet defines a map from $\Sigma$ to a coordinatized flat manifold
which has the property enunciated above. Note that, in this ``development'' of the neighbourhood
of $\mathcal{L}$ within $\Sigma$ onto a flat sheet, the shape (as seen on the flat sheet)
of the ``developped'' curve $\mathcal{L}$ is generically not a straight line. It is only when $\mathcal{L}$ 
was a {\it geodesic} line on $\Sigma$, that its development will be a straight line.
This proof, and its consequences, are valid in any dimension and signature.
 
Here, we have in mind applying this result to the ``center of mass'' worldline $\mathcal{L}$
of an arbitrary
body moving in a background spacetime, or more generally of any sufficiently small laboratory
(containing several bodies, between which we can neglect gravitational effects). 
We assume that we can neglect the backreaction of
the body (or bodies) on the spacetime. In the approximation where we can neglect the tidal effects
(linked to the $\mathcal{O}\lb \vec{x}^2\rb$ terms in $g_{\mu \nu}$ in Fermi coordinates),
we can consider that we have a body, or a small lab, moving in a flat spacetime. 
In other words, the theorem
of Fermi and Cartan tells us that we can essentially ``efface'' (modulo
small, controllable tidal effects) the background gravitational field
$g_{\mu \nu}$ all along the history of a small lab, or a body. 
This ``effacement property'' is telling us, for instance, that the physical properties
we can measure in a small lab will be independent of where the lab is, and when the measurements
are made. In particular, all the (dimensionless) coupling constants\footnote{We assume here that the
cutoff  length scale  $\epsilon = 1/\Lambda$ of any low-energy effective QFT description of the physics in a small lab is
is fixed, when measured in units of $ds = \sqrt{g_{\mu \nu} dx^{\mu} dx^{ \nu}}$.} that enter the
interpretation of local experiments (such as various mass ratios, the fine-structure constant, etc.)
must be independent of where and when they are (locally) measured ({\it constancy of the constants}). A second consequence
of this effacement property is that local physics should be Lorentz $SO\lb 3,1\rb$ invariant, because
this is a symmetry of the (approximate) flat spacetime appearing after one has
effaced the tidal effects ({\it local Lorentz invariance}).

Moreover, in absence of coupling to
other long-range fields (such as electromagnetism for a charged body), the center of mass
of an isolated body (viewed as moving in a flat spacetime)
must follow  a straight worldline  (principle of inertia).
We therefore conclude (by the theorem above)  that $\mathcal{L}$ has to be a geodesic in the 
original curved spacetime.  This is true independently of the internal properties of the object. 
One may thus conclude that isolated neutral bodies fall along geodesics independently of the internal
 properties of the object, since at no point in the demonstration had we to rely on any internal properties 
of the object. This is therefore a proof of the {\it weak equivalence principle}, \ie all bodies  in a 
gravitational field fall with the same acceleration.  Note, once again, that the absence of other long 
range fields besides $g_{\mu \nu}$ that could influence the object considered is  crucial.  

Finally, another universality property, that  of the gravitational redshift, may be shown by a comparison 
of the GR formulation with the Newtonian one. 
In lowest order approximation, the deviation of the $g_{00}$ component of the metric from $\eta_{00}= -1$ 
is twice the Newtonian potential $U(x)$. Indeed, comparing the action for a geodesic,
\be
S_E=-m \int \mrm{d}t \sqrt{-g_{\mu \nu}\dot{x}^{\mu}\dot{x}^{\nu}}
\ee 
with
\be
S_{\mathrm{Newton}}=\int \mrm{d}t \left[ \frac{1}{2}m\dot{x}^2+mU\lb x \rb \right],
\ee
one finds $g_{00}=-1+\frac{2}{c^2}U\lb x \rb +\mathcal{O}\lb \frac{1}{c^4}\rb$ where $U = \displaystyle \sum_a\frac{G m_a}{|\vec{x}-\vec{x_a}|}$.  

Experimentally, one may transfer electromagnetic signals from one clock to another identical 
clock located in a gravitational field. If we are in a stationary situation (\ie if there exists 
a coordinate system w.r.t.~which the physics is  independent of time $x^0= c t$), the time translation invariance
of the background shows that electromagnetic signals will take a {\it constant coordinate time}
to propagate from clock 1 to clock 2. We can then use the link 
$ d \tau_i =\sqrt{ - g_{00} (\vec{x_i})} dt_i$
between the proper time (at the location of clock $i$ ($i=1,2$)) and the corresponding coordinate time,
as well as the (approximate) result above for $g_{00}$. Finally,
we conclude that two identically constructed clocks located at two different positions in a static
external Newtonian potential exhibit, when intercompared by electromagnetic signals, the (apparent)
difference in clock rate
\be
\frac{\tau_1}{\tau_2}=\frac{\nu_2}{\nu_1}=1+\frac{1}{c^2}\left[U\lb \vec{x_1}\rb - U\lb \vec{x_2} \rb\right]+\mathcal{O}\lb \frac{1}{c^4}\rb .
\ee
This gravitational redshift effect is  proportional to the difference in the Newtonian potential 
between the two locations, independently of the constitution of the clocks (say Hydrogen maser,
or Cesium clock, etc.). This is a property known as the {\it universality of the gravitational redshift}.

The various consequences, discussed above, of the universal character of the coupling of matter to gravity  
are usually summarized under the generic name of {\it equivalence principle}.  In the next section, we discuss 
the experimental tests of the equivalence principle and their accuracy.

\subsection{Experimental tests of the coupling of matter to gravity}

\subsubsection{How constant are the constants?}

The best tests of the ``constancy of the constants'' concern the fine structure 
constant $\displaystyle \alpha = \displaystyle e^2/ \hbar c \simeq 1/137.037$, and the ratio of the electron mass to 
that of the proton $\frac{m_e}{m_p}$ (see Ref.~\cite{U2003} for a review).  There exist several types of 
tests, based, for instance, on geological data (\eg measurements made on the nuclear decay products of 
old meteorites), or on measurements (of astronomical origin) of the fine structure of absorption and 
emission spectra of distant atoms,  as, \eg the absorption lines of atoms on the line-of-sight 
of quasars at high redshift. Such kinds of tests all depend on the value of $\alpha$. There exist, 
in addition, several laboratory tests such as, for example, comparisons made between several different  
high-stability clocks. However, the best measurement of the constancy of $\alpha$ to date is the Oklo
 phenomenon\footnote{ The Oklo phenomenon was discovered by scientists at the 
{\it Commissariat \`a l'\'energie atomique} (CEA) in France.  A study of the uranium ore in a 
Gabonese mine revealed an unusual depletion in $U^{235}$ (used in fission reactors)  w.r.t. the 
usual proportion.   Uranium ore is a mix of two isotopes, with, in usual samples, $99.28\%$ $U^{238}$ and $0.72\%$ $U^{235}$. 
By contrast, the Oklo ore had only $\ll 0.72\%$ of $U^{235}$. It was realized that a natural fission 
process took place, prompted by the presence of ground water, in Oklo some two billion years ago, 
and lasted for about two million years.  Scientists analysed in detail the 2 billion year-old 
fission decay products. One can then infer from these measurements the  scattering cross-sections of slow
neutrons on various isotropes. Then, modulo some further assumptions about the dependence of 
various nuclear quantities on $\alpha$, one could constrain the variation of $\alpha$ between 
 the time of the fission reaction (roughly two billion years ago)
and now. For details about the analysis and interpretation of Oklo data see \cite{DD1996}
and references therein.}. It sets the following (conservative) limits on the variation of $\alpha$ 
over a period of two billion years \cite{S1983,DD1996,F2000}
\be
-0.9\times 10^{-7}< \frac{\alpha^{\mathrm{Oklo}}-\alpha^{\mathrm{today}}}{\alpha^{\mathrm{today}}}<1.2 \times 10^{-7}.
\ee
Converting this result into an average time variation, one finds
\be
-6.7\times 10^{-17}\, \mrm{yr}^{-1}<\frac{\dot{\alpha}}{\alpha}<5\times 10^{-17}\, \mrm{yr}^{-1}.
\ee
Note that this variation  is  a factor of $\sim 10^7$ smaller than the Hubble 
scale, which is itself $\sim 10^{-10}\, \mrm{yr}^{-1}$.   Comparably stringent limits were obtained using 
the Rhenium $187$ to Osmium $187$ ratio in meteorites \cite{O2004} yielding an upper bound 
$ \frac{\Delta \alpha}{\alpha} = (8 \pm 8) \times 10^{-7}$ over $4.6\times 10^9$ years. 
Laboratory limits were also obtained from the comparison, over time, of stable atomic clocks. 
More precisely,  given that $\frac{v}{c} \sim \alpha$ for electrons in the first Bohr orbit, 
direct measurements of the variation of $\alpha$ over time can be made by comparing the frequencies 
of atomic clocks that rely on different atomic transitions.  The upper bound on the variation of 
$\alpha$ using such methods is $ \frac{\dot{ \alpha}} {\alpha} = (-0.9 \pm 2.9) \times 10^{-15}\, \mrm{yr}^{-1}$
\cite{clocks}.  It should be mentioned that a few years ago claims were made concerning 
observational evidence of non-zero
time variations  of $\alpha$ and $\frac{m_e}{m_p}$ from analyses of
some astronomical spectra (see Ref.~\cite{U2003}). Other recent astronomical data indicate
no variability of these constants (see Ref.~\cite{U2003} and the chapter 18 of the Review of 
Particle Physics\footnote{Available on http://pdg.lbl.gov/} for references).

\subsubsection{Tests of local Lorentz invariance}

We should first mention that the Michelson-Morley experiment\footnote{First performed as part of 
a series of experiments, beginning in Potsdam in 1881 (by Michelson alone) and then in the US until 
1887 (by both Michelson and Morley) to test  the existence of the {\it aether}.} has been repeated (with 
high accuracy) and strong limits have been obtained on a possible anisotropy of the propagation of 
light.  In its modern realizations (Brillet and Hall, 1979), it has been performed with laser 
technology on rotating platforms.  This experiment is now viewed as a test of the isotropy of 
space on the moving Earth, and thereby as a test of local Lorentz invariance.  There also exists 
another idea for testing the isotropy of space, and although its interpretation is not totally 
clear, it is a conceptually interesting idea. This is why  we choose to outline it in these lectures. 

For simplicity, consider the hydrogen atom. Assuming the isotropy of space, \ie the existence of a  
SO(3) symmetry, we know that there should exist a degeneracy in the energy levels, given by the 
magnetic quantum number $m$.  However, it is interesting to understand how the SO(3) symmetry 
comes about dynamically (and therefore, how it might be dynamically violated). The Hamiltonian 
for the electron is given by
\be
\hat{H}=- \frac{\hbar^2}{2 m}\triangle -\frac{e^2}{r}
\ee
where the first term is the kinetic term  ($\triangle$ being the Laplacian), and the second term is the 
Coulomb potential.  Note that in fact, $\triangle=\delta^{ij}\partial_{ij}$ and $r^2=\delta_{ij}x^i x^j$, 
such that both terms depend on the same spatial structure $\delta_{ij}$, the flat metric, thereby ensuring 
the SO(3) symmetry. However, both terms also come from an underlying field theoretic formulation: (i) the 
non-relativistic electron kinetic energy term  $\propto \triangle=\delta^{ij}\partial_{ij}$
comes from the kinetic term in the Dirac action, $\bar{\psi}\gamma^{\mu}\partial_{\mu}\psi-m\bar{\psi}\psi$, 
with $\left\{ \gamma^{\mu}, \gamma^{\nu} \right\}=\eta_{\mu \nu}$, while (ii) the $e^2/r$ term is the static 
Green's function of the
electromagnetic field, which comes from inverting the kinetic term
of the photon $\eta^{\alpha \mu} \eta^{\beta \nu} F_{\alpha \beta} F_{\mu \nu}$, which manifestly depends, by 
assumption, on the same spacetime metric $\eta_{\mu \nu}$.  Einstein assumed that,
in order to take into account the coupling to gravity, it was sufficient to replace  $\eta_{\mu \nu}$ by 
the same $g_{\mu \nu}$  both for the electron and the photon. 
By contrast, let us consider the possibility that electrons (``matter'') and  photons (``electromagnetism'')  
have a different coupling to gravity, \eg described by saying that they couple to two different (spatial) metrics, say
\be
\begin{array}{ccc}
g_{ij}^{\mrm{matter}} & = & \delta_{ij}\\
g_{ij}^{\mrm{em}} & = & \delta_{ij}+h_{ij},
\end{array}
\ee
Then, computing the new propagators for the electron and the photon in their respective metrics,
one finds that the SO(3) symmetry would be violated by tensor terms, appearing in the Hamiltonian, 
of the form $ \delta H \sim \frac{e^2}{2}h_{ij}\frac{x^i x^j}{r^3}$. This is a violation, at a deep 
level, of the universality discussed in the previous section.  The usual SO(3) symmetry implies that 
all energy levels with magnetic quantum number $m$ are degenerate.  But if tensor terms violating SO(3) 
were to exist, then, observables effects would include potentially measurable quadrupole-type splittings 
in the energy levels, which, applied to the atomic nucleus (whose energy levels are a more sensitive probe 
of anisotropy), are  $\propto \langle I\,M | \hat{Q}_{ij}|I\,M\rangle$, where $I$ and $M$ are the nuclear 
spin quantum numbers, and where $\hat{Q}_{ij}$ is a symmetric tracefree tensor operator that couples to the 
tracefree part of $h_{ij}$. Such types of measurements have been performed on the energy levels of nuclei 
with impressively high accuracy, the current upper bound being
\be
\left| h_{ij}-\frac{h_{k k}}{3} \delta_{ij} \right|\leq 10^{-27}.
\ee
The universality of space is thus valid to one part in $10^{27}$, showing how delicate Einstein's postulate is.

\subsubsection{Universality of free fall}

The most recent limits  on the deviation from the universality of free fall have been obtained by
Eric's Adelberger's group \cite{Adelberger}. In particular, they  compared the acceleration of 
a Beryllium mass and a Copper one in the Earth's gravitational field and found
\be
\lb \frac{\Delta a}{a} \rb _{\mrm{Be}-\mrm{Cu}}=\lb -1.9\pm 2.5\rb \times 10^{-12},
\ee
where $\Delta a=a_{\mrm{Be}}-a_{\mrm{Cu}}$. Other limits exist, such as, for instance, the fractional 
difference in acceleration of  earth-core-like ($\sim$ iron) and moon-mantle-like (silica) bodies,
\be
\lb \frac{\Delta a}{a} \rb _{\mrm{Earth-core-Moon-mantle}}=(3.6 \pm 5.0) \times 10^{-13}.
\ee
There are also excellent limits concerning celestial bodies. In particular the possible difference 
in the accelerations of the Earth and the  Moon  towards the Sun have been measured using laser
 ranging (with 5 mm accuracy)
 with retro-reflectors (corner cubes) placed on the Moon, giving the result \cite{Williams:2004qba}
\be
\lb \frac{\Delta a}{a} \rb _{\mrm{Earth-Moon}}=(-1.0\pm 1.4) \times 10^{-13}.
\ee
One should, however, remember that only a fraction ( $\sim 1/3$) of the Earth mass is 
made of iron, while the rest is mostly silica (which is the main material the Moon is made of).  
As, independently of the equivalence principle, silica must fall like silica,  one looses a factor 
3, so that the resulting bound on a possible violation of the equivalence principle is only 
around the $5 \times 10^{-13}$ level,
which is comparable to laboratory bounds.

\subsubsection{Universality of the gravitational redshift}
We conclude the section on the tests of the coupling of matter to gravity by just mentioning 
that the universality of the gravitational redshift, namely the apparent change in the frequencies 
of two similar clocks in a gravitational field, has been tested by comparing the frequencies of 
hydrogen masers at the Earth surface and in a rocket.  Vessot and Levine (1979) in Ref.~\cite{VL1979} 
verified that the  fractional change in the measured frequencies is consistent with GR to the $10^{-4}$ level:
\be
\frac{\Delta \nu}{\nu}=\lb 1 \pm 10^{-4} \rb \frac{\Delta U}{c^2}.
\ee
The universality of this redshift has also been verified by measurements
involving other types of clocks.

\subsection{Tests of the dynamics of the gravitational field}

\subsubsection{Brief review  of the theoretical background}

Until now we have only considered the  coupling between matter and gravity, and various tests 
of its universality. We now discuss the tests of the {\it dynamics of the gravitational field}, \ie 
tests  probing either the propagator of the gravitational field, or the cubic or  higher order
gravitational vertices (for more detailed reviews see Refs.~\cite{W2001,Dpdg}). We first consider the 
weak field regime, regime in which we can write $g_{\mu \nu}=\eta_{\mu \nu}+h_{\mu \nu}$, where $h_{\mu \nu}$ 
is numerically much smaller than one. For instance, in the solar system, $h_{\mu \nu}\sim 10^{-6}$ on the 
surface of the Sun,  $\sim10^{-8}$ on the Earth orbit around the Sun, 
or $\sim 10^{-9}$ on the Earth surface. With values so small, it is clear that the solar
system will not allow one to test many nonlinear terms in the perturbative expansion of $g_{\mu \nu}$. 
 
 We start by considering the gravitational interaction between two particles of masses $m_A$ and $m_B$.
 At linear order in $h_{\mu \nu}$, we will have an interaction corresponding to the following 
 (classical Feynman-like) graph
\begin{center}
\begin{fmffile}{fig1}
\begin{fmfchar*}(40,25)
  \fmfleft{i1,i2}
  \fmfright{o1,o2}
  \fmf{fermion}{i1,v1,i2}
  \fmf{photon,tension=0,label=$h_{\mu \nu}$}{v1,v2}
  \fmf{fermion}{o1,v2,o2}
\end{fmfchar*}
\end{fmffile}
\end{center}
To compute explicitly what  the preceding graph means, we must start from
the full action describing two gravitationally interacting  bodies $A$ and $B$:
\be
S = -m_A\int \mrm{d}s_A-m_B \int \mrm{d}s_B+\int \frac{\sqrt{g}R}{16 \pi G}.
\ee
Expanding $S$ in the deviations of $g_{\mu \nu}$ away from $\eta_{\mu \nu}$, one obtains
(denoting $h \equiv \eta^{\mu \nu} h_{\mu \nu}$)
\be
\begin{array}{ccc}
\displaystyle S & \displaystyle = & \displaystyle -m_A\int \sqrt{-\eta_{\mu \nu}\mrm{d}x_A^{\mu}\mrm{d}x_A^{\nu}}-m_B\int \sqrt{-\eta_{\mu \nu}
\mrm{d}x_B^{\mu}\mrm{d}x_B^{\nu}}+\\
\displaystyle & \displaystyle & \displaystyle \frac{1}{2}\int h_{\mu \nu}T_{A}^{\mu \nu}+\frac{1}{2}\int h_{\mu \nu}T_{B}^{\mu \nu}+
\int\frac{1}{32 \pi G} h^{\mu \nu}
\Box  \lb h_{\mu \nu} - \frac{1}{2} h \eta_{\mu \nu} \rb + \\
\displaystyle & \displaystyle & \displaystyle \mathcal{O}(h^2 T) + \mathcal{O}(h^3),
\end{array}
\ee
where $T_{A}^{\mu \nu}$ is the (flat-space limit) of the stress-energy tensor of particle $A$ 
(given by a $\delta$-function localized on the worldline of $A$),
and where the kinetic term of $h_{\mu \nu}$ is the one corresponding to the harmonic gauge ({\it i.e.}, $\partial_{\nu}
 \lb \sqrt{g}g^{\mu \nu} \rb=0$).  Inverting this kinetic term yields for $h_{\mu \nu}$
the following lowest-order equation (corresponding to Einstein's equations at  linearized order)
\be
\Box h_{\mu \nu}=-\frac{16 \pi G}{c^4}\lb T_{\mu \nu}-\frac{1}{D-2}T \eta_{\mu \nu}\rb.
\ee
with  $T_{\mu \nu} = \eta_{\mu \alpha}\eta_{ \nu \beta} (T_{A}^{\alpha \beta}+T_{B}^{\alpha \beta})$.
  We can then ``integrate out''  $h_{\mu \nu}$ by solving the latter field equation for $h$,
and replacing the result in the original action. Modulo self-interaction terms $\propto T_A^{\mu \nu}\Box^{-1}P_{ \mu \nu \rho
 \sigma} T_A^{\rho \sigma}$, 
the action then splits into the sum of  three terms, a term  
$-m_A\int \sqrt{-\eta_{\mu \nu}\mrm{d}x_A^{\mu}\mrm{d}x_A^{\nu}} $, describing the free propagation of body $A$, a 
similar term for $B$, and an interaction term,
\be
S^{\mrm{int}}=-\frac{8 \pi G}{c^4} \int T_A^{\mu \nu}\Box^{-1} \lb T^B_{\mu \nu}-\frac{1}{D-2}T^B \eta_{\mu \nu} \rb.
\ee

More explicitly, if we introduce the scalar Green's function $G(x)$, such that $\Box G(x) = - 4 \pi \delta^D(x)$, 
this lowest-order interaction term reads
\be
\begin{array}{ccc}
\displaystyle S^{\mrm{int}} & \displaystyle = & \displaystyle 2 G \int \int \mrm{d}s_A 
\mrm{d}s_B m_A u_A^{\mu} u_A^{\nu}P_{\mu \nu}^{\rho \sigma} \\
\displaystyle & \displaystyle & \displaystyle G\lb x_A \lb s_A \rb- x_B \lb s_B \rb \rb m_B u_{B \rho} u_{B \sigma},
\end{array}
\ee
in which one easily identifies the usual structure of a Feynman graph (namely the one
depicted above), with the coupling constant $G$ in front, and a graviton  propagator (comprising the
scalar Green's function, together with the spin-two projection operator $P_{\mu \nu}^{\rho \sigma}$,
which can be read off the previous explicit result) sandwiched between two  source terms.  

\begin{figure}
\includegraphics[width=10cm]{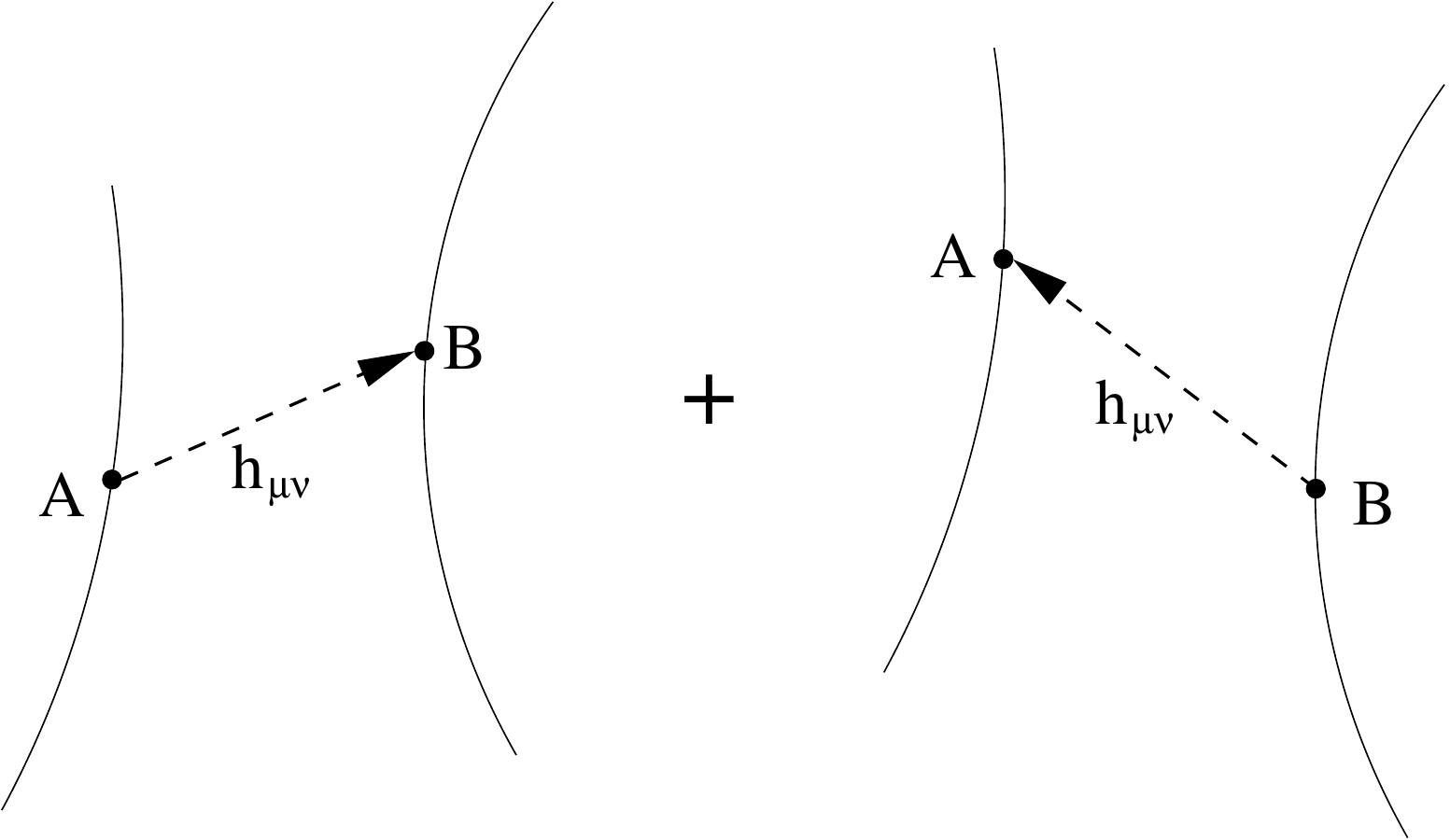}
\caption{Time-symmetric half-advanced half-retarded contributions to
the gravitational interaction between particles A and B.}
\label{RetardedGreenFct}
\end{figure}

In the stationary approximation, the scalar Green's function reduces to the usual Newtonian propagator
$1/r$. If one further neglects the relative velocity of the two worldlines one can replace
the spacetime velocities $u_A^{\mu}$ and $u_B^{\mu}$ by $(1,0,0,0,\cdots)$. This yields the
usual Newtonian interaction term $G \int dt m_A m_B/r$. However, the ``one-graviton exchange''
diagram above contains many Einsteinian effects that go beyond the Newtonian approximation.
To compute them explicitly, we first need the explicit expression of the relativistic
scalar Green's function $G(x)$. As we are deriving here the part of the gravitational
interaction which is ``conservative'' (\ie energy conserving), we must use the {\it time-symmetric} 
(half-advanced half-retarded) Green's function. In four dimensions, 
it is given by $\delta \lb ( x_A-x_B)^2 \rb $. 
It is the sum of two terms (a retarded and an advanced one), as depicted in 
FIG. \ref{RetardedGreenFct}. Note in passing that this classical time-symmetric propagator
corresponds to the real part of the Feynman propagator. Indeed, for a massless scalar particle
in $D=4$ the Feynman propagator  (in $x$ space) is proportional to
\be
\frac{i}{x^2+i \varepsilon}=i PP \frac{1}{x^2}+\pi \delta \lb x^2 \rb,
\ee
where the first term on the r.h.s. is a distributional ``principal part'' (it is pure imaginary and
``quantum''), while the second (real) term is the classical contribution  (classically the interaction 
propagates along the light cone, see FIG. \ref{RetardedGreenFct}).  Note that, contrary to the
Newtonian picture where the interaction is instantaneous, we have here an interaction which depends  both 
on the future and on the past\footnote{This ``acausal'' behaviour is due to our considering
the conservative (``Fokker'') action. If we were computing the ``real'' classical equations of
motion of the two particles, we would use only a {\it retarded} Green's function. The equations
of motion so obtained would then be ``causal'' and would automatically contain some (physically needed),
time-asymmetric ``radiation reaction'' terms. The trick, used here, to employ an acausal
time-symmetric Green's function is a technical shortcut allowing one to derive the action
yielding the conservative part of the equations of motion.}.
When considering the case (of most importance  in applications) where  $A$ and $B$ move slowly relative to $c$, 
the time-symmetric propagator can be expanded in powers of $1/c$. The first term in this
expansion yields the usual Newtonian instantaneous interaction, while all the higher-order terms can be
 expressed in terms of successive derivatives of the positions of $A$ and $B$, (so that the
acausality formally disappears, and is replaced by a dependence of the Lagrangian on
derivatives higher than the velocities, \ie accelerations, derivatives of accelerations, and so forth.
Actually, such higher-derivative terms start appearing only at the so-called 
``second post-Newtonian'' (2PN) order, \ie the
order $\mathcal{O} \lb \frac{1}{c^4} \rb$. Such higher-order post-Newtonian (PN) contributions
are important for some applications (binary pulsars, coalescing black holes),
and have been computed up to the 3PN ($\mathcal{O} \lb \frac{1}{c^6} \rb$),
as well as 3.5 PN ($\mathcal{O} \lb \frac{1}{c^7} \rb$) levels. Here we shall consider only
the first post-Newtonian, 1PN, level, \ie $\mathcal{O} \lb \frac{1}{c^2} \rb$.
At this level, the action can be written entirely in terms of the velocities of $A$
and $B$ (taken at the same instant $t$ in some Lorentz frame). By expanding the 
time-symmetric acausal one-graviton-exchange action written above to order $1/c^2$ one finds the following explicit 
1PN Lagrangian (now considered for an $N$-body system made of 
masses labelled $A,B = 1,2, \cdots, N.$)
\be
\begin{array}{ccc}
\displaystyle \mathcal{L}^{2-\mrm{body}} & \displaystyle = & \displaystyle \frac{1}{2}\sum_{A \ne B} \frac{G m_A m_B}{r_{AB}}\Big[1+\frac{3}{2c^2} \lb \vec{v_A}^2+\vec{v_B}^2
 \rb- \\
\displaystyle & \displaystyle & \displaystyle \frac{7}{2c^2}\vec{v_A}.\vec{v_B}-\frac{1}{2 c^2}\lb \vec{n}_{AB}.\vec{v_A} \rb \lb \vec{n}_{AB} . \vec{v}_B \rb + \mathcal{O}
 \lb \frac{1}{c^4} \rb \Big],
\end{array}
\ee
Note that the coefficients $3/2$, $7/2$, $1/2$, etc., arise from the spin 2 nature of the 
graviton, \ie they are uniquely fixed by  Einstein's propagator.  

When considering a gravitationally bound $N$-body system, we must remember that there is a link
$v^2\sim\frac{GM}{r}$, due to the ``virial theorem''. This link says that
the $\frac{v^2}{c^2}$ contributions in the one-graviton-exchange graph considered above
must be completed by computing non-linear interaction graphs containing more gravitons,
namely the ones of order $G^2/c^2$ involving two powers of the coupling constant $G$.
These contributions correspond to the terms $\mathcal{O}(h^2 T)$ or  $\mathcal{O}(h^3)$ in the $h$-expansion
of the exact Einstein action. In terms of Feynmam-like diagrams, this means the following graphs:
\begin{center}
\begin{fmffile}{fig2}
\begin{fmfgraph}(25,25)
  \fmfleft{i1,i2}
  \fmfright{o1,o2}
  \fmf{plain}{i1,v1,i2}
  \fmf{wiggly,tension=1/6}{v1,v2}
  \fmf{wiggly,tension=1/6}{v1,v3}
  \fmf{plain}{o1,v2,v3,o2}
\end{fmfgraph}
\begin{fmfgraph}(25,25)
  \fmfleft{i1,i2}
  \fmfright{o1,o2}
  \fmf{plain}{i1,v1,i2}
  \fmf{wiggly,tension=1/3}{v1,v2}
  \fmf{wiggly,tension=1/4}{v2,v3}
  \fmf{wiggly,tension=1/4}{v2,v4}
  \fmf{plain}{o1,v3,v4,o2}
\end{fmfgraph}
\begin{fmfgraph}(25,25)
  \fmfleft{i1,i2}
  \fmfright{o1,o2,o3,o4}
  \fmf{plain}{i1,v1,i2}
  \fmf{wiggly,tension=1/6}{v1,v2}
  \fmf{wiggly,tension=1/6}{v2,v3}
  \fmf{wiggly,tension=1/6}{v2,v4}
  \fmf{plain}{o1,v3,o2}
  \fmf{plain}{o3,v4,o4}
\end{fmfgraph}
\begin{fmfgraph}(25,25)
  \fmfleft{i1,i2}
  \fmfright{o1,o2,o3,o4}
  \fmf{plain}{i1,v1,i2}
  \fmf{wiggly,tension=1/6}{v1,v2}
  \fmf{wiggly,tension=1/6}{v1,v3}
  \fmf{plain}{o1,v2,o2}
  \fmf{plain}{o3,v3,o4}
\end{fmfgraph}
\end{fmffile}
\end{center}
Note, in particular, that these terms involve the graviton cubic vertex (whose
structure will therefore be probed by solar-system experiments). Note also that
some of these terms involve only two bodies (being proportional, say, to $m_A m_B^2$),
while others can involve three distinct bodies ($\propto m_A m_B m_C$).
The full $G^2/c^2$ result (containing both two-body and three-body terms) is found to
be equal to
\be
\mathcal{L}^{3-\mrm{body}}=-\frac{1}{2}\sum_{B \ne A \ne C} \frac{G^2 m_A m_B m_C}{r_{AB} r_{AC} c^2},
\ee
where the factor of $1/2$ is a prediction from Einstein's theory. Note that the summation
is restricted by $B \ne A \ne C$ which allows for the two-body terms where $B=C$.

When looking at the nonlinear diagrams above one sees some ``loops'' made by graviton
propagator lines closing up on a matter worldline. This may seem paradoxical because
we are considering here classical gravitational effects, and classical theory is
usually thought of as involving only tree diagrams. Indeed, if we replace all our
``source worldlines'' (drawn above as continuous worldlines) by separate external
sources (\ie by replacing the line describing $T_{A}^{\mu \nu}(x)$ by  separate ``blobs''
on which graviton propagators start or end), we see that all the diagrams above open up
and become tree diagrams. However, the presence of loops in the diagrams used here do
correspond to essentially some of the same physical effects that ``quantum loops''
describe. This is particularly clear for the diagrams below (which are included
in the classical calculations)
\begin{center}
\begin{fmffile}{fig3}
\begin{fmfgraph}(25,25)
  \fmfleft{i1,i2}
  \fmfright{o1,o2}
  \fmf{plain,tension=1/3}{i1,v1,v2,i2}
  \fmf{wiggly,right,tension=0}{v1,v2}
  \fmf{plain}{o1,o2}
\end{fmfgraph}
\hspace{2cm}
\begin{fmfgraph}(25,25)
  \fmfleft{i1,i2}
  \fmfright{o1,o2}
  \fmf{plain,straight}{i1,v1,v2,i2}
  \fmf{wiggly,right,tension=0}{v1,v2}
  \fmf{plain}{o1,v3,o2}
  \fmf{wiggly,tension=0}{v3,v2}
\end{fmfgraph}
\end{fmffile}
\end{center}
It is clear that these diagrams describe the
self-gravity effects of a mass $m_A$ on itself. As such, they do describe the classical
limit of quantum loops such as the simplest one-loop diagram
\begin{center}
\begin{fmffile}{fig4}
\begin{fmfgraph}(24,24)
\fmfleft{i1}
\fmfright{i2}
\fmf{plain,tension=1/3}{i1,v1,v2,i2}
\fmf{wiggly,right,tension=0}{v1,v2}
\end{fmfgraph}
\end{fmffile}
\end{center}
which describes the back action  of the emission and reabsorption of a graviton on a quantum particle.
Another similarity beween ``classical loops'' and quantum ones, is that, in practice, multi-loops are associated 
to the presence of multiple integrals which are increasingly difficult  to compute.\footnote{Note
that classical diagrams must all be computed in $x$-space. An increasing number of loops
signals the presence of intermediate vertices in $x$-space on which one must integrate.}
In addition, the loop diagrams depicted in the penultimate graph lead (like quantum loops)
to formally divergent integrals. The origin of these divergences is that we have been
describing the gravitationally interacting bodies as ``pointlike'', \ie mathematically
described by a $\delta$-function (on a worldline). There are several ways of dealing with this
technical problem. One can complete the formal perturbative calculations done with point-like bodies
by another approximation scheme in which each body is locally viewed (in its own rest frame)
as a weakly perturbed isolated body. The development of such a dual perturbation 
method \cite{DEath75,DamourLH83} shows that the non-point-like, internal structure of
 non-rotating {\it compact} bodies (neutron
stars or black holes) will enter their translational dynamics only at the 5PN level ($\sim G^6/c^{10}$),
corresponding to 5 loops ! Knowing this, one expects that the use of a gauge-invariant regularization
method for treating gravitationally-interacting point masses should give a physically
unique answer up to 5 loops (excluded). Using {\it dimensional regularization}, one finds that
all self-gravity effects are unambiguous and finite at 1PN and 2PN \cite{DamourLH83}. Recent work
has pushed the calculation to the 3PN (\ie 3 loop) level. Again, one finds, either 
(when using a convenient
gauge) a finite answer \cite{Damour:2001bu}, or, when using the harmonic gauge, an equivalent
answer after renormalizing the position of the worldline used in the $\delta$-function
source \cite{Blanchet:2003gy}.

\subsubsection{Experimental tests in the solar system}

The 1PN-level results described above are accurate enough for describing the gravitational
dynamics in the solar system.
Testing the validity of GR's description of the gravitational field's dynamics
 is then achieved by verifying the agreement of the coefficients introduced above with experimental
 measurements.  In this section, we will see that these GR-predicted coefficients agree with their
 experimentally measured value to better than the $10^{-5}$ level for the coefficients
entering the one-graviton-exchange term, and to about the
$10^{-3}$ level for the additional 1PN multi-graviton term. There are many observables that can be used
to test relativistic gravity in
the solar system. One may use the advance of the perihelion of planets, the deflection, by the local
 curvature, of light reaching the Earth from distant stars, the additional time delays suffered by
 electromagnetic signals compared to their flat spacetime counterparts, or also, general relativistic
 corrections to the Moon's motion using the laser ranging technique already mentioned in previous sections.

When testing  Einstein's predictions it is convenient to embed GR within a class of
alternative gravity theories. For instance, one could consider not only the interaction of matter with
 the usual Einstein (pure spin-2) graviton but also an interaction with a long-range scalar field
$\varphi$, \ie a spin-zero massless field, coupled to the trace of $T^{\mu \nu}$ with
strength $ \sqrt{G} \alpha(\varphi)$. This leads to an additional attractive force, so that the
 effective gravitational constant measured in a Cavendish
experiment is $G_{\mrm{eff}}= G (1 + \alpha^2)$. This also modifies the $v^2/c^2$ terms in the two-body
action introduced above by terms proportional to $\alpha^2$. These modifications are often
summarized by writing the 1PN-level metric generated by an $N$-body system in the form
\be
\begin{array}{ccc}
\displaystyle \mrm{d}s^2& \displaystyle = &\displaystyle -\lb 1-\frac{2 U}{c^2}+2\lb 1+\bar{\beta}\rb \frac{U^2}{c^4}\rb c^2 \mrm{d}t^2+\nonumber \\
\displaystyle & \displaystyle & \displaystyle \lb 1+2\lb 1+\bar{\gamma} \rb \frac{U}{c^2}\rb \delta_{ij}\mrm{d}x^i \mrm{d}x^j,
\end{array}
\ee
where $U = G_{\mrm{eff}} \sum_A m_A/r_A$ is the (effective) Newtonian potential and where the
two dimensionless coefficients $\bar{\beta}$ and $\bar{\gamma}$ encode the two possible
(Lorentz-invariant) deviations from GR which enter the 1PN level. For an additional
coupling to a scalar with $\varphi$-dependent coupling strength $\alpha(\varphi)$, one finds
 that the ``post-Einstein''
parameter $\bar{\gamma}$ is given by
$\bar{\gamma} = - 2 \alpha^2/(1 + \alpha^2)$. As for the other ``post-Einstein'' parameter
$\bar{\beta}$ it measures a possible modification of the (cubic-vertex related)
three-body action $\mathcal{L}^{3-\mrm{body}}$ written above, and it is given by
$\bar{\beta} = + \frac{1}{2} \beta \alpha^2/(1 + \alpha^2)^2$ where $\beta$ denotes the
derivative of the scalar coupling $\alpha$ w.r.t. the field $\varphi$.

The most accurate test of GR in the solar system is the one made using the Cassini spacecraft
 by the authors of Ref.~\cite{BIT2003}. This test is (essentially) only sensitive to the
post-Einstein parameter $\bar{\gamma}$ (\ie it depends only on the graviton propagator and 
the coupling to matter but not on nonlinear terms).
This experiment used electromagnetic signals sent from the
Earth to the Cassini spacecraft and transponded back to Earth, and monitored the
ratio between the electromagnetic frequency $\nu + \Delta \nu$ recorded back on Earth to the
 initial frequency $\nu$. This ratio was used  to probe the change in the geometry of
 spacetime in the vicinity of the Sun as the line of sight moved, especially when it was nearly
 grazing the Sun.  The theoretical prediction for the experimental quantity measured in this experiment is
\be
\lb \frac{\Delta \nu}{\nu}\rb^{\mrm{2-way}}=-4 \lb 2+\bar{\gamma}\rb \frac{G M_{\mrm{sun}}}{c^3 b}\frac{\mrm{d}b}{\mrm{d}t}
\ee
where $b$ is the impact parameter, \ie the distance of closest approach of the signal's trajectory
 to the center of the Sun. The experimental data gave the following result for the parameter $\bar{\gamma}$
\be
\bar{\gamma}=\lb 2.1 \pm 2.3 \rb \times 10^{-5}.
\ee
This confirms GR (namely $\bar{\gamma}^{GR}=0$) to the $\mathcal{O}(1) \times 10^{-5}$ level.

The three-graviton vertex can be probed by considering a body, having a non-negligible
gravitational self-binding energy,  
in an external gravitational field. Indeed, as emphasized by Nordtvedt \cite{Nordtvedt:1968qr,Nordtvedt:1968qs},
 the free fall acceleration of
a self-gravitating body is, in most gravity theories (except GR) sensitive to
its gravitational binding energy. For instance, the Earth and the Moon will
have, in a general theory, a slightly different acceleration of free fall towards the Sun.
The effect is proportional to the combination of post-Einstein parameters $4 \bar{\beta}-\bar{\gamma}$.
Lunar laser ranging data have allowed one to put a stringent upper limit on such a possibility
\cite{Williams:2004qba}, namely
\be
4 \bar{\beta}-\bar{\gamma}= (4.4  \pm 4.5) \times 10^{-4}.
\ee
Thus, to date, predictions of Einstein's theory in the linear (one-graviton-exchange)
approximation have been verified to the $10^{-5}$ level, while some of the
cubically nonlinear aspects have been verified to the $10^{-3}$ level. 

The tests discussed up to now concern the quasi-stationary, weak-field regime, as it
can be probed in the solar system. We shall now discuss the tests obtained in binary
 pulsar data, which have gone beyond the solar-system tests in  probing part of the strongly nonlinear regime of gravity.

\subsubsection{Objects with strong self-gravity: binary pulsars}

Binary pulsars were discovered by Hulse and Taylor in 1974.  Such systems are made  of two
 objects going around each other in very elliptical orbits. Both objects are neutron 
stars\footnote{Except in a few cases where the companion is another compact (though less compact) star
remnant, a white dwarf.}, of which one is a pulsar, \ie a rotating, magnetized object
 that emits a beam of electromagnetic noise (which includes radio waves, as well as other parts of the
electromagnetic spectrum). When one looks at the geometry generated by a neutron star, and computes
 deviations from flat space of the metric components, one finds (on the surface of the star)
\be
g_{00}=-1+\frac{2GM}{c^2 R}\simeq -1+0.4
\ee
for a star of (typical) mass $1.4M_{\odot}$, and  radius $R=10$ km.  This is a 40\% deviation from flat space.  
By contrast, we recall that, in the solar system, the largest metric deviation
from flat space occurs on the Sun's surface and is of order $G M/(c^2 R) \sim 10^{-6}$. We should
therefore a priori expect that such objects might provide tests that go beyond the solar system ones
in probing some of the strong-field aspects of relativistic gravity.
In addition,  in the solar system, the time-irreversible radiative aspects of gravity
(\ie radiation reaction ) are negligible (this is why we focussed above on
time-symmetric interactions).  Here, not only are  strongly nonlinear effects relevant, but one must also
 take into account the time-dissymetric  effects linked to using a {\it retarded} 
propagator  $\propto \delta \lb t-\frac{\left| x_A-x_B \right|}{c}\rb / \left| x_A-x_B \right|$. The
 corresponding time delay $\frac{\left| x_A-x_B \right|}{c}$ is  typically $\simeq 1$ sec (since
 typical separations between the two objects are of order $300,000$ km) and plays an essential role in the 
equations of motion of a binary pulsar. As a consequence, binary pulsars have given us firm
experimental evidence for the reality of gravitational radiation, and for the fact that 
on-shell gravitational radiation  is described by two transverse tensorial degrees of freedom travelling at the velocity of light.

In practice, only a subset of the known binary pulsars can be used for testing the strong nonlinear
 regime of GR, and/or its radiative regime.  Among these, the very best ones are PSR1913+16
 (where the numbers 19h13m and $ +16 \deg$ measure angles on the sky), PSR1534+12, PSRJ1141-6545, and
 PSRJ0737-3039, the first {\it double}  binary pulsar (made of two radio
pulsars, simultaneously emitting toward the Earth).

>From the theoretical point of view, methods have been developed to deal with strongly self-gravitating
 objects, both in Einstein's theory and in alternative theories (see \cite{Damour:2007ti}
for a recent review, and references).
To adequately discuss the observations of binary pulsars, one has had to push  the post-Newtonian perturbative
 calculation to the 2.5 PN level, \ie to order $(v/c)^5$. This  odd power of the ratio $v/c$ is linked to
 the time-dissymetric, retarded nature of the propagator (together with
some nonlinear effects). It is the first PN level where radiation reaction effects arise.
Any experimental test of the presence of such $(v/c)^5$ terms in the equations of motion is
a probe of the reality of gravitational radiation. In addition, as mentionned above, one must
carefully treat, and disentangle, the various strong-field effects that are linked to the
self-gravity of each neutron star in the system.

The existing experimental tests are based on the {\it timing} of  binary pulsars.  Each time the beam of
 radio waves sweeps across the Earth, one observes a pulse of electromagnetic radiation. The data
consists in recording the successive arrival times, say $t_N$ (with $N=1,2,3, \cdots$), of these pulses. 
 Were the pulsar fixed in space, these arrival times would be equally spaced in time,
\ie $t_N = t_0 + N P$, where $P$ would be the (fixed) period of the pulsar. However, in a binary
pulsar, the sequence of arrival times is a more complicated function of the integer $N$ than such a simple
 linear dependence. Indeed, one must take into account many effects: the fact that the pulsar
moves on an approximately elliptical orbit, the deviation of this orbit 
from a usual Keplerian ellipse, the deviation of the orbital velocity from the usual Kepler areal velocity
 law, the existence of various additional relativistic effects: gravitational redshift, second-order (relativistic) Doppler
effect, time-delay when the electromagetnic pulse passes near the companion, radiation reaction
effects in the orbital motion, etc.  To compute all these effects, one needs to solve Einstein's equations
 of motion with high ($ \sim (v/c)^5$) accuracy.

The final result of these theoretical calculations is to derive the so-called {\it DD timing formula},
which gives the $N^{\rm th}$ pulse arrival time $t_N$ as an explicit function of various 
``Keplerian'' ($p^{\mrm{K}}$), and ``post-Keplerian'' ($p^{\mrm{PK}}$) parameters, say
\be
t_N-t_0=F\left[ N;p^{\mrm{K}};p^{\mrm{PK}}\right].
\ee
Here, the Keplerian parameters ($p^{\mrm{K}}$) comprise parameters that would exist in a purely
Keplerian description of the timing: the orbital period $P_b$, the eccentricity of the orbit $e$,
the time of passage at some initial
periastron $T_0$, and some corresponding angular position of the periastron $\omega_0$, and, finally,
 the projected semi-major axis $x=\frac{a_{1}\mrm{sin}i}{c}$, where $a_1$ is the semi-major axis of
 the orbit of the observable pulsar\footnote{In general, only one of the two objects, here labelled as $1$, is
 a pulsar.} and $i$ is the inclination angle w.r.t. the plane of the sky.  The post-Keplerian parameters ($p^{\mrm{PK}}$) then correspond
 to many relativistic effects that go beyond
a Keplerian description, namely: a dimensionless parameter $k$ measuring the 
progressive advance of the periastron $k=\langle \dot{\omega} \rangle P_b/2 \pi$, a parameter
$\gamma_{\mrm{t}}$ measuring the combined second-order Doppler  and gravitational redshift effects,
possible secular variations in Keplerian
parameters $\dot{e}$, $\dot{x}$, $\dot{P}_b$,  two parameters $r$, $s$ measuring the ``range'' and the
 ``shape'' of the additional time delay that appears when the radio waves pass near the
companion, and finally a parameter $\delta_{\theta}$ measuring the distortion of the
orbit w.r.t an ellipse.  By least-squares fitting the observed arrival times $t_N^{\rm obs}$ to the above 
general theoretical timing formula one can accurately determine the numerical values of all the
 Keplerian parameters, as well as some of the post-Keplerian ones.
At this stage, the determination of these phenomenological parameters is (in great part)
independent of the choice of a theory of gravity. On the other hand, in any specific
theory of gravity, each post-Keplerian parameter is predicted to be some well-defined function of the
 Keplerian parameters and of the two masses, $m_1$ and $m_2$, of the pulsar and its companion. 
 For instance, {\it within GR} the advance of the periastron is given by
\be
k^{\rm GR}\lb p^{\mrm{K}},m_1,m_2 \rb=\frac{3}{c^2}\frac{\lb G M n \rb ^{2/3}}{1-e^2},
\ee
where $n=2\pi /P_b$ and $M=m_1+m_2$, while the secular variation of the orbital period (caused
by radiation reaction effects) is given by
\be
\dot{P}_b ^{\rm GR}\lb p^{\mrm{K}},m_1,m_2 \rb =-\frac{192 \pi}{5 c^5}\frac{1+\frac{73}{24}e^2+\frac{37}{96}e^4}{\lb 1-e^2 \rb ^{7/2}}
 \lb G M n \rb ^{5/3}\frac{m_1 m_2}{M^2}.
\ee
Note that while $k$ is proportional to $1/c^2$ (1PN level), the secular variation
of the orbital period is proportional to $1/c^5$ (and is indeed numerically of
order $(v/c)^5$). The GR-predicted value for $\dot{P}_b$ is a direct reflection of the presence of $\mathcal{O}\lb(v/c)^5\rb$ 
time-asymmetric radiation damping terms in the equations of motion. Numerically, $\dot{P}_b$
(which is dimensionless) is predicted to be of typical order of magnitude 
 $\dot{P}_b \sim 10^{-12}$, which seems very small, but happens to be large enough to be
 measured with good accuracy in several binary pulsars. 
   
The crucial point to notice is that the GR predictions (of which two are given here
as examples) for the link between the post-Keplerian parameters and the masses are
specific to the structure of GR, and will be replaced, in other theories of gravity,
by different functions $k^{\rm theory}\lb p^{\mrm{K}},m_1,m_2 \rb, \dot{P}_b ^{\rm theory}\lb p^{\mrm{K}},m_1,m_2 \rb$, 
etc.  In particular, it has been explicitly shown in various cases
(and notably in the case of generic tensor-scalar theories where gravity is mediated both
by a spin-2 field and a spin-0 one) that the large self-gravity of neutron stars would
generically enter these functions, and drastically modify the usual prediction of GR,
see \cite{DEF1996}.

To see which theory of gravity is in agreement with pulsar timing data, one can proceed
as follows. Within each theory of gravity, the measurement of each post-Keplerian parameter
defines a corresponding curve in the $m_1$, $m_2$ plane. Therefore, in general, the
measurement of two post-Keplerian parameters is sufficient to determine the (a priori unknown)
numerical values of the two masses $m_1$ and $m_2$ (as the location where the two
curves intersect). Then, the measurement of any additional  post-Keplerian parameter
yields a clear test of the validity of the theory considered:  the corresponding
third curve should pass  precisely through  the intersection point of the first two curves.
If it does not, the theory is invalidated by the binary pulsar data considered. By the same
reasoning, the measurement of $n$ different  post-Keplerian parameters yields $n-2$
tests of the underlying theory of gravity.  Many such stringent tests have been obtained
in  binary pulsar observations (more precisely, nine different tests in all have been
obtained when considering the data from four binary pulsars). 
Remarkably, {\it GR has been found to be consistent
with all these tests}. Many alternative gravity theories have fallen by the wayside, or
their parameters have been constrained so as to make the theory extremely close to GR
in all circumstances (including strong-field ones).
 
 Let us just give two 
impressive examples of the beautiful agreement between GR and pulsar data.
In the case of the original Hulse-Taylor pulsar PSR1913+16 the ratio between the observed value 
of $\dot{P_b}$ to that predicted by GR is given by
\be
\left[ \frac{\dot{P}_b^{\mrm{obs}}-\dot{P}_b^{\mrm{gal}}}{\dot{P}_b^{\mrm{GR}}\left[k^{\mrm{obs}},\gamma^{\mrm{obs}}\right]}
\right]=1.0026 \pm 0.0022,
\ee
where $\dot{P}_b^{\mrm{gal}}$ is a Galactic correction.  The fact that this ratio is close to one
 corresponds to a confirmation of the relativistic force law acting on the
pulsar, of the symbolic form  $F=\frac{GM}{r^2} \lb 1+\dots+\lb \frac{v}{c} \rb ^5 \rb$, where
the crucial last term $\sim (v/c)^5$ (\ie an effect of order $10^{-12}$) has been verified with a
 fractional accuracy of order $10^{-3}$. Note that this corresponds to an absolute accuracy
of order  $10^{-15}$ compared to the leading Newtonian term $\sim GM/r^2$ !
 
The timing data from the recently discovered double binary pulsar  
 PSRJ0737-3039 led to the following ratio between the observed, and GR-predicted,
 values of the post-Keplerian parameter $s$
\be
\left[ \frac{s^{\mrm{obs}}}{s^{\mrm{GR}}\left[ k^{\mrm{obs}},R^{\mrm{obs}}
\right]}\right]=0.99987 \pm 0.00050.
\ee
The agreement for this parameter is at the $5 \times 10^{-4}$ level.

Summarizing: binary pulsar timing data have led to accurate confirmations of the
strong-field and radiative structure of GR. Roughly speaking, these confirmations
exclude any alternative theory containing long-range fields\footnote{By which, one really
means here fields with range larger than the distance between the two pulsars,
\ie $\sim 300 \, 000$ km.} coupled to bulk (hadronic) matter.

\subsubsection{Tests of gravity on very large scales}

So far, we have mainly focussed on tests of GR on spatial scales of several astronomical units
 (the size of the solar system), and on scales of $300,000$ km (the typical separation between
 two neutron stars).  We conclude this section on experimental tests of gravity by  mentioning 
the existence of tests made on very large spatial and temporal scales. {\it Gravitational lensing effects} by
galaxy clusters allow one to probe some aspects of relativistic gravity on scales $\sim 100$ kpc.
Here, one is talking of the effect of the curved spacetime metric generated by the cluster 
on light emitted by very distant quasars and passing near a galaxy cluster  containing (in
addition to visible galaxies) a lot of dark matter, as well as some  X-ray gas.  Data on the
 temperature distribution of the X-ray gas allows one to directly probe the Newtonian gravitational 
potential $U(x)$ of the cluster (without having to assume much about the (dark)
matter distribution). In turn, the potential $U(x)$ determines the relativistic lensing
of light, via the spacetime metric predicted by Einstein's theory, \ie
$-g_{00}=1-\frac{2 U}{c^2}$, $g_{ij}=\lb 1+\frac{2 U}{c^2} \rb \delta_{ij}$.
According to Ref.~\cite{Dar92}, the agreement is of the order of 30\%. This confirms the
validity of GR on scales $\sim 100$ kpc.  

Primordial nucleosynthesis of light elements (\eg Helium, Lithium, Deuterium) in the early universe 
depends on both the expansion rate and on the weak-interaction reaction rate for the conversion between
 neutrons and protons.  Given that the Hubble parameter $H^2 \propto G \rho \propto G T^4$, 
the creation of  light elements at early times (and high temperatures $T$) depends on Newton's constant.
The comparison between theoretical predictions and observations of the abundance of light
elements typically constrains the value of $G$ at the time of Big Bang nucleosynthesis,
say $G^{\mrm{BB}}$ to differ by less than $\mathcal{O}\lb 10 \% \rb$ from its current value
$ G^{\mrm{now}}$ (see \eg Chapter 18 of the Review of Particle Physics, http://pdg.lbl.gov/).

\section{String-inspired phenomenology of the gravitational sector}

\subsection{Overview}

>From the previous sections, one can conclude that GR is a very well confirmed theory so that one might be 
tempted to require of any future theory (and especially string theory) that it lead to essentially no
observable deviations from usual 4-dimensional GR. 
For instance, one might require that all the a priori massless
scalar fields that abound in (tree level, compactified) 
string theory acquire large masses. However, as there is yet no clear understanding of how to fit our
 world within string theory, it is
phenomenologically interesting to keep an open mind and explore whether there exist possibilities
for deviations of GR that have naturally escaped detection so far.
  
String theory predicts the existence of an extended mass spectrum ($g_{\mu \nu}\lb x \rb$, $\Phi \lb x \rb$, $B_{\mu \nu}$, 
moduli fields, etc.) from which there could result some long range or short range modification of gravity.  The existence
 of branes and large extra dimensions could also be sources of modified gravity (\eg KK gravity).  There could exist short
 distance effects at scales of order the string scale $\ell_s$ which are observable in cosmology or in high energy astrophysics.  
We shall also consider possible gravitational wave signals from string-cosmology models.
Finally, we refer the reader to the lectures by Juan Maldacena for a discussion of  non-gaussianities in CMB data.

A phenomenologically interesting idea (though it is not supported by precise theoretical arguments) 
is a possible breakdown of Lorentz invariance, on large scale physics,
linked to string-scale cutoff-related effects. An example of this is a modified dispersion relation of the type
\be
E^2=m^2+\vec{p}^2+\beta_1 \frac{E^3}{m_P}+\beta_2 \frac{E^4}{m_P^2}+\dots
\ee
where $m_P$ denotes the Planck mass.
One could think that because of the large value of the Planck mass, any such corrections to the usual dispersion
 relation are unobservable. However, there exist astrophysical phenomena, such as high energy cosmic rays, for
 instance high energy $\gamma$-rays, for which such a small change in this relation could be observed.  For example,
 by comparing the times-of-arrival of $\gamma$-rays of different energies, one has been able to place strong limits on
 the parameter $\beta_1$.  Such modifications of the dispersion relation have also been used in the analysis of the CMB,
 since, in the standard inflationary model, initial quantum fluctuations (the seeds of today's large scale structures)
 arise in the deep ultraviolet \ie at transplanckian scales.  Note that there exist theoretical difficulties
\footnote{Linked to its proportionality to $1/m_P$, while most theoretical
models suggest a proportionality to $1/m_P^2$.} with the inclusion of the $\beta_1 \frac{E^3}{m_P}$ term 
(the one which is  severely constrained experimentally), while the more conventional fourth order term
 would be too small to be observed.  Note that in the case of the photon, a modification on short scales
 could imply a birefringence of the vacuum as $\omega_{\pm}=\left| k \right| \lb 1 \pm \beta \frac{\left| k \right|}{m_P}\rb$. 
For references on these issues see \cite{Jacobson,AmelinoCamelia:2004hm}.
Speaking of string-inspired astrophysical effects, let us mention the suggestion
of Ref.~\cite{Gimon:2007ur} that string theory might imply a violation of the
usual Kerr  bound on the spin of rotating black holes: $ J \leq G M^2$.
\begin{figure}
\includegraphics[width=10cm]{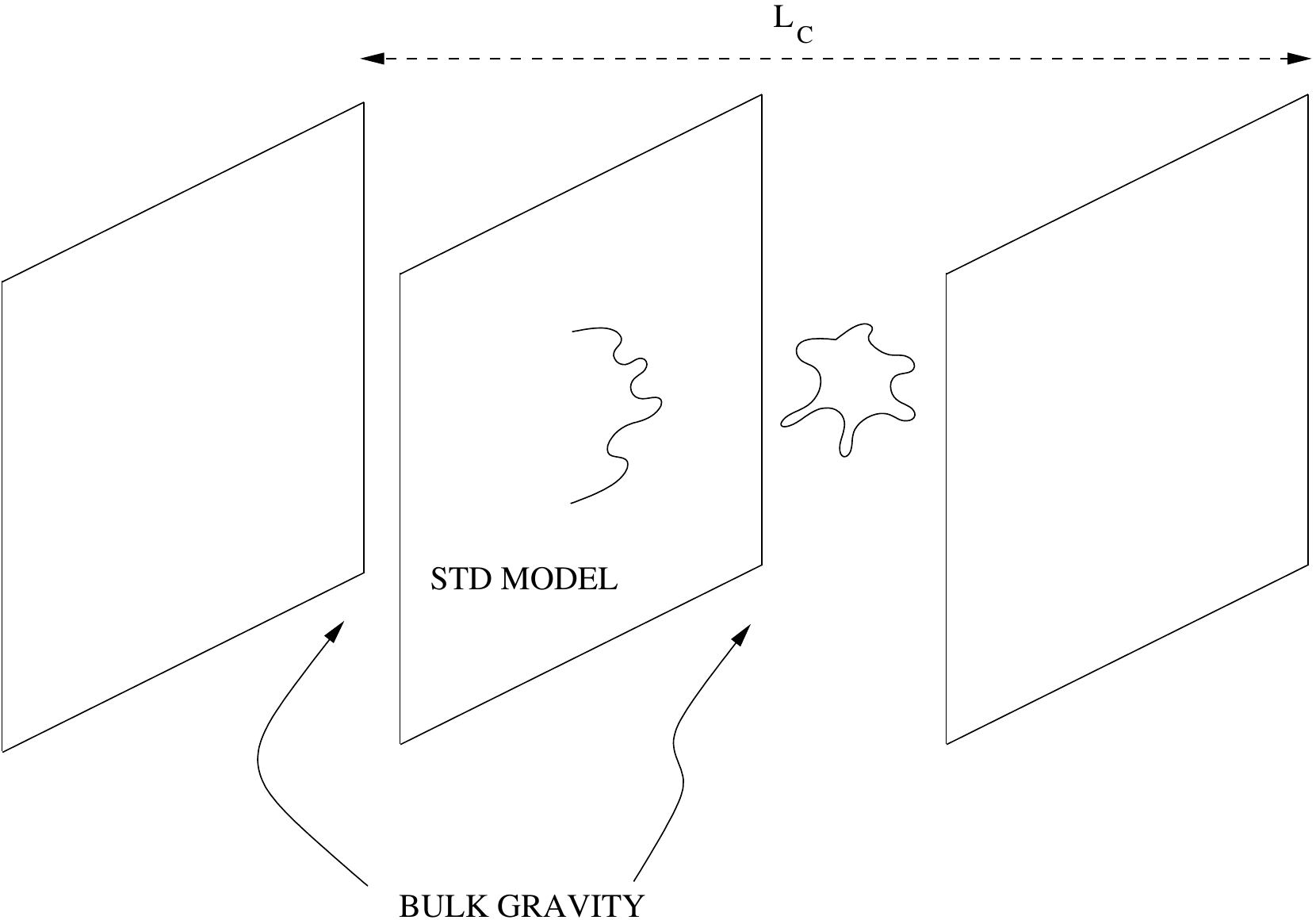}
\caption{The ends of open strings are attached to a brane, giving rise to SM particles, while closed strings are free to 
propagate in the bulk.}
\label{BraneWorlds}
\end{figure}

Other possible predictions of string theory arise from the  picture in which one considers the existence of branes
 on which (open string) SM particles are confined, while (closed string) gravitons are free to propagate in the
 bulk (FIG. \ref{BraneWorlds}).  The extra dimensions of the bulk can then be compactified, on a Calabi-Yau or
 simply on a torus (thereby ``localizing'' gravity around
the SM brane). Constraints on the size of the compactified dimensions then come either from the gravitational
 phenomenology, or from effects on SM particles. This is the ``large'' extra dimensions idea \cite{AAHD1998}
 which could be tested at the LHC, and so is of interest today.  Other realizations include models with ``very large''
 extra dimensions \cite{RS1999}, but it is less clear how they are realized in string theory.  In the
 Randall-Sundrum model \cite{RS1999}, a brane can be like a defect in a bulk with a negative
 cosmological constant, in which case the zero mode of bulk gravitational waves behaves as a
 surface wave localized on the brane due to the discontinuity located at the interface of the brane with the bulk.  
In the  DGP model \cite{DGP2000}, the approximate localization of bulk
gravity on the SM brane is achieved through the interplay of two dynamics for the gravitational
sector: a 5D Einstein action, plus a 4D ``induced'' Einstein action, with a different value of Newton's 
constant, on the brane.  Combining the two inverse propagators, the global propagator drastically modifies
 gravity on large length scales $r$:
\be
r \geq L=\frac{G_5}{G_4}.
\ee
In addition, even on length scales $r \leq L$ there exist modifications of usual gravity.
Indeed, the claim is that Newton's potential is modified as \cite{DGZ2003}
\be
U \simeq \frac{GM}{r}\left[ 1-\frac{1}{L}\sqrt{\frac{r^3 c^2}{GM}}\right].
\ee
At the phenomenological level, it is interesting that ({\it Newtonian}) gravity be modified in this way.  Estimates indicate 
that effects are small enough to have escaped detection so far,
but could be seen in refined  solar system experiments (\eg Lunar Laser Ranging). Some authors have argued that such models may 
have acausal behaviours, with, for instance, the appearance of closed timelike curves \cite{AAHDNR2006}.
 
Another conceptually interesting idea involves the possible existence  of several (parallel) Randall-Sundrum branes.  The confining 
mechanism of gravity in the Randall-Sundrum model is such that the wavefunction of surface gravitons is exponentially decaying away 
from the brane. If two branes are nearby, such quasi-confined gravitational effects can tunnel from one brane to the other via 
exponentially small effects.  As a consequence, the effective Lagrangian would contain two metric tensors with two gravitons, one 
massless, the other massive \cite{MultiBranes}. There are,
however, theoretical difficulties with any  massive gravity theory, in relation with the van Dam-Veltman-Zakharov discontinuity 
(see, \eg \cite{Damour:2002gp} and references therein).

\subsection{Long range modifications of gravity}

It is well known that, at tree level in string theory, there exist many massless scalar fields with gravitational strength coupling, 
the so-called moduli fields.  Phenomenologically, one would expect that having massless scalar fields at low energies is undesirable 
(Would a theory containing such massless fields not immediately fail the GR tests discussed in the section above?). General
arguments suggest that such scalar fields should not be expected to remain massless after supersymmetry breaking \cite{Douglas:2006es}. 
Recently, a large ``industry'' has been devoted to try to construct
explicit compactification models where all moduli are fixed, and actually
acquire very heavy masses, which is needed if inflation is to happen in the usual way.  
Here however, in the spirit of keeping an open mind, we will instead assume that a scalar field
remains massless  in the low-energy effective theory, and discuss ways in which it might not disagree with existing tests of general 
relativity. In other words, we suppose there exists a flat or almost flat direction in the total scalar potential $V\lb \varphi \rb$,
 such that there remains a massless field after supersymmetry breaking. Let us mention in this respect the idea suggested, in particular,
by Eliezer Rabinovici \cite{Rabinovici:2007hz} that the ultimate explanation for the
smallness of the cosmological constant might be a mechanism of spontaneous breaking of
an underlying scale invariance. In that case, we would expect to have an associated 
massless Goldstone boson (the ``dilaton'', in the original sense of the word).

\subsubsection{The cosmological attractor mechanism}

Let us discuss here the idea of the {\it least coupling principle}, realized via a
cosmological attractor mechanism (see \eg Refs \cite{DP1994,DPV2002}),  which can reconcile the existence of a massless scalar field 
in the low energy world with existing tests of GR  (and with cosmological inflation).  Note that, to date, it is not known whether this 
mechanism can be realized in string theory. We  assume the existence of a massless scalar field $\Phi$ (\ie of a flat direction in 
the  potential),  with gravitational-strength coupling to matter.  {\it A priori}, this looks phenomenologically forbidden but we are
 going to see that the cosmological attactor mechanism (CAM)
tends to  drive  $\Phi$ towards a  value where its coupling to matter becomes naturally $ \ll 1$.  In the string frame, we start with 
an effective action of the generic form
\be
\begin{array}{ccc}
\displaystyle S_{\mrm{eff}} & \displaystyle = & \displaystyle \int \mrm{d}^4 x \sqrt{\hat{g}} \Big[ B_{g}\lb \Phi \rb \frac{\hat{R}}{\alpha '}+\frac{B_{\Phi} \lb \Phi
 \rb}{\alpha '}\lb 4 \Box \Phi-4\lb \nabla \Phi \rb ^2 \rb \nonumber \\
\displaystyle & \displaystyle & \displaystyle -B_F \lb \Phi \rb \frac{k}{4}F_{\mu \nu}^2-B_{\Psi}\lb \Phi \rb \Psi \mrm{D} \bar{\Psi}-\frac{1}{2}B_{\chi}\lb \Phi \rb
 \lb \hat{\nabla}\hat{\chi}\rb ^2 \nonumber \\
\displaystyle & \displaystyle & \displaystyle -\frac{1}{2} m_{\chi}^2 \lb \Phi \rb \chi^2 \Big],
\end{array}
\ee
where $\Phi$ is the massless dilaton field, $\chi$ is the inflaton, to which has been associated a simple chaotic-inflation-type potential 
term, with the exception that here $m_{\chi}$ is a function of $\Phi$. In heterotic string theory for instance, $
B_{g}$, $B_{\Phi}$, $B_F$, $B_{\Psi}$ and $B_{\chi}$ are given by  expansions in powers of the string coupling $g_s=e^{\Phi}$, as
\be
B_{i}=e^{-2 \Phi}+c_0^{(i)}+c_1^{(i)} e^{2 \Phi}+\dots
\ee
where the first term is the tree level term, followed by an infinite series of correction terms involving positive powers of $g_s$ 
(or non-perturbative functions of $g_s$). Switching to the Einstein frame, and  redefining $\hat{g}_{\mu \nu}$ and the nonstandard
 $\Phi$ kinetic terms according to
\be
\begin{array}{ccc}
\displaystyle  \hat{g}_{\mu \nu}& \displaystyle  \rightarrow & \displaystyle  g_{\mu \nu}=C B_{g} \lb \Phi \rb \hat{g}_{\mu \nu},\\
\displaystyle  \varphi & \displaystyle  = & \displaystyle  \int d\Phi \left[ {3\over 4}\left({B'_g\over B_g}\right)^2
 + 2 {B'_\Phi\over B_g} + 2 {B_\Phi\over B_g}  \right]^{1/2},
\end{array}
\ee
(where a prime denotes $\mrm{d}/\mrm{d}\Phi$), the effective action turns into
\be
\begin{array}{ccc}
\displaystyle S_{\mrm{eff}} & \displaystyle = & \displaystyle \int \mrm{d}^4 x \sqrt{g} 
\Big[ \frac{\tilde{m}_p^2}{4}R-\frac{\tilde{m}_p^2}{2}\lb \nabla \varphi
 \rb ^2-\frac{\tilde{m}_p^2}{2} F \lb \varphi \rb \lb \nabla \chi \rb ^2-\\
\displaystyle & \displaystyle & \displaystyle \frac{1}{2}m_{\varphi}^2\lb \chi \rb \chi ^2 \Big]  + \cdots,
\end{array}
\ee
with $\tilde{m}_P^2 \equiv \frac{1}{4 \pi G}$, and in which the $\chi$ terms are important during inflation while additional 
terms that include the gauge fields and ordinary matter such as
\be
\begin{array}{c}
\displaystyle -\frac{1}{4}B_F\lb \varphi \rb F_{\mu \nu}^2-\sum_A \int m_A \left[ B_F \lb \varphi \lb x_A \rb \rb \right]\times \\
\displaystyle \sqrt{-g_{\mu \nu} \lb x_A \rb \mrm{d}x_A^{\mu}\mrm{d}x_A^{\nu}}-V_{\mrm{vac}}\lb \varphi \rb 
\end{array}
\ee
are relevant in the matter dominated era. 

As we shall see, the CAM leads to some generic predictions even without knowing the specific structure of the various coupling 
functions, such as \eg $m_{\chi}(\varphi), m_A(B_F(\varphi)), \cdots$. 
The basic assumption one has to make is that the string-loop corrections are such  that there exists a {\it minimum} in (some of) 
the functions $m\lb \varphi \rb$ at some (finite or infinite) value, $\varphi_m$.  During inflation, the dynamics is governed by a
 set of coupled differential equations for the scale factor, $\chi$ and $\varphi$. In particular,  the equation of motion for  $\varphi$
 contains a term $\propto -\frac{\partial}{\partial \varphi}m_{\chi}^2\lb \varphi \rb \chi^2$. During inflation (\ie
when $\chi$ has a large vacuum expectation value, this coupling drives  $\varphi$ towards the special point $\varphi_m$
where  $m_{\chi}(\varphi)$ reaches a minimum. Once $\varphi$ has been so attracted near $\varphi_m$,
$\varphi$ essentially (classically) decouples from $\chi$ (so that inflation proceeds as if 
$\varphi$ was not there). A similar attractor mechanism exists during the other phases
of cosmological evolution, and tends to decouple $\varphi$ from the dominant cosmological matter.
For this mechanism to efficiently decouple $\varphi$ from all types of matter, one needs the
special point $\varphi_m$ to approximately minimize all the important coupling functions.
This can be naturally realized  by assuming that $\varphi_m$ is a special point in field space:
for instance it could be the fixed point of some $Z_2$ symmetry of the $T$- or $S$-duality type
(so that one could say that ``symmetry is attractive'').
An alternative way of having such a special point in field space is to assume that
$\varphi_m = + \infty$\footnote{This is viewed as a strong-(bare-)coupling limit, by contrast to
the usual weak-coupling limit  $\varphi \to - \infty$ and $\Phi \to - \infty$.}
is a limiting point where all coupling functions have finite limits. This leads to the so-called
{\it runaway dilaton} scenario \cite{DPV2002}. In that case the mere assumption that
$B_i\lb \Phi \rb \simeq c_1^{i}+\mathcal{O}\lb e^{-2 \Phi}\rb$ as $\Phi \to + \infty$ 
implies that  $\varphi_m = + \infty$ is an attractor where all couplings vanish.

\subsubsection{Observable consequences of the Cosmological Attractor Mechanism}

Before discussing the observational predictions of the CAM, let us remind the reader of 
a few facts that are relevant for studying the possible effects of a string-inspired modification of gravity.  The main source 
of modification of gravity comes from the fact that the ``moduli''
field  $\varphi$ will influence the values of the masses of the (low-energy) particles and nuclei.
This means that the classical action of, say an atom $A$, will be
\be
-\int m_A(\varphi) \mrm{d}s_A=- \int m_A(\varphi) \sqrt{-g_{\mu \nu} \mrm{d}x_A^{\mu} \mrm{d}x_A^{\nu}}
\ee
where $g_{\mu \nu}$ is the Einstein-frame metric. Then, one finds that the scalar field $\varphi$
will be coupled to the atom $A$ with the strength $\alpha_{A}\sqrt{G}$, where the
dimensionless coupling strength $\alpha_{A}$ (with the same normalization as the one 
discussed above for usual tensor-scalar theories\footnote{In particular, the effective
Newton constant for a Cavendish experiment between a body made of atoms $A$ and another one
made of atoms $B$ is $G^{\rm eff}_{A B} = G( 1 +\alpha_{A} \alpha_{B})$.}) is simply given by
\be
\alpha_{A}=\frac{\partial}{\partial \varphi}\ln m_A \lb \varphi \rb.
\ee

To see better the various ways in which $\varphi$ might enter into $m_A$,
let us consider for instance the various parts constituting the mass of an atom:
\be
m_A \lb \varphi \rb=Z m_p + N m_n + Z m_e +E_{\mrm{SU}3}^{\mrm{nucleus}}+E_{\mrm{U}1}^{\mrm{nucleus}},
\ee
where $Z$ is the atomic number, $m_p$ the mass of the proton, $N$ the neutron number, $m_n$ the mass of the neutron, $m_e$ the mass 
of the electron, $E_{\mrm{SU}3}^{\mrm{nucleus}}$ and $E_{\mrm{U}1}^{\mrm{nucleus}}$ the nuclear and Coulomb interaction energies of the 
nucleus, respectively. In addition, one must note that the mass of the proton is given by
\be
\begin{array}{ccc}
\displaystyle m_p \lb \varphi \rb & \displaystyle = & \displaystyle a \Lambda_{\mrm{QCD}}\lb g_3^2\lb \varphi \rb \rb + b_u m_u \lb \varphi \rb +\\
\displaystyle & \displaystyle & \displaystyle b_d m_d \lb \varphi \rb + c_p \Lambda_{\mrm{QCD}} \alpha_{\mrm{em}} \lb \varphi \rb.
\end{array}
\ee
The main scale that determines the mass of the proton is $\Lambda_{\mrm{QCD}}$.  It depends on all the moduli including the massless 
field $\varphi$ and is roughly of the form
\be
\Lambda_{\mrm{QCD}} \lb \varphi \rb=C_g^{1/2}\lb \varphi \rb B_g^{-1/2} \lb \varphi \rb \mrm{exp} \left[ -\frac{8 \pi B_F \lb \varphi \rb}{b_3}
 \right] \tilde{M}_{\mrm{string}}.
\ee
Here, $C_g$ is the conformal factor  from the string to the Einstein frame. The most important contribution to the $\varphi$ dependence 
of $\Lambda_{\mrm{QCD}}$
is that given by the $\varphi$ dependence of the exponential term. This dependence comes
from the well-known running (via the $\beta$-function of  $\mrm{SU}(3)$) of some
(unified) gauge coupling constant between its value $ 1/g_3^2 \propto B_F(\varphi)$ considered at
 a GUT-scale cut-off (here approximately related to $\tilde{M}_{\mrm{string}}$), to
a value of order unity at the confining scale $\Lambda_{\mrm{QCD}}$. The other contributions to the mass of the proton are the quark masses, 
which are determined by the vev of the Higgs boson and by the Yukawa coupling constants, which, again, are 
expected to be functions of $\varphi$ at high energy.  There also exists a contribution from the electromagnetic sector since part of 
the mass of the proton is a function of the fine structure constant $\alpha_{\mrm{em}}\lb \varphi \rb$.  Finally the nuclear binding 
energy of a nucleus is quite important and must also be expressed as a function of basic scales.  In an approximate form it reads
\be
E_{\mrm{SU}3}^{\mrm{nucleus}} \simeq \lb N+Z \rb a_3 + \lb N+Z \rb ^{2/3} b_3 \ee where \be a_3 \simeq a_3^{\mrm{chiral \, limit}}
 +\frac{\partial a_3}{\partial m_{\pi}^2}m_{\pi}^2 \lb \varphi \rb.
\ee
In the chiral limit ({\it i.e.}, taking the quark masses to zero) one gets a non-zero limit $a_3^{\mrm{chiral \, limit}}$ to which must 
be added a term approximately proportional to the squared pion mass. In turn, $m_{\pi}^2$  is proportional to the product of $\Lambda_{\mrm{QCD}}$
and $m_u+m_d$, both of which are expected to be functions of  $\varphi$. Incidentally, let us note that there exists a delicate 
balance between attractive and repulsive nuclear interactions \cite{SW1997},
which implies a strong sensitivity of the binding energy of nuclei to the value of the quark
masses \cite{D2006}. A recent result shows that if the quark masses were to increase by 50\% 
(at one $1 \sigma$, or 64\% at $ 2 \sigma$), all heavy nuclei would fall apart because
there would be no nuclear binding \cite{Damour:2007uv}.

At leading order, the mass of any nucleus is a pure number times $\Lambda_{\mrm{QCD}}$.
In this approximation, $m_A$ would depend universally on $\varphi$ (via $\Lambda_{\mrm{QCD}}(\varphi)$),
and the scalar coupling strength $\alpha_A$ would be independent of the atomic
species $A$ considered. As a consequence, there would be no violation of the universality of free fall.
This shows that the violations of the universality of free fall will depend
on the small fractional corrections in $m_A$ proportional to the ratios
\be
\frac{m_u}{\Lambda_{\mrm{QCD}}},\,\,\, \frac{m_d}{\Lambda_{\mrm{QCD}}}, \,\,\, \mrm{and} \,\,\, \alpha_{\mrm{em}}.
\ee
When differentiating the mass of an atom w.r.t.~$\varphi$, say
\be
m_A \lb \varphi\rb = \mathcal{N} \Lambda_{\mrm{QCD}}  
\lb 1+\varepsilon_A^{\sigma}\frac{m_u+m_d}{\Lambda_{\mrm{QCD}}}
+\varepsilon_A^{\delta}\frac{m_d-m_u}{\Lambda_{\mrm{QCD}}}
+\varepsilon_A^{\mrm{em}}\alpha_{\mrm{em}}\rb,
\ee
where $\cal{N}$ is a pure number (which depends on $N$ and $Z$), one obtains
 for the scalar coupling strength 
 $\alpha_A \lb \varphi\rb = \frac{\partial}{\partial \varphi}\ln m_A \lb \varphi \rb$ 
 an (approximate) expression of the form
\be
\alpha_A \lb \varphi\rb \simeq \alpha_{\mrm{had}}\lb \varphi \rb +  \varepsilon_A^{\sigma}\frac{\partial}{\partial \varphi} \lb
\frac{m_u+m_d}{\Lambda_{\mrm{QCD}}}\rb+\varepsilon_A^{\delta}\frac{\partial}{\partial \varphi}\lb\frac{m_d-m_u}{\Lambda_{\mrm{QCD}}}\rb
+\varepsilon_A^{\mrm{em}}\frac{\partial}{\partial \varphi}\alpha_{\mrm{em}},
\ee
where $\displaystyle \alpha_{\mrm{had}}\equiv\frac{\partial}{\partial \varphi}\ln \Lambda_{\mrm{QCD}}\lb \varphi \rb$. When the CAM has 
attracted $\varphi$ near a
value  $\varphi_m$ which minimizes all the separate coupling functions entering
the various ingredients of $m_A \lb \varphi\rb$, each term in the above expression for
$\alpha_A \lb \varphi\rb$ will be (approximately) proportional to the 
small difference $\varphi - \varphi_m$. As a consequence all the contributions
to $\alpha_A \lb \varphi\rb$ will be  small, so that all the observable deviations from GR
will be naturally small.
 
Let us  describe more precisely the possible observable consequences of the CAM.  In this
mechanism, the couplings of the massless scalar field to the various physical sectors are
not assumed to be initially small (they are given by the various coupling
functions $B_i(\varphi)$ entering the Lagrangian, and these functions are ``of order unity'').
However, via its coupling to cosmological evolution, the scalar field is driven towards a point where the couplings to matter become 
small, but not exactly zero. Indeed, one can analytically
estimate the ``efficiency'' of the cosmological evolution in driving $\varphi$ towards
$\varphi_m$, and one finds some expression for the difference\footnote{When $\varphi_m$ is infinite,
$\delta \varphi\equiv\varphi - \varphi_m$ is replaced, \eg by $e^{- c \varphi}$.}
$\delta \varphi\equiv\varphi - \varphi_m$ \cite{DP1994,DPV2002}. The deviations from GR are all proportional to the small quantity
$\delta \varphi^2$ because the scalar coupling strengths $\alpha_{A}, \alpha_{B} $ are
proportional to $\delta \varphi$, and all ``post-Einstein'' observables contain two
scalar couplings , say $\alpha_A \alpha_B $ when talking about the scalar exchange between
$A$ and $B$ (for instance the modified gravitational constant for a Cavendish
experiment involving two bodies made of atoms $A$ and $B$ is $G_{AB}=G \lb 1+\alpha_A \alpha_B \rb$). 
In addition to  predicting small values for the (approximately composition-independent) 
``post-Einstein'' parameters $\bar{\gamma}$ and $\bar{\beta}$   this mechanism also predicts
various (small) violations of the equivalence principle.
 
For instance, the above expressions for the
ingredients entering $m_A$ and $\alpha_A$ lead to generic predictions about the
type of violation of the universality of free fall that one might expect
in string theory. Indeed, one finds that the fractional difference in the
free fall acceleration of two bodies (made of atoms $A$ and $B$) takes the form
\be
\begin{array}{ccc}
\label{EP}
\displaystyle \frac{a_A-a_B}{\langle a \rangle} & \displaystyle \simeq & \displaystyle 2 \times 10^{-5} \alpha_{\mrm{had}}^2 \Big[ \Delta \lb \frac{E}{M} \rb _{AB}+c_B \Delta \lb
 \frac{N+Z}{M}\rb _{AB}+ \\
\displaystyle & \displaystyle & \displaystyle c_D \Delta \lb \frac{N-Z}{M} \rb _{AB} \Big],
\end{array}
\ee
with
\be
\frac{E}{M}=\frac{Z \lb Z-1 \rb}{\lb N+Z \rb ^{1/3}}.
\ee
where $ \lb \Delta Q \rb_{AB}\equiv Q_A - Q_B$, and where  the first, second and third terms 
in the brackets are contributions from the Coulomb energy of the nucleus ($\propto \alpha_{\mrm{em}}$), 
and from the $\varphi$-dependence of the sum and difference of the quarks masses, \ie $m_u+m_d$ and $m_u-m_d$.  
 
 This mechanism also predicts (approximately composition-independent) values for the post-Einstein
 parameters $\bar{\gamma}$ and $\bar{\beta}$ parametrizing 1PN-level deviations
 from GR. They are of the form
\be
\bar{\gamma}=-2 \frac{\alpha_{\mrm{had}}^2}{1+\alpha_{\mrm{had}}^2}\simeq -2\alpha_{\mrm{had}}^2,
\ee
and
\be \bar{\beta}=\frac{1}{2}\frac{\alpha_{\mrm{had}}^2 \frac{\partial \alpha_{\mrm{had}}}{\partial \varphi}}{(1+\alpha_{\mrm{had}}^2)^2}
\simeq \frac{1}{2}\alpha_{\mrm{had}}^2\frac{\partial \alpha_{\mrm{had}}}{\partial \varphi}.
\ee
In this model, one in fact violates all tests of GR.  However, all these violations are correlated. 
For instance, using the numerical value $\Delta \lb \frac{E}{M} \rb \simeq 2.6$ (which applies both to 
the pair Cu--Be and to the pair Pt--Ti), one finds the following link between equivalence-principle violations and
solar-system deviations
\be
\lb \frac{\Delta a}{a}\rb \simeq -2.6 \times 10^{-5} \bar{\gamma}.
\ee
Given that present tests of the equivalence principle place a limit on the ratio $\Delta a/a$ of 
the order of $10^{-12}$, one finds $\left| \bar{\gamma}  \right| \leq 4 \times 10^{-8}$. Note that the 
upper limit given on $\bar{\gamma}$ by the Cassini experiment was $10^{-5}$, so that in this case the 
necessary sensitivity has not yet been reached to test the CAM. 

As another example, one can compute the evolution of the fine structure constant w.r.t. time. Given that 
it is a function of $\varphi$, and that $\varphi$ evolves as a function of cosmological evolution due to 
its coupling to matter, $\alpha_{\mrm{em}}$ is indeed a function of time, and its time derivative
can be written as
\be
\frac{\mrm{d}}{\mrm{d}t} \ln \alpha_{\mrm{em}}\sim \pm 10^{-16}\sqrt{1+q_0-\frac{3 \Omega_m}{2}}\sqrt{10^{12} \frac{\Delta a}{a}} \,\, 
\mrm{yr}^{-1}.
\ee
The first square root on the r.h.s. of this equation can also be written as

\be
\frac{\Omega_m \alpha_m+4 \Omega_v \alpha_v}{\Omega_m+2\Omega_v}.
\ee
where $\Omega_m$ denotes the fraction of the cosmological closure density due to dark
matter, and $ \alpha_m$ the scalar coupling to dark matter, while $\Omega_v$ and $ \alpha_v$
denote the corresponding quantities for ``dark energy'' (or ``vacuum energy'').  
For instance, if we assume $\alpha_v \sim 1$ (so that $\varphi$ is a kind of ``quintessence'')
while $\alpha_m \ll 1$, we see from the result above that the current
experimental limit $\frac{\Delta a}{a}< 10^{-12}$, implies the following upper bound
on a possible time variation of the fine-structure constant: $\frac{\mrm{d}}{\mrm{d}t} \lb \ln \alpha_{\mrm{em}}\rb \leq 10^{-16}\, \, 
\mrm{yr}^{-1}$. This upper bound is below the current
laboratory limits on $\dot {\alpha}/\alpha$, but comparable to the Oklo limit
mentioned above.
When working out the generic predictions of the {\it runaway dilaton} version of the cosmological attractor mechanism, one finds that it 
naturally predicts (when assuming an inflationary
potential $\propto \chi^2$) a level of deviation from GR of order
$- \bar{\gamma} \sim 4 C  \times 10^{-8}$, corresponding, for instance, to a violation of
the equivalence principle at the level $\Delta a/a \sim C \times10^{-12}$. Here, $C$ is a
combination of model-dependent dimensionless parameters, which are generically expected to be
 ``of order unity''. This suggests (if $C$ is smaller, but not much smaller than 1)
that the current sensitivity of equivalence principle experiments may be close to what is needed 
to test the deviations from GR predicted by such a runaway dilaton. Let us note in this respect that
ongoing improved lunar laser ranging experiments will probe $\Delta a/a$ to better than the
$\sim 10^{-13}$ level, and that the CNES satellite mission  MICROSCOPE (to be launched 
in the coming years) will reach $\Delta a/a \sim 10^{-15}$. Another more ambitious satellite mission (which
is not yet approved), STEP (Satellite Test of the Equivalence Principle),
plans to probe violations of the equivalence principle down to the
$10^{-18}$ level\footnote{Let us also mention the suggestion \cite{Dimopoulos:2006nk} that
atom interferometry might be used for testing the equivalence principle down to the 
$\Delta a/a \sim 10^{-17}$ level.}. In addition, post-Newtonian solar system experiments 
at the $10^{-7}$ level would be of interest. The approved micro-arcsecond global astrometry 
experiment GAIA will probe $\bar{\gamma}\sim 10^{-7}$, while the planned laser experiment LATOR 
might reach $\bar{\gamma}\sim 10^{-9}$. In addition, the comparison of cold-atom clocks might soon reach
the  interesting level  $\frac{\mrm{d}}{\mrm{d}t} \lb \ln \alpha_{\mrm{em}}\rb  \sim 10^{-16}\,\,\mrm{yr}^{-1}$.

Finally, let us mention that one can combine the basic mechanism of the CAM (which consists
in using the coupling of $\varphi$ to matter, \ie the presence of a term of the form 
$a(\varphi) \rho_{\rm matter}$ in the action) with the presence of a ``quintessence''-like
potential $V(\varphi) \propto 1/\varphi^p$. This yields the ``chameleon'' mechanism \cite{Khoury:2003aq}
in which both the value $\varphi_m$ towards which $\varphi$ is attracted, and the
effective mass (or inverse range) of $\varphi$, depend on the local
matter density $\rho_{\rm matter}$. 
Whatever be one's opinion concerning the a priori plausibility of having
some nearly massless moduli field surviving in the low-energy physics of string theory,
it is clear that such experiments are important and could teach us something new about reality.\footnote{For instance, if one considers
 it very unlikely that such a field can exist,
these experiments are important because they can {\it falsify} string theory. By contrast, if
one  finds a violation of the equivalence principle which is nicely consistent
with the prediction (\ref{EP}) for the composition dependence of a moduli field,
this might be viewed as a {\it confirmation} of string theory.}
\section{String-related signals in cosmology}
\subsection{Alternatives to slow-roll inflation}
In the usual inflationary scenario, the period of exponential expansion is based on the slow roll mechanism, \ie one has to assume a 
sufficiently flat potential  so that the scalar field, the inflaton, slowly rolls down to its minimum in such a way that the approximate 
equality $p_{\varphi} \simeq -\rho_{\varphi}$ lasts sufficiently long, say for a minimum of 60--70 e-folds.  The simplest inflationary 
Lagrangian reads
\be
\label{slowroll}
\mathcal{L}=-\frac{1}{2}\lb \partial \varphi \rb^2-V\lb \varphi \rb
\ee
with a usual  kinetic term and a potential $V\lb  \varphi \rb$. This simple 
inflationary framework leads to specific predictions such as a relation between the ratio of tensor to scalar primordial perturbations
and the ``distance'' in field space over which $\varphi$ runs during inflation (the so-called Lyth bound, see the lectures by Juan Maldacena 
in these proceedings). Let us, however, emphasize that
these predictions (which lead to constraints on the model) do depend on the assumption that inflation is realized
by the simple  action  (\ref{slowroll}) with a slow-roll potential. There are, however, other ways of
realizing inflation, in which these constraints might be relaxed. Let us note in this respect
that inflation can be realized even if the potential $V\lb \varphi \rb$ in (\ref{slowroll})
is {\it not} of the slow-roll type  \cite{DM1998}. Moreover, one may  have inflation without a potential at all if the Lagrangian is a 
complicated enough function of 
$X \equiv - \lb \partial \varphi \rb^2$. Indeed, if one has an action of the type
$\mathcal{L}=p( X)$, one finds that there can exist attractors toward a de Sitter expansion phase,
corresponding to a line where the effective equation of state deduced from $\mathcal{L}=p( X)$
 is $p = - \rho$ (\eg k-inflation \cite{APDM1999}; ghost inflation \cite{AHCMZ2004}). 
 To have a ``graceful exit'' from this de Sitter phase one needs, for instance, to
 introduce some additional $\varphi$ dependence in $\mathcal{L}$. It has been
 suggested in \cite{ST2004} that such a mechanism might be realized  in string theory, via
 a Dirac-Born-Infeld-type action, say  
\be
p(X,\varphi)=-\frac{\varphi^4}{\lambda^2}\lb \sqrt{1-\frac{\lambda \dot{\varphi}^2}{\varphi^4}}-1\rb-V\lb \varphi \rb.
\ee
In such a ``DBI inflation'', the use of non-standard kinetic terms greatly relaxes the restrictions imposed on the flatness of the potential
 $V\lb  \varphi \rb$ which must be imposed in the usual
case of  (\ref{slowroll}). It also tends to produce larger non-gaussianities in the CMB \cite{AST2004}.
Let us also point out that the use of a non-linear kinetic term might significantly affect the Lyth bound. For instance, if one considers
 the action 
\be
\mathcal{L}= p\lb X, \varphi \rb = K\lb X \rb -V\lb  \varphi \rb, 
\ee
where $K\lb X \rb$ is a non-linear function of the kinetic term 
$X \equiv - \lb \partial \varphi \rb^2$,  one finds the following modified
form of the relation between the ratio $r$ of tensor to scalar primordial perturbations
and the derivative of $\varphi$ w.r.t.~the number of efolds $N$:
\be
\frac{r}{8}=\lb \frac{\mrm{d}\varphi}{\mrm{d}N}\rb ^2 a
\ee
Here $a$ is an additional amplification factor, which is given by the following
expression in terms of the kinetic function $K\lb X \rb$
\be
\displaystyle a=\frac{2 K'}{\sqrt{1+2 X
\frac{K''}{K'}}}=\left\{\begin{array}{l}
\displaystyle 1\,\, \mrm{for} \,\,K=\frac{X}{2},\\
\displaystyle 1\,\, \mrm{for\,\, DBI\,\,type:}\,\,-\sqrt{1-X},\\
\displaystyle \gg 1\,\, \mrm{for},\,\,\mrm{\eg}\,-\frac{1}{2
\alpha}\lb 1-X\rb^{\alpha},\,\, \mrm{with}\,\, \alpha<\frac{1}{2}.\nonumber
\end{array}\right.
\ee
A large amplification factor $ a \ll 1$ would (formally) correspond to a relaxed Lyth bound
on the excursion of $\varphi$, given a minimum number $N$ of efolds.
It is interesting to note that $a=1$ (unchanged Lyth bound)  in the  DBI-type square root model.
However, we note that a more general power $\alpha$, with $\alpha<1/2$, would formally
relax the Lyth bound. [A more detailed study is, however, necessary for seeing whether the
bound is {\it physically} relaxed.]

The present section presented only a very partial and sketchy picture. It was only intended as
an illustration that folklore results and constraints on inflationary models  do depend on 
using the standard slow-roll action (\ref{slowroll}), and that there  exist other mechanisms
in which those results and constraints might be different and possibly relaxed.

\subsection{Cosmic superstrings}

\subsubsection{Phenomenological origin}

\begin{figure}
\includegraphics[width=5cm]{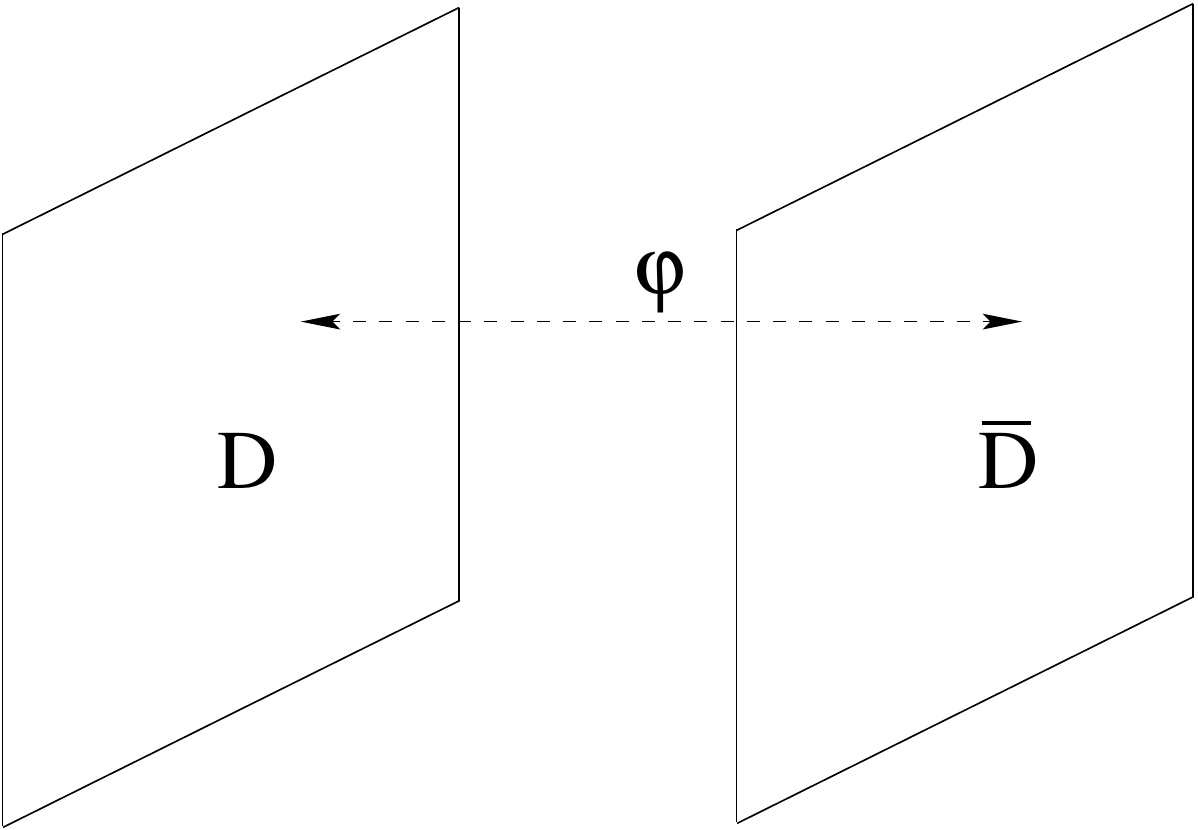}
\hspace{0.2cm}
\includegraphics[width=5cm]{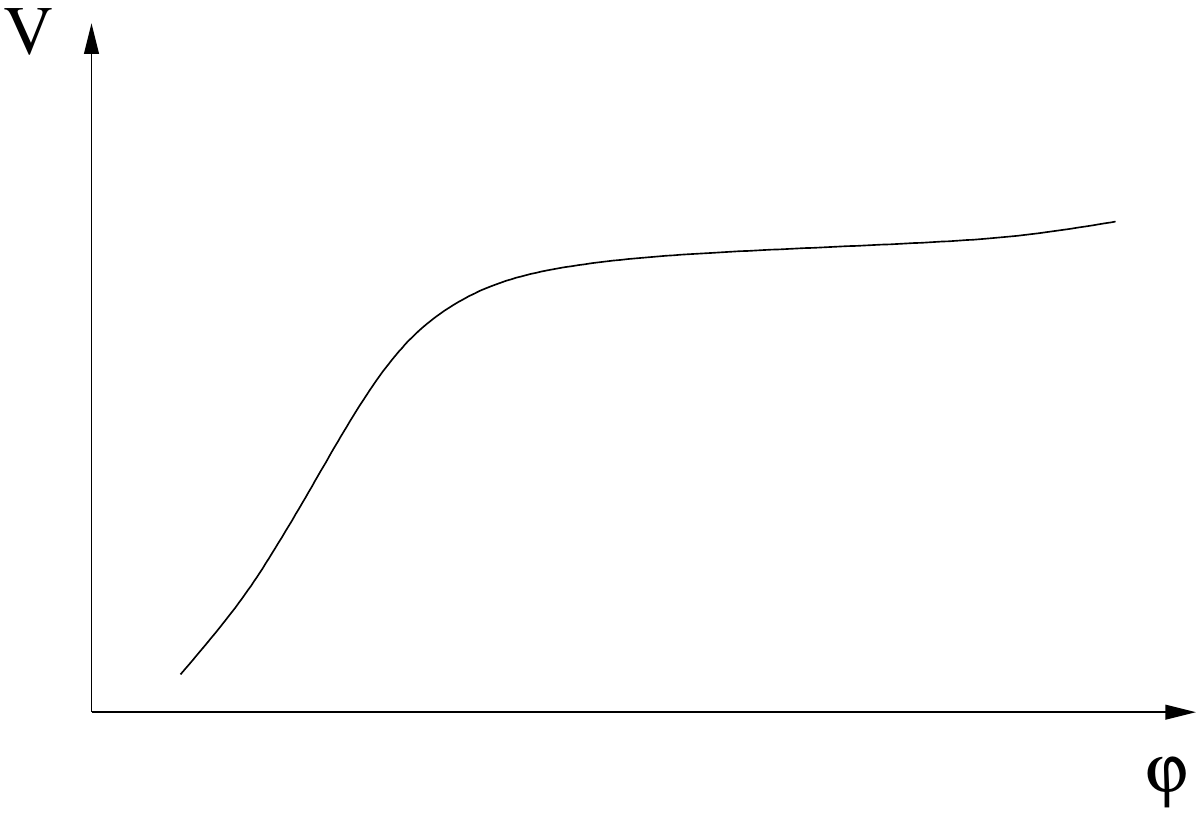}
\caption{Left: The brane-antibrane distance as a scalar field
$\varphi$. Right: $V\lb \varphi \rb$ behaves as $c_1 - c_2 \varphi^{-4}$ for
large brane-antibrane separations} \label{BraneInflation}
\end{figure}

The existence and detection of {\it cosmic superstrings} is an exciting possibility that was first suggested in Ref.~\cite{W1985}, then 
kept alive for a number of years, and recently revived dramatically notably in Ref.~\cite{KKLMMT2003} and in other papers 
\cite{ST2002,DV2004,CMP2004}. They arise in brane antibrane scenarios where the inflaton is the brane-antibrane separation (see \eg \cite{DT1999}). In such scenarios, for large enough brane-antibrane separations, the potential behaves like $c_1 - c_2/ \varphi^4$ such that it satisfies 
the slow roll conditions (FIG. \ref{BraneInflation}). When the branes are near, some of the modes connecting the two branes 
become tachyonic, \ie a complex field (with kinetic term $-\partial T  \partial \bar{T} $) having a potential $V(|T|^2)$  with wrong-sign
curvature near $T=0$.  This instability can generate topological defects since the phase of the vev of $T$ need not be uniformly the same
 all over space. Contrary to the situation in which strings are created at the beginning of inflation and then diluted away, this scenario
 naturally produces strings at the end of inflation so that they are not diluted by the expansion.  For causality reasons, the value of the
 field's vev in a given Hubble patch should be uncorrelated to that in other Hubble patches.  This is what creates a network of strings and
 one can then compute the initial density and correlation length of the string network.
The string tension $\mu$ in Planck units, \ie  the dimensionless parameter $G\, \mu$, where $G$ is Newton's constant, was initially thought
 to be high, of the order of $10^{-3}$ at best, because of the string theoretic origin of these objects and of the then expected relation between $\alpha '$
 and the Planck length. However, in models with warping factors and large fluxes, the string tension can be lowered to much smaller values. 
In practice, the string tension is tuned to fit current CMB data.   Tye and collaborators \cite{ST2002} find a window of the type 
$10^{-12}< G\, \mu < 10^{-6}$, while in the more
detailed  KKLMMT model \cite{KKLMMT2003} one finds $G \mu \sim 10^{-10}$.

In trying to gain insight into the observational predictions that can be made from cosmic superstring models, one must consider not only the
 stretching by the cosmological expansion of an initial network of cosmic strings with a correlation length of the order of the Hubble scale
, but also string interactions.  A string can for instance self-intersect or two strings can intersect and reconnect. The Hubble
 expansion tends to locally straighten out the strings while interconnections tend to produce loops and small-scale structure.  Given 
an initial correlation length and reconnection probability $p$, working out the time evolution of a string network is essentially a
 classical problem. Two types of strings develop, long strings with correlation length of the order of the time scale $t$, and small loops
 that loose energy by gravitational radiation \cite{V1981,VilenkinBook,Cusps}.

In order to define the typical size  of string loops (at the time they are formed)
 we introduce a dimensionless parameter $\alpha$
such that $\ell_{\mrm{loop}} \lb t \rb = \alpha\, c\, t$.  It was initially thought that $\alpha = 50\, G\, \mu$, an estimate linked to the
 idea that gravitational damping is the essential
mechanism wich  determines the lifetime of loops. More recently, is has been suggested that $\alpha$
might be significantly different from $50\, G\, \mu$. There is, however, no consensus on the
``correct'' value of $\alpha$. Estimates vary between $\alpha \sim (50\, G\, \mu)^{\beta}$,
with $\beta > 1$ (leading to ``small loops'') and   $\alpha \sim 0.1$
(leading to large loops). [For an introduction to
this problem, and references, see \eg the talk of Joe Polchinski at the 2007 String meeting in 
Madrid.] 
 Happily, some of the predictions we shall discuss below (notably those concerning
the observability of
gravitational waves from a cosmic string network) are rather insensitive to the value of $\alpha$.

Several numerical simulations confirm the tendency of string networks to display a scale-invariant 
behavior \cite{VOV0506,MS2006}. There have been recent attempts at refining the theoretical description 
of string networks \cite{PR2007,DR2007}. However, there is, to date, no consensus among experts as to
 the typical size distribution of loops (\ie the dominant gravitational wave-emitting string type).  
In several simulations, the distribution of the size of loops is bimodal, with one peak at $\alpha \sim 0.1$ 
and another peak at the UV cutoff. It has been argued by Vilenkin and collaborators
that only the ``large loop'' part, \ie $\alpha \sim 0.1$, will survive.

In the following, assuming KKLMMT-type brane inflation and the stability of strings over cosmologically 
interesting time scales, we discuss the phenomenological predictions made by treating $p$ and $\alpha$ 
as free parameters and their possible observable signals.

\subsubsection{Observational signatures}

Partly for historical reasons, the phenomenology of cosmic strings has been studied mostly in the context 
of CMB observations.  Slow-roll inflation generates a random $\frac{\delta T}{T}$ 
angular distribution on the sky that fits well the observations. Adding a random network of cosmic strings 
generates  additional (non-Gaussian) fluctuations in the CMB 
which have less angular structure (the string has a lensing effect proportional to its velocity $v$ over 
the sky, $\delta T /T\sim 8 \pi G \mu v \gamma$). CMB observations can then be used to place an upper 
bound on $G \mu$, of order $5 \times 10^{-7}$. Much smaller values of the string tension will not lead
to any observable signature in the CMB. Let us also mention that cosmic (super)strings
might be detected via their  gravitational lensing of galaxies, or   microlensing of stars.

By contrast to the CMB (or  lensing) observations, which are only sensitive to
string tensions $G \mu > 10^{-7}$, existing or planned gravitational wave interferometers
could detect cosmic (super)strings with tensions in the much wider range $10^{-15} < G\, \mu < 10^{-6}$.
Let us recall that a gravitational wave (GW) detector is actually measuring tidal forces,
and more precisely a component of the Riemann tensor projected ``along'' a detector having
a quadrupolar structure\footnote{This is the spin-2 analog of saying that electromagnetic
antennas are sensitive to the projection of the electric field along the direction of
a dipolar antenna.}. In other words, a GW antenna measures the second time derivative
$\ddot{h}\lb t \rb$ of a projection of the metric fluctuation $h_{\mu \nu}$. Current
detectors are sensitive down to the level $h \sim 10^{-22}$, for frequencies $ f \sim 100$ Hz.

The GW signal from a string network is  an incoherent background of GWs
made of the superposition of all GWs ever emitted by string loops
(from zero to very large redshifts). This signal is distributed over a very
large spectrum of frequencies (including wavelengths of the size of the universe, as
well as very short ones). In order to determine the frequency distribution, the number of loops and how they 
evolve, one needs to know the evolution of the universe during the inflationary, radiation and matter dominated eras.

Besides detecting the GWs from a string network in a man-made interferometer, 
another observational possibility lies in the timing of isolated millisecond pulsars.  
In a stationary spacetime, the pulses emitted by an isolated pulsar would be observed on Earth (after correcting
for the Earth motion) at very regular intervals.  By contrast, in presence of a fluctuating
background of GWs, the times of arrival of successive pulses would fluctuate,
and exhibit some red noise. Pulsar timing over some time interval $T$ (which is typically
several years) is most sensitive to the part of the GW frequency spectrum with frequencies
$ f\sim 1/T$. Therefore, pulsar timing is most likely to detect long wavelength GWs (several light years long).

\begin{figure}
\includegraphics[width=10cm]{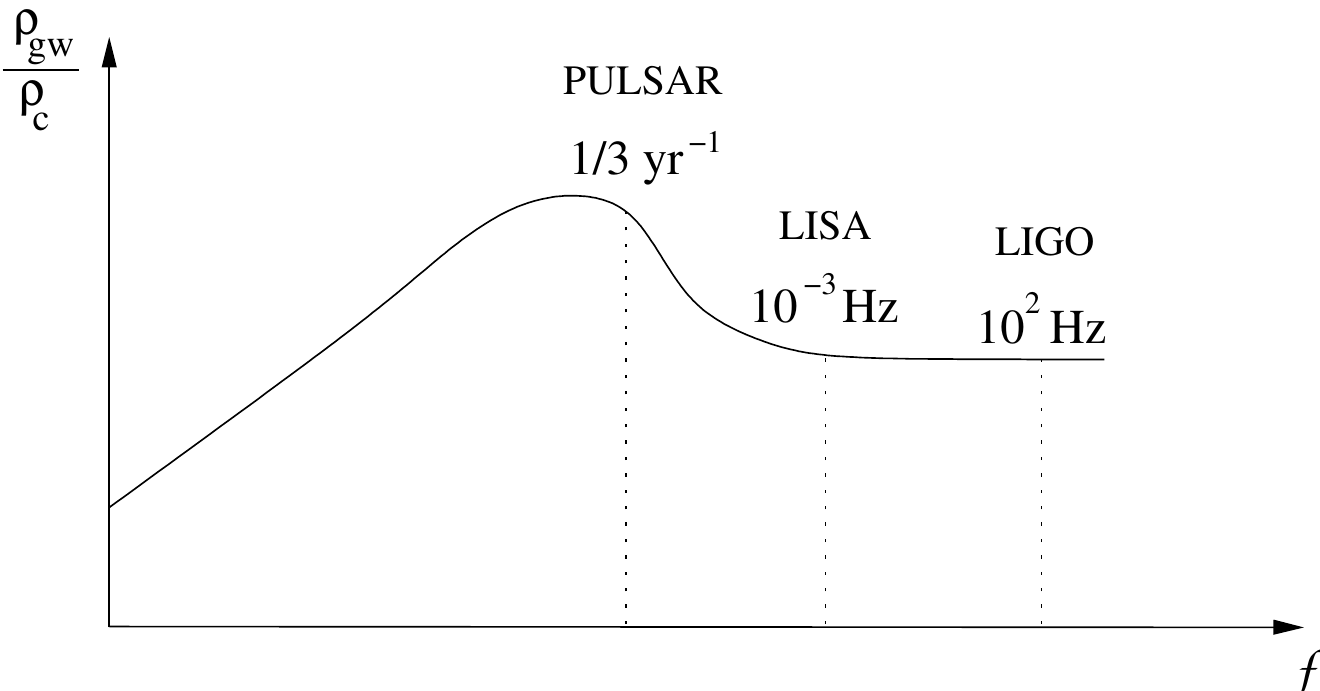}
\caption{Expected frequency distribution of the ratio
$\Omega_{GW}=\frac{\rho_{GW}}{\rho_c}$ for the  stochastic gravitational wave
background of cosmic strings.}
\label{GWSpectrum}
\end{figure}

Along with LIGO-type ground based interferometers, a space-based one, the Laser Interferometer 
Space Antenna (LISA) has been conceived,  with arm lengths of the order of $10^6$ km instead 
of the 3 or $4$ km ones constructed on the ground.  LISA can therefore explore much smaller 
frequencies.  The best achievable sensitivity for LIGO-type instruments is reached for frequencies $f \sim 100$ Hz, \ie rather fast events
lasting $\sim 10^{-2}$ seconds, while space experiments may probe events with periods $\sim 1000$ 
secs, which are quite slow events (see FIG. \ref{GWSpectrum}).
In Ref.~\cite{Cusps}, the possible existence of sharp gravitational wave bursts above the background 
caused by string cusps was pointed out.  For a typical oscillating loop, there occurs a cusp once or twice 
per oscillation with the extremity of the cusp going at the velocity of light and emitting a strong gravitational 
wave signal in the direction in which the cusp is moving.  Statistically, these events are random. A GW burst 
will be  detected if it happens to be emitted towards the detector.  Under some conditions, these cusps can 
create signals which stand much above the quasi-Gaussian
random mean square background ``GW noise''. This raises the exciting possibility that LIGO/VIRGO/GEO
or LISA might detect GW signals emitted by giant superstrings at cosmological distances.

Let us now give an introduction to the physics behind the occurrence of those cusps, and the associated 
emission of GW bursts.

\subsubsection{String dynamics}

We consider the string position $X^{\mu}$ as a function of the worldsheet coordinates $\tau$ and $\sigma$. 
 We treat the string dynamics in a locally flat spacetime.  Introducing the lightcone coordinates in conformal gauge,
\be
\sigma_{\pm}=\tau\pm \sigma,
\ee
$X^{\mu} \lb \tau, \sigma \rb$ satisfies
\be
\frac{\partial}{\partial \sigma_+}\frac{\partial}{\partial \sigma_-}X^{\mu}\lb \tau, \sigma \rb=0
\ee
such that the generic string solution is the sum of left and right movers
\be
X^{\mu}\lb \tau , \sigma \rb = \frac{1}{2} \left[ X_+^{\mu}\lb \sigma_+ \rb+X_-^{\mu} \lb \sigma_- \rb \right]
\ee
in which the factor $1/2$ is introduced for convenience. The Virasoro constraints read
\be
\begin{array}{ccc}
\label{vir}
\displaystyle \lb \partial_+X_+^{\mu}\rb ^2 & \displaystyle = & \displaystyle 0\\
\displaystyle \lb \partial_-X_-^{\mu}\rb ^2 & \displaystyle =  & \displaystyle  0
\end{array}
\ee
In the time gauge, the worldsheet is sliced by constant coordinate time
hyperplanes $X^0 = x^0 = \tau$, so that 
\be
X^0\lb \tau, \sigma \rb =\tau=\frac{1}{2}\lb \sigma_+ + \sigma_- \rb.
\ee
We thus have $X^0_+=\sigma_+$ and $X^0_-=\sigma^-$. Then $\partial_{\pm} X^0=1$ contributes a $-1$
in the Virasoro constraints, so that
\be
\lb \partial_+X_+^{\mu}\rb ^2 = -1+\lb \partial_+ X^i\rb ^2,
\ee
and similarly for the $-$ equation. This means that the derivatives (w.r.t.~their argument)
of the spatial components $X_{\pm}^i(\sigma_{\pm})$ of the left and right modes
are constrained to be {\it unit euclidean vectors}:
\be
\lb \dot{X}_{\pm}^i \rb ^2=1.
\ee
The $X_{\pm}^i$ are periodic and the time derivative of the spatial component of $X$ are unit vectors.  
We may now use a representation first introduced in Ref.~\cite{KT1982}.  The derivatives $\dot{X}_{\pm}^i$ 
can be seen as drawing two curves on the unit sphere (the Turok-Kibble sphere). In addition, as
 $X_{\pm}^i$ is  periodic in three-dimensional space  (there is no winding), we have 
 $\int \mrm{d\sigma_{\pm}}  \dot{X_{\pm}^i}=0$.  
As a result, the ``center of mass'' of both left and right moving curves must be at the center of the sphere. 
This implies that the two curves generically\footnote{There exist, however, specially contrived curves that can avoid
intersecting.} intersect twice \cite{T1984}.
\begin{figure}
\includegraphics[width=10cm]{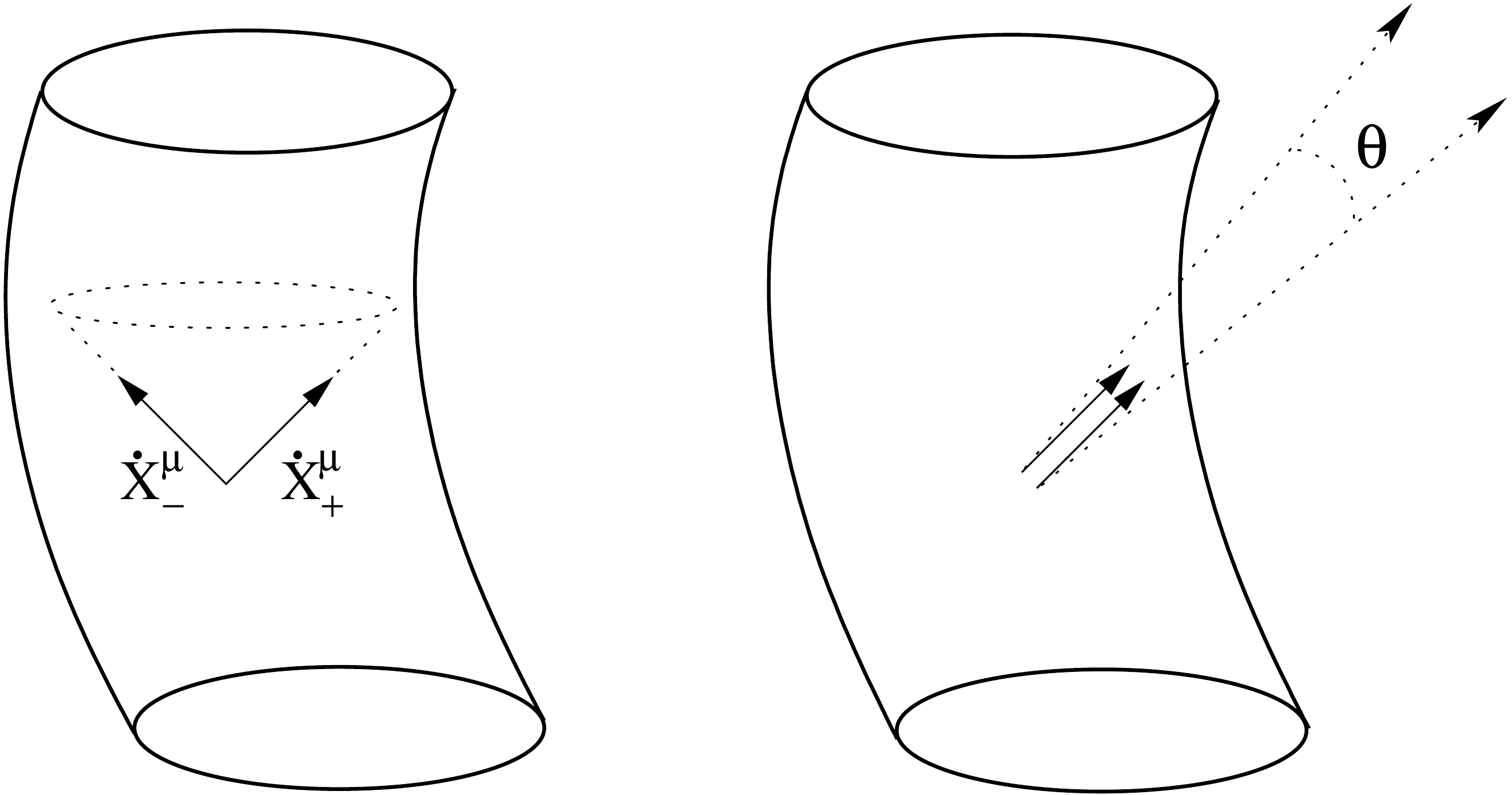}
\caption{Left: The lightcone generically intersects the worldsheet into two separate null
directions corresponding to the velocity of the left and right movers. Right: When a cusp occurs, the 
two null vectors are parallel and the worldsheet is tangent to the lightcone.  There results a large
 burst of outgoing gravitational radiation.}
\label{GWBurst}
\end{figure}
Now, the main point is that an intersection between the two curves represents a {\it cusp}.  More technically, 
such an intersection corresponds to particular points on the string worldsheet at which the two null (see (\ref{vir}))
tangent vectors $\dot{X}_+^{\mu}$ and $\dot{X}_-^{\mu}$ are parallel in spacetime. In general, the string worldsheet 
intersects locally the light cone along the
two distinct directions $\dot{X}_+^{\mu}$ and $\dot{X}_-^{\mu}$.  The cusps are special points where the 
worldsheet is {\it tangent to the lightcone} (see FIG. \ref{GWBurst}).  This is a singularity of the classical 
worldsheet at which a strong gravitational wave signal is emitted along the common null vector. Let us now 
indicate how one computes the emission of
gravitational wave bursts from cuspy strings. We consider Einstein's theory in the linearized approximation,
\be
g_{\mu \nu}\lb x \rb = \eta_{\mu \nu}+h_{\mu \nu}\lb x \rb.
\ee
We use the harmonic gauge, $\partial^{\nu}\bar{h}_{\mu \nu}=0$, so that Einstein's equations simplify to
\be
\label{boxh}
\Box\bar{h}_{\mu \nu}=-16 \pi G T_{\mu \nu}\lb x \rb,
\ee
where
\be
\bar{h}_{\mu \nu}=h_{\mu \nu}-\frac{1}{2}h \eta _{\mu \nu}.
\ee
The stress-energy tensor is obtained by differentiating the Nambu action w.r.t. $g_{\mu \nu}$. Taking its Fourier transform, one finds
\be
T^{\mu \nu}\lb k^{\lambda} \rb=\frac{\mu}{T_{\ell}}\int_{\Sigma_{\ell}}\mrm{d}\tau \mrm{d}\sigma \dot{X}_+^{\left( \mu \right.}
 \dot{X}_-^{\left. \nu \right)} e^{-\frac{i}{2}k.\lb X_+ + X_-\rb}
\ee
where $(\mu \nu)$ indicates symmetrization over the indices  $\mu,\nu$,
and where, in the exponential, we have replaced $X^{\mu}$ by the half-sum of the left and right movers.  
The fundamental period of a loop of length $\ell$ is $T_{\ell}=\frac{\ell}{2}$.   Note that $\ell$ is 
the invariant total length  $\frac{E_0}{\mu}$, where $\mu$ is the string tension.  In string theory, one usually uses 
a worldsheet gauge where $\ell$ is either $1$ or $2 \pi$ but here one finds it more convenient  
to use a gauge where $\sigma$ and $\tau$ are connected to an external definition of time (namely $X^0 = x^0=\tau$).

We wish to compute the integral giving $T^{\mu \nu}\lb k^{\lambda} \rb$  over a periodic domain $\Sigma_{\ell}$ 
in the $\tau$ , $\sigma$ plane. We can rewrite the integral as an integral over $\mrm{d} \sigma_+ \mrm{d} \sigma_{-}$.  
This yields the famous left-right factorization
of closed string amplitudes\footnote{Though we are doing here a classical calculation,
one recognizes that the result is given by the graviton vertex operator.} and the 
Fourier transform of the string stress-energy tensor  reads
\be
T^{\mu \nu}\lb k  \rb = \frac{\mu}{\ell}I_+^{\left(\mu\right.}I_-^{\left. \nu \right)},
\ee
where
\be
I_{\pm}^{\mu}=\int_0^{\ell} d\mrm{\sigma}_{\pm}\dot{X}_{\pm}^{\mu}\lb
\sigma_{\pm} \rb e^{-\frac{i}{2}k.X_{\pm}}.
\ee
By solving Einstein's equation (\ref{boxh}), one finds that the spacetime-Fourier transform of
the source on the r.h.s. actually gives the time-Fourier transform of the asymptotic
GW amplitude (emitted in the direction $n^i =k^i/k^0$), \ie the time-Fourier transform of
 the quantity $\kappa_{\mu \nu} \lb t-r,\vec{n}\rb$ appearing in the asymptotic
expansion 
\be
\bar{h}_{\mu \nu}\lb t,\vec{x} \rb=\frac{\kappa_{\mu \nu} \lb t-r,\vec{n}\rb}{r}+\mathcal{O}\lb \frac{1}{r^2}\rb.
\ee
Here the $1/r$ decrease in amplitude as a function of distance away from the string is caused by the retarded 
Green's function in 3+1 dimensions. $\kappa_{\mu \nu}$ is a function of both the time variable and the angle of emission.   
As we just said, the time-Fourier transform of   $\kappa^{\mu \nu}$
is proportional to the spacetime-Fourier transform $ T^{\mu \nu}\lb k  \rb$ of the source,
and is explicitly given by
\be
\begin{array}{ccc}
\displaystyle \kappa_{\mu \nu}\lb f,\vec{n}\rb & \displaystyle = & \displaystyle \left| f \right| 
\int\mrm{d}t e^{2 \pi i f \lb t-r \rb} \kappa^{\mu \nu}\lb t-r,\vec{n} \rb \\
\displaystyle & \displaystyle = & \displaystyle 2 G \mu \left| f \right|I_+^{\left( \mu \right.}\lb \omega,\omega \vec{n} \rb I_-^{\left. \nu \right)}\lb \omega,\omega \vec{n} \rb.
\end{array}
\ee
This formula shows that we
can compute  what is observed in a GW detector as a function of string tension,  frequency, and the
product of two integrals involving left and right moving modes. 

We can then estimate the generic features of the
 GW burst emitted by a cusp by noticing that, in the Fourier domain,
each integral $I^{\mu}_{\pm}$ is dominated (when considering large frequencies: $f \gg T_{\ell}^{-1}$) by the singular
behaviour of the two integrands $ \dot{X}_{\pm}^{\mu}\lb
\sigma_{\pm} \rb e^{-\frac{i}{2}k.X_{\pm}}$ near a cusp. The calculation proceeds by (Taylor) 
expanding the vectors $X_{\pm}^{\mu}$ and $\dot{X}_{\pm}^{\mu}$ in powers of ${\sigma}_{\pm}$.    
One finds that the first 
few leading terms in this expansion can be gauged away,  so that the signal amplitude 
is much smaller than what could have (and had) been initially thought. After Fourier transforming back to the
time domain, it is finally found that \cite{Cusps}
\be
\begin{array}{ccc}
\displaystyle \kappa\lb t \rb & \displaystyle \propto & \displaystyle \left| t-t_c \right| ^{1/3},\\
\displaystyle \ddot{\kappa} \lb t \rb & \displaystyle \propto & \displaystyle \left| t-t_c \right| ^{-5/3} \, .
\end{array}
\ee
As this result seems to crucially depend on the presence of a mathematically singular
behaviour of the {\it classical} string worldsheet at a cusp, one might worry that
quantum effects could blur away the sharp cusp, and make the above classical burst
signal disappear. It was checked that this is not the case \cite{CD2006} (the basic reason
being that, finally, the strong GW burst signal is emitted by a large segment of the
string around the cusp).

\subsubsection{Gravitational waves from a cosmological string network}

In order to understand the observational signature of a cosmic string network and not just a 
single string, one must combine the analysis of the previous section with the cosmological expansion of a 
Friedman-Lema\^itre universe and with an integration over redshift.  A crucial point is then to 
estimate the number density of string loops. This density can be analytically estimated as a 
function of the string parameters, such as the reconnection probability $p$, and the string 
tension $G \mu$.   Note that the reconnection probability is expected to be quite different for cosmic supertrings
compared to the traditionally considered field-theory strings. Field theory strings are expected
to reconnect, when they cross, with essentially unit probability ($p\simeq 1$), while
fundamental or $D$-strings are expected to reconnect with a smallish probability,
$ 10^{-3} < p < 1$ \cite{Jackson:2004zg}
(because of the presence of the string coupling, and other factors).

The loop number density can be approximately estimated as \cite{DV2005}
\be
n_{\ell} \sim \frac{1}{p\, 50\, G \mu \, t^3}+\dots
\label{LoopNumberEq}
\ee
where the first term on the r.h.s. comes from loops that were created at redshifts $ \leq 1$, 
while the `$\dots$' denotes a possible additional contribution from high-redshift strings.
When the loop-size parameter $\alpha$ is smaller or equal to the ``traditionally expected''
value $50 G \mu$, the contribution from high-redshift strings is negligible (because
strings decay in less than a Hubble time). By contrast, when  $\alpha \gg 50 G \mu$,
the strings survive over many Hubble times, and the contribution of high-redshift strings starts to
dominate the loop density. Note the somewhat unexpected feature displayed by the first
term in $n_{\ell}$, namely that it increases both as $G \, \mu$ and/or $p$ are decreased. This feature
is one of the features which allow GW signals from strings to be detected down to
very small values of the string tension (contrary to CMB effects). Indeed, as  $G \, \mu$
is decreased, though each individual string signal will decrease proportionally to $G \, \mu$,
there will be more emitting loops. After integrating over redhifts, one finds that the
observable signal is a complicated, {\it non monotonic} function of $G \, \mu$.   The numerical 
estimates of Ref.~\cite{DV2005} considered the case in which the loop size parameter  $\alpha < 50 G \mu$, 
in which case the first term in (\ref{LoopNumberEq}) is dominant.  If, on the other hand,  one 
assumes  $\alpha \sim 0.1$ (as is suggested by some numerical simulations \cite{VOV0506}) , strings 
survive longer so that higher redshift contributions are non-negligible.   It has been found that 
in such cases these contributions increase the number of loops (which increases the GW signal) 
but tend to drown the cusp signal within the quasi-Gaussiam random-mean-square GW background \cite{H2006}.

Based on current detector capabilities and on the sensitivity estimates for future detectors, 
one finds that if $\alpha \leq 50 G \mu$, LIGO could detect $G \, \mu \ge 10^{-12}$ while LISA 
could detect $G \, \mu \ge 10^{-14}$. On the other hand, if $\alpha \gg 50 G \mu$ LISA could 
reach $G \, \mu \ge 10^{-16}$. One has looked in the current LIGO data for the possible presence of a background
of GW's, but without success so far  \cite{Abbott:2006zx}.
The best current bound on  $G \, \mu$ comes from pulsar timing \cite{Jenet:2006sv} and  is roughly at the  $G \, \mu \leq  10^{-9}$ level 
(which is about three orders of magnitude more stringent than the limits than
can be obtained from CMB data).

Gravitational wave detectors are thus excellent probes of cosmic (super)strings.  There is therefore the 
possibility that they could confirm or refute KKLMMT-type scenarios in a large domain of parameter space.  
However, there are large uncertainties in string network dynamics which prevent one from
being able to make reliable analytical estimates. If one is in a region of parameter space
where the rather specific cusp-related signals are well above the r.m.s. background one might
find rather direct experimental evidence for the existence of cosmic strings.  There would however 
remain the task of discriminating between string theoretic strings and field theoretic ones.  
One way would be (assuming one could strongly reduce the string network uncertainties)
to determine the reconnection probability $p$ from its influence on the loop number density,
and, thereby, on the recurrence rate of observed signals.  Another more ambitious possibility would be to 
exploit the presence of two populations of strings, namely D and F strings, in D-brane anti D-brane 
annihilation, and attempt to measure two different values of $G \, \mu$, the ratio of which satisfies $\mu _D=\mu_F /g_s$.

\section{Conclusion}

We hope that these lectures have shown that gravity phenomenology is a potentially
interesting arena for eventually confronting string theory to reality.

\section*{Acknowledgement}
TD wishes to thank the organizers of this Les Houches session 
for putting together  a  timely and interesting programme. He is especially grateful
to Nima Arkani-Hamed, Michael Douglas, Igor Klebanov and Eliezer Rabinovici for informative
discussions.  ML wishes to thank the organizers of the school and the lecturers for a very 
stimulating time in les Houches.


\begin{thebibliography}{1}
\bibitem{P1969}
\bauthor{\fnm{R.} \snm{Penrose}},
\bseries{\btitle{Riv. Nuov. Cimento. } \bvolumeno{1}}\bdate{1969}
\bfirstpage{252}.
\bibitem{C1970}
\bauthor{\fnm{D.} \snm{Christodoulou}},
\bseries{\btitle{Phys. Rev. Lett.} \bvolumeno{25}}\bdate{1970}
\bfirstpage{1596}.
\bibitem{CR1971}
\bauthor{\fnm{D.} \snm{Christodoulou}}, \bauthor{\fnm{R.} \snm{Ruffini}}
\bseries{\btitle{Phys. Rev. D} \bvolumeno{4}}\bdate{1971}
\bfirstpage{3552}.
\bibitem{H1971}
\bauthor{\fnm{S.W.} \snm{Hawking}},
\bseries{\btitle{Phys. Rev. Lett.} \bvolumeno{26}}\bdate{1971}
\bfirstpage{1344}.
\bibitem{BCH1973}
\bauthor{\fnm{J.} \snm{Bardeen}}, \bauthor{\fnm{B.} \snm{Carter}} and \bauthor{\fnm{S.W.} \snm{Hawking}},
\bseries{\btitle{Comm. Math. Phys.} \bvolumeno{31}}\bdate{1973}
\bfirstpage{161}.
\bibitem{HH1972}
\bauthor{\fnm{S.W.} \snm{Hawking}}, and \bauthor{\fnm{J.B.} \snm{Hartle}},
\bseries{\btitle{Comm. Math. Phys.} \bvolumeno{27}}\bdate{1972}
\bfirstpage{283}.
\bibitem{HR1973}
\bauthor{\fnm{R.S.} \snm{Hanni}}, and \bauthor{\fnm{R.} \snm{Ruffini}},
\bseries{\btitle{Phys. Rev. D} \bvolumeno{8}}\bdate{1973}
\bfirstpage{3259}.
\bibitem{D1978}
\bauthor{\fnm{T.} \snm{Damour}},
\bseries{\btitle{Phys. Rev. D} \bvolumeno{18}}\bdate{1978}
\bfirstpage{3598}.
\bibitem{D1979}
T. Damour,
in: ``Quelques propri\'et\'es m\'ecaniques, \'electromagn\'etiques, thermodynamiques
et quantiques des trous noirs''; Th\`ese de Doctorat d'Etat, Universit\'e Pierre et Marie Curie, Paris VI, 1979.
available (see files these1.pdf to these6.pdf) on http://www.ihes.fr/\~{}damour/Articles/
\bibitem{D1982}
T. Damour,
in: ``Surface Effects in Black Hole Physics'';
Proceedings of the Second Marcel Grossmann Meeting on General Relativity,
(edited by R. Ruffini, North Holland, 1982) pp 587-608;
available (see file surfaceeffects.pdf) on 
 http://www.ihes.fr/\~{}damour/Articles/
\bibitem{Z1978}
\bauthor{\fnm{R.L.} \snm{Znajek}},
\bseries{\btitle{MNRAS} \bvolumeno{185}}\bdate{1978}
\bfirstpage{833}.
\bibitem{Thorne:1986iy}
  K.~S.~.~Thorne, R.~H.~.~Price and D.~A.~.~Macdonald,
 ``Black Holes: The Membrane Paradigm,''
{\it  New Haven, USA: Yale Univ. Press (1986) 367p}.
\bibitem{KSS2005}
\bauthor{\fnm{P.K.} \snm{Kovtun}}, \bauthor{\fnm{D.T.} \snm{Son}} and \bauthor{\fnm{A.O.} \snm{Starinets}},
\bseries{\btitle{Phys. Rev. Lett.} \bvolumeno{94}}\bdate{2005}
\bfirstpage{111601}.
\bibitem{Son:2007vk}
  D.~T.~Son and A.~O.~Starinets,
  Ann.\ Rev.\ Nucl.\ Part.\ Sci.\  {\bf 57} (2007) 95
  [arXiv:0704.0240 [hep-th]].
\bibitem{C1973}
 B.~Carter,
 in: \textit{Black Holes},
 Proceedings of 1972 Les Houches Summer School,
(edited by C.~DeWitt and B.~S.~DeWitt, Gordon and Breach, NY, 1973).
\bibitem{B1973}
\bauthor{\fnm{J.} \snm{Bekenstein}},
\bseries{\btitle{Phys. Rev. D} \bvolumeno{7}}\bdate{1973}
\bfirstpage{2333}.
\bibitem{H1975}
\bauthor{\fnm{S.W.} \snm{Hawking}},
\bseries{\btitle{Comm. Math. Phys.} \bvolumeno{43}}\bdate{1975}
\bfirstpage{199}.
\bibitem{DR1976}
\bauthor{\fnm{T.} \snm{Damour}}, and \bauthor{\fnm{R.} \snm{Ruffini}},
\bseries{\btitle{Phys. Rev. D} \bvolumeno{14}}\bdate{1976},
\bfirstpage{332}.
\bibitem{Gourgoulhon:2005ch}
  E.~Gourgoulhon,
  Phys.\ Rev.\  D {\bf 72}, 104007 (2005)
  [arXiv:gr-qc/0508003].
\bibitem{Gourgoulhon:2005ng}
  E.~Gourgoulhon and J.~L.~Jaramillo,
  Phys.\ Rept.\  {\bf 423}, 159 (2006)
  [arXiv:gr-qc/0503113].
\bibitem{U2003}
\bauthor{\fnm{J.P.} \snm{Uzan}},
\bseries{\btitle{Rev. Mod. Phys.} \bvolumeno{75}}\bdate{2003}
\bfirstpage{403}.
\bibitem{DD1996}
\bauthor{\fnm{T.} \snm{Damour}}, and \bauthor{\fnm{F.} \snm{Dyson}},
\bseries{\btitle{Nucl. Phys. B} \bvolumeno{480}}\bdate{1996}
\bfirstpage{37}.
\bibitem{S1983}
\bauthor{\fnm{A.I.} \snm{Shlyakhter}},
in: ATOMKI Report A/1 (Debrecen, Hungary) 1983.
\bibitem{F2000}
\bauthor{\fnm{Y.} \snm{Fujii}},
\bseries{\btitle{Nucl. Phys. B} \bvolumeno{573}}\bdate{2000}
\bfirstpage{337}.
\bibitem{O2004}
\bauthor{\fnm{K.A.} \snm{Olive}}, \bauthor{\fnm{M.} \snm{Pospelov}}, \bauthor{\fnm{Y.-Z.} \snm{Qian}}, 
\bauthor{\fnm{G.} \snm{Manhes}}, \bauthor{\fnm{E.} \snm{Vangioni-Flam}}, \bauthor{\fnm{A.} \snm{Coc}}, and \bauthor{\fnm{M.} \snm{Casse}},
\bseries{\btitle{Phys. Rev. D} \bvolumeno{69}}\bdate{2004}
\bfirstpage{027701}.
\bibitem{clocks}
 S.~Bize {\it et al.},
  Phys.\ Rev.\ Lett.\  {\bf 90}, 150802 (2003)
H.~Marion {\it et al.},
  Phys.\ Rev.\ Lett.\  {\bf 90}, 150801 (2003);
M.~Fischer {\it et al.},
  Phys.\ Rev.\ Lett.\  {\bf 92}, 230802 (2004).
\bibitem{Adelberger}
\bauthor{\fnm{C.D.} \snm{Hoyle}}, \bauthor{\fnm{U.} \snm{Schmidt}}, \bauthor{\fnm{B.R.} \snm{Heckel}}, \bauthor{\fnm{E.G.} 
\snm{Adelberger}}, \bauthor{\fnm{J.H.} \snm{Gundlach}}, \bauthor{\fnm{D.J.} \snm{Kapner}}, and \bauthor{\fnm{H.E.} \snm{Swanson}}
\bseries{\btitle{Phys. Rev. Lett.} \bvolumeno{86}}\bdate{2001}
\bfirstpage{1418};
\bauthor{\fnm{E.G.} \snm{Adelberger}}, \bauthor{\fnm{B.R.} \snm{Heckel}}, and  \bauthor{\fnm{A.E.} \snm{Nelson}},
\bseries{\btitle{Ann. Rev. Nucl. Part. Sci.} \bvolumeno{53}}\bdate{2003}
\bfirstpage{77};
\bauthor{\fnm{C.D.} \snm{Hoyle}}, \bauthor{\fnm{D.J.} \snm{Kapner}},\bauthor{\fnm{E.G.} \snm{Adelberger}}, \bauthor{\fnm{J.H.} 
\snm{Gundlach}}, \bauthor{\fnm{U.} \snm{Schmidt}}, and \bauthor{\fnm{H.E.} \snm{Swanson}},
\bseries{\btitle{Phys. Rev. D} \bvolumeno{70}}\bdate{2004}
\bfirstpage{042004};
\bauthor{\fnm{D.J.} \snm{Kapner}}, \bauthor{\fnm{T.S.} \snm{Cook}}, \bauthor{\fnm{E.G.} \snm{Adelberger}}, \bauthor{\fnm{J.H.} 
\snm{Gundlach}}, \bauthor{\fnm{B.R.} \snm{Heckel}}\bauthor{\fnm{C.D.} \snm{Hoyle}}, and \bauthor{\fnm{H.E.} \snm{Swanson}},
\bseries{\btitle{Phys. Rev. Lett.} \bvolumeno{98}}\bdate{2007}
\bfirstpage{021101}.
\bibitem{Williams:2004qba}
  J.~G.~Williams, S.~G.~Turyshev and D.~H.~Boggs,
  Phys.\ Rev.\ Lett.\  {\bf 93}, 261101 (2004)
  [arXiv:gr-qc/0411113].
\bibitem{VL1979}
\bauthor{\fnm{R.F.C.} \snm{Vessot}}, \bauthor{\fnm{M.W.A.} \snm{Levine}},
\bseries{\btitle{General Relativity and Gravitation} \bvolumeno{10}}\bdate{1979}
\bfirstpage{181}.
\bibitem{W2001}
\bauthor{\fnm{C.} \snm{Will}},
\bseries{\btitle{Living Rev. Relativity} \bvolumeno{4}}\bdate{2001}.
\bibitem{Dpdg}
See chapter 18 (``Experimental tests of gravitational theory'', by T. Damour) in
W.-M.Yao et al. (Particle Data Group), J. Phys. G 33, 1 (2006);
a 2007 partial update for the 2008 edition is available on the PDG WWW pages (URL: http://pdg.lbl.gov/).
 \bibitem{DEath75}
  P.~D.~D'Eath,
  Phys.\ Rev.\  D {\bf 12}, 2183 (1975).
\bibitem{DamourLH83}
T. Damour,
Gravitational radiation and the motion of compact bodies.
 in {\it Gravitational Radiation}, 
 edited by N. Deruelle and T. Piran, North-Holland, Amsterdam, 1983, pp 59-144.
\bibitem{Damour:2001bu}
  T.~Damour, P.~Jaranowski and G.~Schafer,
  Phys.\ Lett.\  B {\bf 513}, 147 (2001)
  [arXiv:gr-qc/0105038].
\bibitem{Blanchet:2003gy}
  L.~Blanchet, T.~Damour and G.~Esposito-Farese,
  Phys.\ Rev.\  D {\bf 69}, 124007 (2004)
  [arXiv:gr-qc/0311052].
\bibitem{BIT2003}
\bauthor{\fnm{B.} \snm{Bertotti}}, \bauthor{\fnm{L.} \snm{Iess}}, and \bauthor{\fnm{P.} \snm{Tortora}},
\bseries{\btitle{Nature} \bvolumeno{425}}\bdate{2003}
\bfirstpage{374}.
\bibitem{Nordtvedt:1968qr}
  K.~Nordtvedt,
  Phys.\ Rev.\  {\bf 169} (1968) 1014.
\bibitem{Nordtvedt:1968qs}
  K.~Nordtvedt,
  Phys.\ Rev.\  {\bf 169} (1968) 1017.
\bibitem{Damour:2007ti}
  T.~Damour,
  arXiv:0705.3109 [gr-qc].
\bibitem{DEF1996}
\bauthor{\fnm{T.} \snm{Damour}}, and \bauthor{\fnm{G.} \snm{Esposito-Farese}},
\bseries{\btitle{Phys. Rev. D} \bvolumeno{54}}\bdate{1996}
\bfirstpage{54}.
\bibitem{Dar92}
A. Dar, Nucl. Phys. (Proc. Supp.) {\bf B28},321(1992).
\bibitem{Jacobson}
T.~Jacobson, S.~Liberati and D.~Mattingly,
  Annals Phys.\  {\bf 321}, 150 (2006).
\bibitem{AmelinoCamelia:2004hm}
  G.~Amelino-Camelia,
  Lect.\ Notes Phys.\  {\bf 669}, 59 (2005)
  [arXiv:gr-qc/0412136].
\bibitem{Gimon:2007ur}
  E.~G.~Gimon and P.~Horava,
  arXiv:0706.2873 [hep-th].
\bibitem{AAHD1998}
\bauthor{\fnm{I.} \snm{Antoniadis}},  \bauthor{\fnm{N.} \snm{Arkani-Hamed}}, 
\bauthor{\fnm{S.} \snm{Dimopoulos}}, and \bauthor{\fnm{G.} \snm{Dvali}},
\bseries{\btitle{Phys. Lett. B} \bvolumeno{436}}\bdate{1998}
\bfirstpage{257}.
\bibitem{RS1999}
\bauthor{\fnm{L.} \snm{Randall}}, and \bauthor{\fnm{R.} \snm{Sundrum}},
\bseries{\btitle{Phys. Rev. Lett.} \bvolumeno{83}}\bdate{1999}
\bfirstpage{3370}.
\bibitem{DGP2000}
\bauthor{\fnm{G.} \snm{Dvali}}, \bauthor{\fnm{G.} \snm{Gabadadze}}, and \bauthor{\fnm{M.} \snm{Porrati}},
\bseries{\btitle{Mod. Phys. Lett.} \bvolumeno{83}}\bdate{2000}
\bfirstpage{1717}.
\bibitem{DGZ2003}
\bauthor{\fnm{G.} \snm{Dvali}}, \bauthor{\fnm{A.} \snm{Gruzinov}}, and \bauthor{\fnm{M.} \snm{Zaldarriaga}},
\bseries{\btitle{Mod. Phys. D} \bvolumeno{68}}\bdate{2003}
\bfirstpage{024012}.
\bibitem{AAHDNR2006}
A.~Adams, N.~Arkani-Hamed, S.~Dubovsky, A.~Nicolis and R.~Rattazzi,
  JHEP {\bf 0610}, 014 (2006)
  [arXiv:hep-th/0602178].
\bibitem{MultiBranes}
\bauthor{\fnm{I.I.} \snm{Kogan}}, \bauthor{\fnm{S.} \snm{Mouslopoulos}}, \bauthor{\fnm{A.} \snm{Papazoglou}}, 
\bauthor{\fnm{G.G.} \snm{Ross}}, \bauthor{\fnm{J.} \snm{Santiago}},
\bseries{\btitle{Nucl. Phys. B} \bvolumeno{584}}\bdate{2000}
\bfirstpage{313};
\bauthor{\fnm{R.} \snm{Gregory}}, \bauthor{\fnm{V.A.} \snm{Rubakov}}, and \bauthor{\fnm{S.M} \snm{Sibiryakov}},
\bseries{\btitle{Phys. Rev. Lett} \bvolumeno{84}}\bdate{2000}
\bfirstpage{5928};
  T.~Damour and I.~I.~Kogan,
  Phys.\ Rev.\  D {\bf 66}, 104024 (2002)
  [arXiv:hep-th/0206042].
\bibitem{Damour:2002gp}
  T.~Damour, I.~I.~Kogan and A.~Papazoglou,
  Phys.\ Rev.\  D {\bf 67}, 064009 (2003)
  [arXiv:hep-th/0212155].
\bibitem{Douglas:2006es}
  M.~R.~Douglas and S.~Kachru,
  Rev.\ Mod.\ Phys.\  {\bf 79}, 733 (2007)
  [arXiv:hep-th/0610102].
\bibitem{Rabinovici:2007hz}
  E.~Rabinovici,
  arXiv:0708.1952 [hep-th].
\bibitem{DP1994}
\bauthor{\fnm{T.} \snm{Damour}}, and \bauthor{\fnm{A.M.} \snm{Polyakov}},
\bseries{\btitle{Nucl. Phys. B} \bvolumeno{423}}\bdate{1994}
\bfirstpage{532}.
\bibitem{DPV2002}
\bauthor{\fnm{T.} \snm{Damour}}, \bauthor{\fnm{F.} \snm{Piazza}}, and \bauthor{\fnm{G.} \snm{Veneziano}},
\bseries{\btitle{Phys. Rev. D} \bvolumeno{66}}\bdate{2002}
\bfirstpage{046007};
\bauthor{\fnm{T.} \snm{Damour}}, \bauthor{\fnm{F.} \snm{Piazza}}, and \bauthor{\fnm{G.} \snm{Veneziano}},
\bseries{\btitle{Phys. Rev. Lett.} \bvolumeno{89}}\bdate{2002}
\bfirstpage{081601}.
\bibitem{SW1997}
\bauthor{\fnm{B.D.} \snm{Serot}}, and \bauthor{\fnm{J.D.} \snm{Walecka}},
\bseries{\btitle{Int. J. Mod. Phys. E} \bvolumeno{6}}\bdate{1997}
\bfirstpage{515}.
\bibitem{D2006}
\bauthor{\fnm{J.F.} \snm{Donoghue}},
\bseries{\btitle{Phys. Rev. C} \bvolumeno{74}}\bdate{2006}
\bfirstpage{515}.
\bibitem{Damour:2007uv}
  T.~Damour and J.~F.~Donoghue,
  arXiv:0712.2968 [hep-ph].
\bibitem{Dimopoulos:2006nk}
  S.~Dimopoulos, P.~W.~Graham, J.~M.~Hogan and M.~A.~Kasevich,
  Phys.\ Rev.\ Lett.\  {\bf 98}, 111102 (2007)
  [arXiv:gr-qc/0610047].
\bibitem{Khoury:2003aq}
  J.~Khoury and A.~Weltman,
  Phys.\ Rev.\ Lett.\  {\bf 93}, 171104 (2004)
  [arXiv:astro-ph/0309300].
\bibitem{DM1998}
\bauthor{\fnm{T.} \snm{Damour}}, and \bauthor{\fnm{V.F.} \snm{Mukhanov}},
\bseries{\btitle{Phys. Rev. Lett.} \bvolumeno{80}}\bdate{1998}
\bfirstpage{3440}.
\bibitem{APDM1999}
\bauthor{\fnm{C.} \snm{Armendariz-Picon}}, \bauthor{\fnm{T.} \snm{Damour}}, and \bauthor{\fnm{V.F.} \snm{Mukhanov}},
\bseries{\btitle{Phys. Lett. B} \bvolumeno{458}}\bdate{1999}
\bfirstpage{209}.
\bibitem{AHCMZ2004}
\bauthor{\fnm{N.} \snm{Arkani-Hamed}}, \bauthor{\fnm{P.} \snm{Creminelli}},  \bauthor{\fnm{S.} \snm{Mukohyama}}, 
and \bauthor{\fnm{M.} \snm{Zaldarriaga}},
\bseries{\btitle{JCAP} \bvolumeno{0404}}\bdate{2004}
\bfirstpage{001}.
\bibitem{ST2004}
\bauthor{\fnm{E.} \snm{Silverstein}}, and \bauthor{\fnm{D.} \snm{Tong}},
\bseries{\btitle{Phys. Rev. D} \bvolumeno{70}}\bdate{2004}
\bfirstpage{103505}.
\bibitem{AST2004}
\bauthor{\fnm{M.} \snm{Alishahiha}}, \bauthor{\fnm{E.} \snm{Silverstein}}, and \bauthor{\fnm{D.} \snm{Tong}},
\bseries{\btitle{Phys. Rev. D} \bvolumeno{70}}\bdate{2004}
\bfirstpage{123505}.
\bibitem{W1985}
\bauthor{\fnm{E.} \snm{Witten}},
\bseries{\btitle{Nucl. Phys. B} \bvolumeno{249}}\bdate{1985}
\bfirstpage{557}.
\bibitem{KKLMMT2003}
\bauthor{\fnm{S.} \snm{Kachru}}, \bauthor{\fnm{R.} \snm{Kallosh}}, and \bauthor{\fnm{A.} \snm{Linde}}, \bauthor{\fnm{J.} 
\snm{Maldacena}}, \bauthor{\fnm{L.} \snm{McAllister}}, and \bauthor{\fnm{S.P.} \snm{Trivedi}},
\bseries{\btitle{JCAP} \bvolumeno{0310}}\bdate{2003}
\bfirstpage{013}.
\bibitem{ST2002}
\bauthor{\fnm{S.} \snm{Sarangi}}, and \bauthor{\fnm{S.-H.} \snm{Tye}}
\bseries{\btitle{Phys. Lett. B} \bvolumeno{536}}\bdate{2002}
\bfirstpage{185}.
\bibitem{DV2004}
\bauthor{\fnm{G.} \snm{Dvali}}, and \bauthor{\fnm{A.} \snm{Vilenkin}}
\bseries{\btitle{JCAP} \bvolumeno{0403}}\bdate{2004}
\bfirstpage{010}.
\bibitem{CMP2004}
\bauthor{\fnm{E.J.} \snm{Copeland}}, \bauthor{\fnm{R.C.} \snm{Myers}}, and \bauthor{\fnm{J.} \snm{Polchinski}},
\bseries{\btitle{JHEP} \bvolumeno{0406}}\bdate{2004}
\bfirstpage{013}.
\bibitem{DT1999}
\bauthor{\fnm{G.} \snm{Dvali}}, and \bauthor{\fnm{S.-H} \snm{Tye}},
\bseries{\btitle{Phys. Lett. B} \bvolumeno{450}}\bdate{1999}
\bfirstpage{72}.
\bibitem{V1981}
\bauthor{\fnm{A.} \snm{Vilenkin}},
\bseries{\btitle{Phys. Rev. D} \bvolumeno{23}}\bdate{1981}
\bfirstpage{852}.
\bibitem{VilenkinBook}
A. Vilenkin and E. P. S. Shellard,
{\it Cosmic Strings and Other Topological Defects}
Cambridge University Press, Cambridge, 2000.
\bibitem{Cusps}
\bauthor{\fnm{T.} \snm{Damour}}, and \bauthor{\fnm{A.} \snm{Vilenkin}},
\bseries{\btitle{Phys. Rev. Lett.} \bvolumeno{85}}\bdate{2000}
\bfirstpage{3761};
\bauthor{\fnm{T.} \snm{Damour}}, and \bauthor{\fnm{A.} \snm{Vilenkin}},
\bseries{\btitle{Phys. Rev. D} \bvolumeno{64}}\bdate{2001}
\bfirstpage{064008}.
\bibitem{VOV0506}
\bauthor{\fnm{V.} \snm{Vanchurin}}, \bauthor{\fnm{K.} \snm{Olum}}, and \bauthor{\fnm{A.} \snm{Vilenkin}},
\bseries{\btitle{Phys. Rev. D} \bvolumeno{72}}\bdate{2005}
\bfirstpage{063514};
\bauthor{\fnm{V.} \snm{Vanchurin}}, \bauthor{\fnm{K.} \snm{Olum}}, and \bauthor{\fnm{A.} \snm{Vilenkin}},
\bseries{\btitle{Phys. Rev. D} \bvolumeno{74}}\bdate{2006}
\bfirstpage{063527}.
\bibitem{MS2006}
\bauthor{\fnm{C.} \snm{Martins}}, and \bauthor{\fnm{E.} \snm{Shellard}},
\bseries{\btitle{Phys. Rev. D} \bvolumeno{73}}\bdate{2006}
\bfirstpage{043515}.
\bibitem{PR2007}
\bauthor{\fnm{J.} \snm{Polchinski}}, and \bauthor{\fnm{J.V.} \snm{Rocha}},
\bseries{\btitle{Phys. Rev. D} \bvolumeno{75}}\bdate{2007}
\bfirstpage{123503}.
\bibitem{DR2007}
\bauthor{\fnm{F.} \snm{Dubath}}, and \bauthor{\fnm{J.V.} \snm{Rocha}},
\bseries{\btitle{Phys. Rev. D} \bvolumeno{76}}\bdate{2007}
\bfirstpage{024001}.
\bibitem{KT1982}
\bauthor{\fnm{T.W.B.} \snm{Kibble}}, and \bauthor{\fnm{N.} \snm{Turok}},
\bseries{\btitle{Phys. Lett. B} \bvolumeno{116}}\bdate{1982}
\bfirstpage{141}.
\bibitem{T1984}
\bauthor{\fnm{N.} \snm{Turok}},
\bseries{\btitle{Nucl. Phys. B} \bvolumeno{242}}\bdate{1984}
\bfirstpage{520}.
\bibitem{CD2006}
\bauthor{\fnm{D.} \snm{Chialva}}, and \bauthor{\fnm{T.} \snm{Damour}},
\bseries{\btitle{JCAP} \bvolumeno{0608}}\bdate{2006}
\bfirstpage{003}.
\bibitem{Jackson:2004zg}
  M.~G.~Jackson, N.~T.~Jones and J.~Polchinski,
  JHEP {\bf 0510}, 013 (2005)
  [arXiv:hep-th/0405229].
\bibitem{DV2005}
\bauthor{\fnm{T.} \snm{Damour}}, and \bauthor{\fnm{A.} \snm{Vilenkin}},
\bseries{\btitle{Phys. Rev. D} \bvolumeno{71}}\bdate{2005}
\bfirstpage{063510}.
\bibitem{H2006}
\bauthor{\fnm{C.J.} \snm{Hogan}},
\bseries{\btitle{Phys. Rev. D} \bvolumeno{74}}\bdate{2006}
\bfirstpage{043526}.
\bibitem{Abbott:2006zx}
  B.~Abbott {\it et al.}  [LIGO Collaboration],
  Astrophys.\ J.\  {\bf 659}, 918 (2007)
  [arXiv:astro-ph/0608606].
\bibitem{Jenet:2006sv}
  F.~A.~Jenet {\it et al.},
  Astrophys.\ J.\  {\bf 653}, 1571 (2006)
  [arXiv:astro-ph/0609013].
\end{thebibliography}
\end{document}